\documentclass[a4paper,12pt,english]{article}
\pdfoutput=1
\usepackage[bindingoffset=0.3cm,textheight=22.5cm,hdivide={2.7cm,*,2.7cm}, vdivide={*,22cm,*}]{geometry}
\usepackage[bookmarksnumbered=true,breaklinks=true]{hyperref}
\usepackage{cite}
\usepackage{amsmath,amsfonts,amssymb,babel,slashed,graphicx,color,empheq}
\usepackage{youngtab}
\usepackage[matrix,arrow]{xy}


\newcommand{\del}{\partial}

\newcommand{\eq}[1]{(\ref{#1})}
\def\nn{\nonumber} 
\def\obar{\overline}

\numberwithin{equation}{section}

\def\a{\alpha}  \def\b{\beta}
 \def\g{\gamma} 
 \def\d{\delta}

 \def\L{\Lambda}  
    \def\r{\rho}
\def\s{\sigma} \def\S{\Sigma}

\def\cA{{\cal A}} \def\cB{{\cal B}} \def\cC{{\cal C}} 
\def\cD{{\cal D}}  \def\cF{{\cal F}} 
\def\cG{{\cal G}} \def\cH{{\cal H}} \def\cI{{\cal I}} 
  \def\cL{{\cal L}} 
\def\cM{{\cal M}} \def\cN{{\cal N}}  
\def\cP{{\cal P}} \def\cQ{{\cal Q}} \def\cR{{\cal R}} 
\def\cS{{\cal S}}

\def\R{{\mathbb R}} \def\C{{\mathbb C}} 

 \def\one{\mbox{1 \kern-.59em {\rm l}}}

\def\mg{\mathfrak{g}}

\def\msu{\mathfrak{su}}
\def\mso{\mathfrak{so}}
\def\hs{\mathfrak{hs}}

\def\Tr{\mbox{Tr}}\def\tr{\mbox{tr}}

\def\({\left(} \def\){\right)}    

\def\und{\underline}

\newcommand{\im}{\mathrm{i}}
\newcommand{\ad}{\mathrm{ad}}

\newcommand{\diag}{\mathrm{diag}}
\newcommand{\End}{\mathrm{End}}

\newcommand{\vol}{\mathrm{Vol}}
\newcommand{\odd}{\mathrm{odd}}

 \allowdisplaybreaks[3]

\begin{document}

\makeatother


\parindent=0cm

\renewcommand{\title}[1]{\vspace{10mm}\noindent{\Large{\bf#1}}\vspace{8mm}} 
\newcommand{\authors}[1]{\noindent{\large #1}\vspace{5mm}}
\newcommand{\address}[1]{{\itshape #1\vspace{2mm}}}


\begin{titlepage}
\begin{flushright}
 UWThPh-2018-16 
\end{flushright}
\begin{center}
\title{ {\Large  The fuzzy 4-hyperboloid $H^4_n$ and higher-spin  \\[1ex]
 in Yang--Mills matrix models}  }

\vskip 3mm

\authors{Marcus Sperling\footnote{\texttt{marcus.sperling@univie.ac.at}} and 
Harold C.\ Steinacker\footnote{\texttt{harold.steinacker@univie.ac.at}}}

\vskip 3mm

 \address{ 
\textit{ Faculty of Physics, University of Vienna\\
Boltzmanngasse 5, A-1090 Vienna, Austria  }  
  }

\bigskip

\vskip 1.4cm

\textbf{Abstract}
\vskip 3mm

\begin{minipage}{14cm}%

We consider the $SO(4,1)$-covariant
fuzzy hyperboloid $H^4_n$ as a solution of Yang--Mills matrix models, 
and study the resulting higher-spin gauge theory.
The degrees of freedom can be identified with
functions on classical $H^4$ taking values in a higher-spin algebra associated 
to $\mso(4,1)$.
We develop a suitable calculus to classify the higher-spin modes,
and show that the tangential modes are stable.
The metric fluctuations encode one of the spin 2 modes, however 
they do not propagate in the classical matrix model. Gravity  is argued to arise  
upon taking into account induced gravity terms. 
This formalism can be applied to the cosmological FLRW space-time
solutions of \cite{Steinacker:2017bhb}, which arise
as projections of $H^4_n$. We establish a one-to-one correspondence between the 
tangential fluctuations of these spaces.

\end{minipage}

\end{center}

\end{titlepage}

{\small{
\tableofcontents}}

\section{Introduction}

In the present paper we continue the exploration of 4-dimensional covariant 
fuzzy spaces and their associated higher-spin gauge theories, as started in 
\cite{Sperling:2017gmy,Steinacker:2016vgf}.
These are non-commutative spaces which allow to reconcile a quantum structure of 
space(-time) with covariance under the maximal isometry. 
In contrast, quantized Poisson manifolds such as  $\R^4_\theta$ 
\cite{Douglas:2001ba,Szabo:2001kg} are not fully covariant, as an explicit 
tensor $\theta^{\mu\nu}$ breaks the covariance.
In previous work \cite{Sperling:2017gmy,Steinacker:2016vgf}, gauge theory on the fuzzy 4-sphere $S^4_N$ was studied in detail, 
starting from the observation that $S^4_N$ is a solution of Yang--Mills matrix 
models supplemented by a mass term, cf.\ \cite{Kimura:2002nq}. 
Here we extend this analysis to fuzzy $H^4_n$, 
which is a non-compact quantum space preserving an $SO(4,1)$ isometry, also known as Euclidean $AdS^4$. 
For other related work on covariant quantum spaces see e.g.\  
\cite{Steinacker:2015dra,Heckman:2014xha,Grosse:2010tm,deMedeiros:2004wb,Ramgoolam:2001zx,Grosse:1996mz,Medina:2002pc,Castelino:1997rv}. 

The motivation for this work is two-fold: first, we want to develop a formalism 
to study gauge theory on $H^4_n$ along the lines of usual calculus and 
field theory, in order to facilitate the interpretation of the resulting 
models. 
While $S^4_N$ allows to use a clean but less intuitive 
organization of fields into polynomials corresponding to Young diagrams,
the non-compact nature of $H^4$ requires to develop a calculus as 
well as field formalism reminiscent of the conventional treatment. 
We will achieve this goal, and obtain results analogous to 
the compact case but in a more transparent manner. 

The second  motivation is to set the stage for a similar analysis of the cosmological fuzzy space-time solutions $\cM^{3,1}_n$ 
found in \cite{Steinacker:2017vqw,Steinacker:2017bhb}.
These FLRW-type space-times have very interesting physical properties such as a regularized Big-Bang-like 
initial singularity and a finite density of microstates.
$\cM^{3,1}_n$ can obtained from the present 
$H^4_n$ via a projection, which not only leads to a Minkowski signature, 
but also reduces the symmetry to $SO(3,1)$. 
Since the group theory becomes weaker, it seems advisable to consider first the simpler 
(Euclidean) case of fuzzy $H^4_n$. 
We establish the relevant formalism in this paper, and  moreover 
provide some explicit links between the modes on $H^4_n$ and $\cM^{3,1}_n$.

One of the most interesting features of 4-dimensional covariant fuzzy spaces 
is the natural appearance of higher spin theories. This can be understood by recalling
that these spaces are quantized equivariant $S^2$-bundles 
over the base space (i.e.\ $S^4$ or $H^4$ here),
where the fiber is given by the variety of self-dual 2-forms on the base.
The equivariant structure implies that would-be Kaluza--Klein modes transmute 
into higher-spin modes. 
Taken as background solution in matrix models, such as the IKKT model, 
one obtains a higher-spin gauge theory as effective theory around the 
4-dimensional covariant fuzzy spaces.  
As a remark, the structure is reminiscent of twistor constructions, 
see also \cite{Valenzuela:2015gia}. 

Let us describe the results of this paper in some detail.
Starting from the classical as well as fuzzy geometry of the hyperboloid $H^4$, 
we develop a calculus, solely based on the Poisson structure, to organize the 
fuzzy algebra of functions on $H_n^4$ into $SO(4,1)$ irreducible components.  
We further establish a map between the modes in the irreducible components, 
suggestively called spin $s$ fields, and conventional (rank $s$) tensor fields 
on $H^4$.

Having understood the ``functions'' on $H^4_n$, we proceed by considering 
$H^4_n$ as background in the IKKT matrix model. As a first result, we 
classify all (tangential) fluctuation modes at a given spin level and exhibit 
their algebraic features. Subsequently, we are able to diagonalize the kinetic 
term in the action governing the fluctuations. Remarkably, the kinetic terms for 
all tangential fluctuations are non-negative such that no instabilities in the 
tangential sector exist.

Having in mind emergent gravity scenarios, we derive the associated 
graviton modes for spin 0, 1 and 2 fields. The spin 0 and spin 2 contributions
satisfy the de Donder gauge, and at spin 2 one graviton mode emerges from 
the tangential sector. 
However, while the underlying modes do propagate, the graviton  turns out to behave 
like an auxiliary field, and does not propagate 
at the classical level. The reason is that the field redefinition 
required for the graviton cancels the propagator, similar as in on $S^4_N$ \cite{Sperling:2017gmy}.

Nevertheless, our results are  interesting and useful.
First of all, since  classical GR  is not renormalizable, it should presumably be viewed as 
a low-energy effective theory. Then the starting point of an underlying quantum 
theory should be quite different from GR at the classical level, as in 
our approach, and gravity may be induced by quantum effects 
\cite{Sakharov:1967pk,Visser:2002ew}. This is the idea of emergent gravity. The 
present model may well realize this idea,
since the basic framework is non-perturbative and well suited for quantization 
(in particular the maximally supersymmetric 
IKKT model), and the required spin 2 fluctuations do arise naturally. The extra 
degrees of freedom may or may not help, but certainly covariance provides a 
significant advantage compared to other related frameworks, cf.\ 
\cite{Steinacker:2010rh}.
In particular, it is remarkable that no negative or ghost-like modes appear in the 
tangential modes.  

Perhaps
the most interesting perspective is the extension to the cosmological 
space-times $\cM^{3,1}$.
We will establish a one-to-one correspondence of the tangential modes on $H^4_n$ to 
the full set of fluctuations on $\cM^{3,1}$.
Since the tangential modes on  $H^4_n$ are  stable and free of pathologies 
(in contrast to off-shell GR), 
it seems likely that the Minkowski setting on $\cM^{3,1}$ provides a good 
model, too. 
In fact, the presence of negative radial modes on $H^4_n$ would require to 
implement a constraint in the matrix model, which may spoil supersymmetry. 
This is not needed for $\cM^{3,1}$, which provides 
further motivation for including a discussion of $\cM^{3,1}$ here. However,
to keep the paper within bounds, we postpone the details for this  case to future work.

The paper is organized as follows:
We start with a discussion of the classical geometry underlying  $H^4_n$  in section 
\ref{sec:class_geometry}, 
before discussing  fuzzy  $H^4_n$ in detail in section \ref{sec:fuzzy_geometry}. In 
particular, we introduce a calculus suitable for decomposing the  algebra 
of functions into modules of equal spin. The details of the 
decomposition and the properties of the irreducible modes are provided in
section \ref{sec:irred-tensors-in-C}. Having established the fundamentals of 
fuzzy $H^4_n$, we explore the fluctuations around an $H^4_n$ background 
in the IKKT matrix model in section \ref{sec:matrix_model}. We pay particular 
attention to the classification of tangential fluctuations, and explicitly 
diagonalize their kinetic term. Subsequently, the gravition modes are 
identified and their equation of motions are derived. 
Before concluding we briefly explore the projection of $H^4_n$ to the 
Minkowskian $\cM^{3,1}_n$ in section \ref{sec:projection-Lorentzian}. 
Finally, section \ref{sec:conclusion} concludes and provides an outlook for 
future work.
Relevant notation and conventions as well as auxiliary identities and 
derivations are collected in appendices 
\ref{app:so42}--\ref{sec:semiclass-id-appendix}.

%
%
\section{Classical geometry underlying \texorpdfstring{$H^4_n$}{H4n}}
\label{sec:class_geometry}
The classical geometry underlying fuzzy $H^4_n$ is $\C P^{1,2}$,
which is an $S^2$-bundle over the 4-hyperboloid $H^4$. 
More precisely, $\C P^{1,2}$  is an 
 $SO(4,1)$-equivariant bundle over $H^4$ as well as a coadjoint orbit of 
$SO(4,2)$.
This means that the local stabilizer group $SO(4)$ acts non-trivially on the 
fiber $S^2$, leading to  higher-spin fields on $H^4$, and a canonical 
quantization exists.
The construction is similar to twistor constructions for Minkowski space.

\subsection{\texorpdfstring{$\C P^{1,2}$}{CP12} as 
\texorpdfstring{$SO(4,1)$}{SO(4,1)}-equivariant bundle over the hyperboloid 
\texorpdfstring{$H^4$}{H4}}
\label{sec:CP12-H4}
Let $\psi \in \C^4$ be a spinor of  $\mso(4,1)$  with $\bar \psi \psi = 1$.
Consider the following Hopf map: 
 \begin{align}
 \begin{aligned}
  H^{4,3} &\to H^4 \ \subset \R^{1,4}   \\
 \psi\ &\mapsto  x^a = \frac r2\bar{\psi} \g^{a} \psi , \qquad a = 0,1,2,3,4 \, 
,
 \end{aligned}
  \label{Hopf}
\end{align}
where $r$ introduces a length scale, and
$H^{4,3}$ is  the  7-hyperboloid
\begin{align}
 H^{4,3} = \{\psi\in \C^4| \  \bar{\psi} \psi = \psi^\dagger \g^0 \psi = 1 \} \ 
.
\end{align}
The $\g^a$, $a=0,\ldots,4$  are $SO(4,1)$ gamma matrices, see 
appendix \ref{app:gamma_matrices} for details.
The map \eqref{Hopf} is a non-compact version of the Hopf map $S^7 \to S^4$, 
which respects 
$SO(4,1)$ and in which the $x^a$ transform as $SO(4,1)$ vectors.
By using \eqref{gamma-tensor-id} one can verify that 
\begin{align}
 \sum_{a,b=0}^4 \eta_{ab} x^a x^b = -\frac {r^2}4 \eqqcolon -  R^2
\end{align}
so that the right-hand side is indeed in $H^4$; note that $x_a \in \R$ due to 
\eqref{gamma-conj}. 
Since the overall phase of $\psi$ drops out, we can  re-interpret 
\eqref{Hopf} as a map 
\begin{align}
x^a:  \quad \C P^{1,2}  \ &\to H^4 \subset \R^{1,4}   
 \label{Hopf-cp3}
\end{align}
where  $\C P^{1,2} = H^{4,3}/U(1)$ is defined as  space of 
unit spinors $\bar\psi \psi = 1$ modulo $U(1)$. In other words, $\C P^{1,2}$ is 
a $S^2$-bundle over $H^4$.
To exhibit the fiber, consider an arbitrary spinor $\psi$ with $\bar{\psi} \psi 
= 1$. Since
\begin{align}
 x^0 = \frac r2 \psi^\dagger \psi \ > 0, 
\end{align}
there exists a suitable $SO(4,1)$ transformation such that
\begin{align}
 x^a|_\xi = R(1,0,0,0,0) ,
 \label{eq:reference_point}
\end{align}
which defines a reference point $\xi \in H^4$.
Its stabilizer group is
\begin{align}
 H = \{h; [h,\g_0] = 0\} \ = SU(2)_R\times SU(2)_L   \subset SO(4,1) 
\end{align}
where $SU(2)_L$ acts on the $+1$ eigenspace of  $\g^0$. 
By introducing complex parameters 
\begin{align} 
 \psi^T = \begin{pmatrix}
         a_1^*, a_2^*,b_1, b_2 
        \end{pmatrix}, \qquad
 1 = \bar\psi \psi = -|a_1|^2 - |a_2|^2 + |b_1|^2 + |b_2|^2  = \psi^\dagger\psi
\end{align}
it follows that $|b_1|^2 + |b_2|^2 =1$ and $a_1 = a_2= 0$. Thus
after an appropriate $SU(2)_L$ transformation we can assume 
\begin{align}
 \psi^T = (0,0,0,1) \; ,
 \label{eq:reference_spinor}
\end{align}
which will be a reference spinor over $\xi$ throughout the remainder.
Hence $\C P^{1,2}$ is a $S^2$-bundle over $H^4$, and
the $S^2$ fiber is obtained by acting with $SU(2)_L$ on $\psi$.
This is analogous to the well-known fact that $\C P^3$ is an $S^2$-bundle over 
$S^4$.
Note that the  metric on the hyperboloid induced via
\begin{align}
 x^a: \quad H^4 \hookrightarrow \R^{1,4}
\end{align}
is  Euclidean, despite the $SO(4,1)$ metric on target space. 
This is obvious at the point 
$\xi=(R,0,0,0,0)$, where the tangent space is $\R^4_{1234}$.

\paragraph{$SO(4,2)$ formulation and embedding functions.}
It is useful to view $\C P^{1,2}$ as a 6-dimensional coadjoint orbit of $SU(2,2)$
\begin{align}
 \C P^{1,2} \  \cong \ \{U^{-1} Z U, \quad U\in SU(2,2) \} \ \hookrightarrow \ \msu(2,2) 
 \label{SU22-orbit}
\end{align}
through the rank one $4\times 4$ matrix
\begin{align}
 Z = \psi\bar\psi, \qquad Z^2 = Z, \qquad \tr(Z) = 1 , \qquad
 Z^\dagger = \g^0 Z {\g^0}^{-1} \ . 
 \label{Z-relations}
\end{align}
The embedding \eqref{SU22-orbit} is  described by 
the  embedding functions 
\begin{align}
 \begin{aligned}
 m^{ab} &= \tr(Z \Sigma^{ab}) = \obar \psi\Sigma^{ab}\psi = (m^{ab})^*, \\
 x^a &= r\, \tr(Z \Sigma^{a5})  = \frac r2 \obar\psi\gamma^{a}\psi =  r \,
m^{a5}, \qquad a,b=0,\ldots,4
 \label{Mab-trace-Z}
\end{aligned}
\end{align}
noting that $\frac{1}{2}\g^a =  \Sigma^{a5}$, see \eqref{Sigma-explicit}.
Upon restricting to $\mso(4,1)\subset\mso(4,2) \cong \msu(2,2)$, we recover 
\eqref{Hopf-cp3}, which reflects that the $SO(4,1)$ action is transitive on $\C 
P^{1,2}$.
The last equation in \eqref{Mab-trace-Z} amounts to a group-theoretical 
definition of the Hopf map, which will generalize to the non-commutative case.
The $SO(4,2)$ structure is often useful, but it does not respect the projection to $H^4$.

We can  compute the invariant functions
\begin{align}
\sum_{0\leq a<b\leq 4} m^{ab} m_{ab} 
  &= \sum_{0\leq a<b\leq 4} \bar{\psi}\bar{\psi}\Sigma^{ab}\otimes \Sigma_{ab} 
\psi \psi
   = \frac{1}{2} \; ,
 \label{mm-id-so41} \\
\sum_{0\leq a<b\leq 5} m^{ab} m_{ab} 
  &= \sum_{0\leq a<b\leq 5} \bar{\psi}\bar{\psi}\Sigma^{ab}\otimes \Sigma_{ab} 
\psi \psi
 = \frac{3}{4} \; ,
  \label{mm-id-so42}
\end{align}
 using the identities \eqref{SO41-sigma-sigma-id} and \eqref{SO42-sigma-sigma}.
 Here, the indices are raised and lowered with $\eta_{ab} = 
\diag(-1,1,1,1,1,-1)$.
 Combining the two identities 
\eqref{SO41-sigma-sigma-id}--\eqref{SO42-sigma-sigma} and recalling $x^a = r 
m^{a5}$, we recover
\begin{align}
  x_a x^a  = -\frac {r^2}4 = - R^2 .
\end{align}
Remarkably, the $SO(4,1)$-invariant $x^a x_a$ is constant on  $\C P^{1,2}$.
Similarly, \eqref{Sigma-Sigma-contract} together with the above relations 
imply\footnote{This is just a manifestation of 
the relation $Z^2 = Z$, see \eqref{Z-relations}.} the $SO(4,2)$ identities
\begin{align}
  \eta_{cc'}m^{ac} m^{bc'} = \frac{1}{4} \eta^{ab}  , \qquad a,b=0,\ldots,5
   \label{so42-MM-rel-class}
\end{align}
which reduces to the $SO(4,1)$ relation
\begin{align}
\eta_{cc'}m^{ac} m^{bc'}  - r^{-2} x^a x^b   &= \frac{1}{4} \eta_{ab}  , \qquad 
a,b=0,\ldots,4 \ .
   \label{so41-MM-rel-class}
\end{align}  
 In particular, this implies that $m^{ab}$ is orthogonal to $x^a$, 
\begin{align}
 x_a m^{ab} = 0 \ .
 \label{m-x-orthogonal}
\end{align}
Furthermore, the following $SO(4,2)$ identities hold:
\begin{align}
\epsilon_{abcdef}m^{ab} m^{cd} &=  \bar{\psi}\bar{\psi}\epsilon_{abcdef}\Sigma^{ab}\otimes \Sigma^{cd}\psi\psi \\
  &=  2 \bar{\psi}\bar{\psi}(\Sigma^{ef}\otimes 1 + 1 \otimes \Sigma^{ef})\psi\psi \nn\\
   &=  4\bar{\psi}\Sigma^{ef}\psi  =  4 m_{ef} \,,
  \label{epsilonMM-id-1}
 \end{align}
 using \eqref{so6-id-sigma}; this can also be seen from \eqref{sigma-sigma-id}.
 Reduced to $SO(4,1)$, this implies
 \begin{align}
  \epsilon_{abcde} m^{ab} m^{cd} = -\frac 4r\, x_e \ , \qquad e =0,\ldots,4  .
   \label{epsilonMM-id-2}
 \end{align}
Finally, there exists a self-duality relation 
\begin{align}
 \epsilon_{abcde}m^{ab} x^{c} &=  \bar{\psi}\bar{\psi}\epsilon_{abc5de}\Sigma^{ab}\otimes \Sigma^{c5}\psi\psi \\
 &= \frac 12 \bar{\psi}\bar{\psi}(\Sigma_{de}\otimes 1 + 1 \otimes \Sigma_{de})\psi\psi\nn\\
 &=  \bar{\psi}\Sigma_{de}\psi = m_{de}
 \label{selfdual-class}
\end{align}
 using \eqref{so6-id-sigma}. Thus $m^{ab}$ is a tangential self-dual rank 2 
tensor on $H^{4}$,
 in complete analogy to $S^4_N$  \cite{Sperling:2017dts}.
At the reference point \eqref{eq:reference_point}, one can express $m^{ab}$ 
in terms of the $SO(4)$ t'Hooft symbols
\begin{align}
  m^{\mu\nu} = \eta_{\mu\nu}^i\, J_i ,  \qquad\quad   J_i J^i = 1
\label{tHooft-classical}
\end{align}
where $ J_i$ describes the internal $S^2$. 
This exhibits the structure of $\C P^{1,2}$ is an $SO(4,1)$-equivariant bundle over $H^4$. 
The fiber $S^2$ is generated by the local $SU(2)_L$, while $SU(2)_R$ acts trivially.

\subsection{\texorpdfstring{$\C P^{1,2}$}{CP12} as 
\texorpdfstring{$SO(3,2)$}{SO(3,2)}-equivariant bundle over the hyperboloid 
\texorpdfstring{$H^{2,2}$}{H22}}
\label{H22-class}
Equivalently, the homogeneous space $\C P^{1,2}$ of $SO(4,2)$ can be 
viewed as $S^2$-bundle over $H^{2,2}$, which arises from  a different Hopf map 
\begin{align}
 H^{4,3} \to \C P^{1,2} \to H^{2,2} \subset \R^{2,3}
\end{align}
as follows, cf. \cite{Hasebe:2012mz}:
 \begin{align}
  t^a &=  \frac 1R  \obar\psi \Sigma^{a4}\psi   =  \frac{1}{R}  m^{a4}, \qquad 
a=0,1,2,3,5 \ .
 \label{Mab-trace-p}
\end{align}
This map is compatible with $SO(3,2)$, and
the reference spinor \eqref{eq:reference_spinor} is now projected to $t^a = 
r^{-1}(0,0,0,0,1) \in \R^{3,2}$,  which transforms as $SO(3,2)$ vector. Then 
$t^a$ defines a hyperboloid $H^{2,2}\subset \R^{3,2}$
with intrinsic signature $(+,+,-,-)$. Using analogous identities as before, we 
obtain the constraints 
\begin{align}
\begin{aligned}
 \tilde\eta_{ab} t^a t^b &= r^{-2} , \qquad \tilde\eta_{ab} = 
\diag(-1,1,1,1,-1)  \; ,\\
 t_a x^a &= 0 = t_\mu x^\mu \; .
 \label{p-normalization-class}
 \end{aligned}
\end{align}
The last relation follows from the $SO(4,2)$ relation 
\eqref{so42-MM-rel-class}, noting that $t^4 \equiv 0$.
More generally, we can consider
\begin{align}
  x^a = m^{ab} \a_b, \qquad t^a = m^{ab} \b_b 
\end{align}
where $\a, \b \in \R^{2,4}$ are two linearly independent vectors with\footnote{The case of light-like $\a$ is also 
interesting, see section \ref{sec:poincare}.}
 $\a_b \b^b = 0$.
Then the previous constructions are recovered for $\a=e^5, \ \b=e^4$.
The common symmetry group which preserves both $\a_b$ and $\b_b$ is $SO(3,1)$. 
Note that $t^5 \propto x^4$  on $\C P^{1,2}$.

\subsection{\texorpdfstring{$SO(3,1)$}{SO(3,1)}-invariant projections and 
Minkowski signature}
\label{sec:so31-proj-class}
\begin{figure}
 \includegraphics[width=0.8\textwidth]{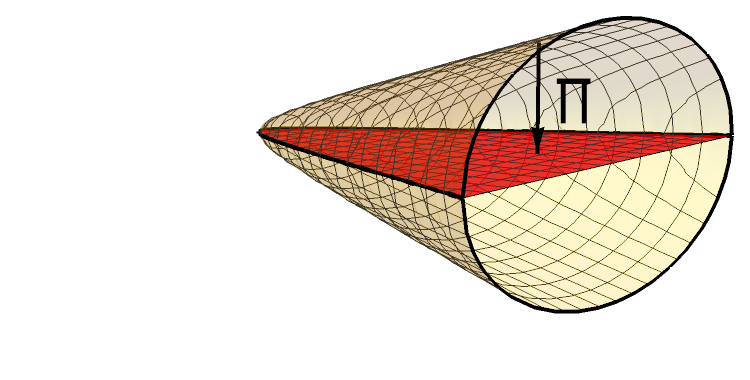}
 \caption{Sketch of the projection $\Pi_x$ from $H^4$ to $\cM^{3,1}$  with 
Minkowski signature.}
 \label{fig:projection}
\end{figure}

So far we have constructed $H^4$ and $H^{2,2}$, but not a space with 
Minkowski signature yet. 
Space-times with Minkowski signature can be obtained by $SO(3,1)$-covariant 
projections of the above hyperboloids. 
Explicitly, consider the projections  
\begin{align}
\begin{aligned}
 \Pi_x: \quad \C P^{1,2} \ &\to \R^{3,1}, \\
               m &\mapsto x^\mu = m^{\mu b} \a_b  \\
\Pi_t: \quad \C P^{1,2} \ &\to \R^{3,1}, \\
               m &\mapsto t^\mu = m^{\mu b} \b_b 
 \label{proj-class}
 \end{aligned}
 \qquad \text{with } \mu = 0,1,2,3,
\end{align}
which respect $SO(3,1)$. A sketch of $\Pi_x$ is displayed in figure 
\ref{fig:projection}.
In section  \ref{sec:projection-Lorentzian},
the image $\cM^{3,1}\subset \R^{3,1}$ of $\Pi_x$ serves as cosmological FLRW 
space-time with $k=-1$, as discussed in \cite{Steinacker:2017bhb}. In contrast, 
$t^\mu$ is interpreted as internal space related to translations.

\section{The fuzzy hyperboloid \texorpdfstring{$H^4_n$}{H4n}}
\label{sec:fuzzy_geometry}
Now we turn to the central object of this paper: the fuzzy hyperboloid $H^4_n$. 
$H^4_n$ is a quantization of the bundle $\C P^{1,2}$ over $H^4$, which 
respects the $SO(4,2)$ structure and the projection to the base space $H^4$. 
This is natural because $\C P^{1,2}$ is a coadjoint orbit of $SO(4,2)$ via 
\eqref{SU22-orbit}. 
As such $\C P^{1,2}$ is equipped with a canonical $SO(4,2)$-invariant Poisson  
(symplectic) structure; whereas on $H^4$ no such structure exists.
$H^4_n$ was first discussed in \cite{Hasebe:2012mz}, and it serves as 
starting point for a quantized cosmological space-time in
\cite{Steinacker:2017vqw}.

As for any coadjoint orbit, fuzzy $H^4_n$ can be defined in terms of the 
operator algebra $\End(\cH_n)$,
where $\cH_n$  is a suitable unitary irrep of $SU(2,2)\cong SO(4,2)$. 
The representation is chosen such that 
the Lie algebra generators $\cM_{ab} \in \End(\cH_n)$ generate a 
non-commutative algebra of functions, interpreted as quantized or fuzzy $\C 
P^{1,2}_n$. 
The $\cM^{ab}$ are naturally viewed as quantized coordinate 
functions $m^{ab}$ \eqref{Mab-trace-Z} on $\C P^{1,2}$.
Fuzzy $H^4_n$ is then generated by Hermitian generators $X^a \sim x^a$, 
which transform as vectors under $SO(4,1)\subset SO(4,2)$, and are interpreted 
as quantized embedding functions \eqref{Hopf-cp3}.
This will be made more explicit through an oscillator construction, which allows 
to derive all the required properties.

To define fuzzy $H^4_n$ explicitly, let 
$\eta^{ab} = \diag(-1,1,1,1,1,-1)$ be the invariant metric of $SO(4,2)$, and
let $\cM^{ab}$ be the Hermitian generators of $SO(4,2)$, which satisfy
\begin{align}
  [\cM_{ab},\cM_{cd}] &=\im(\eta_{ac}\cM_{bd} - \eta_{ad}\cM_{bc} - 
\eta_{bc}\cM_{ad} + \eta_{bd}\cM_{ac}) \ .
 \label{M-M-relations-noncompact}
\end{align}
We choose a particular type of (discrete series) 
positive-energy unitary irreps\footnote{Strictly speaking there are two versions 
$\cH_{n}^L$ or $\cH_{n}^R$ with opposite ``chirality'', 
but this distinction is irrelevant in the present paper and therefore dropped.} $\cH_{n}$ 
known as \emph{minireps} or \emph{doubletons} \cite{Mack:1975je,Govil:2013uta}.
Remarkably, the $\cH_n$ remain irreducible\footnote{This follows from the 
minimal oscillator construction of $\cH_n$, where all 
$SO(4,2)$ weight multiplicities 
are at most one, cf.\ \cite{Mack:1969dg,Mack:1975je,Heidenreich:1980xi}.}
under $SO(4,1) \subset SO(4,2)$. Moreover, the minireps have positive discrete 
spectrum 
\begin{align}
 {\rm spec}(\cM^{05}) = \{E_0, E_0+1, \ldots \}, \qquad E_0 = 1+\frac{n}2 
\end{align}
where the eigenspace with lowest eigenvalue of $\cM^{05}$ is an $n + 
1$-dimensional irreducible
representation of either $SU (2)_L$ or $SU (2)_R$. Then the Hermitian generators
\begin{align}
\begin{aligned}
 X^a &\coloneqq r\cM^{a 5}, \qquad a = 0,\ldots,4  \\
   [X^a,X^b] &= -\im r^2\cM^{ab}  \eqqcolon \im \Theta^{ab} 
  \label{X-X-CR}
  \end{aligned}
\end{align} 
(note the signs!) transform as $SO(4,1)$ vectors, i.e.\
\begin{align}
\boxed{
\begin{aligned}
 [\cM_{ab},X_c] &=  \im (\eta_{ac} X_b - \eta_{bk} X_a) , \\
    [\cM_{ab},\cM_{cd}] &=\im (\eta_{ac}\cM_{bd} - \eta_{ad}\cM_{bc} - 
\eta_{bc}\cM_{ad} + \eta_{bd}\cM_{ac}) \ .
\end{aligned}
  }
 \label{M-M-relations} 
\end{align}
Because the restriction to $SO(4,1) \subset SO(4,2)$ is irreducible, it follows that the $X^a$ live on a hyperboloid, 
\begin{align} 
\boxed{
 \eta_{ab} X^a X^b = X^i X^i - X^0 X^0 \eqqcolon - R^2 \one \ 
 }
 \label{hyperboloid-constraint}
\end{align}
with some $R^2$ to be determined below.
Since $X^0 = r \cM^{05} > 0$ has positive spectrum, this describes a one-sided 
hyperboloid in $\R^{1,4}$, denoted as $H^4_n$.
Analogous to fuzzy $S^4_N$, the semi-classical geometry underlying  
$H^4_n$ is $\C P^{1,2}$ \cite{Hasebe:2012mz},
which is an $S^2$-bundle over $H^4$ carrying a canonical symplectic structure. 
In the fuzzy case, this  fiber is  a fuzzy 2-sphere $S^2_n$.
We work again in the semi-classical limit.
We also note the following commutation relations
\begin{align}
\Box_X X^b =  [X_a,[X^a,X^b]] = -4 r^2 X^b \ .
\label{X-Box}
\end{align}
The negative sign arises from $\eta=\diag(-1,1,1,1,1,-1)$, and
 $\Box_X$ is not positive definite.

\subsection{Fuzzy \texorpdfstring{$H^{2,2}_n$}{H22n} and momentum space}
As in the classical case \eqref{Mab-trace-p} and
for later purpose, we also define 
\begin{align}
 T^a = \frac{1}{R}\cM^{a4}, \qquad a = 0,\ldots,3,5 \ 
 \label{P-def-fuzzy}
\end{align}
where $R \, r\, T^5 = -X^4$.
As the restriction of $\cH_n$ to $SO(3,2) \subset SO(4,2)$ is irreducible,
the operators \eqref{P-def-fuzzy} satisfy the constraint 
\begin{align}
 \tilde\eta_{ab} T^a T^b &= - T^0 T^0  + \sum_{i=1,2,3}  T^i T^i - T^5 T^5  =  \frac{1}{r^2}\one \ 
 \label{H22-constraint}
\end{align}
cf. \eqref{p-normalization-class}.
This is the quantization of the hyperboloid $H^{2,2} \subset \R^{3,2}$ 
with intrinsic signature $(+,+,-,-)$  of section \ref{H22-class} and becomes 
Lorentzian via the projection \eqref{sec:so31-proj-class}.
The commutation relations are 
\begin{align}
\begin{aligned}
 [T^a,T^b] &= \im \frac{1}{R^2}\, \cM^{ab}    \qquad a,b = 0,\ldots,3,5\,,   \\
 [T^\mu,X^\nu] &= \im\frac{1}{R} \eta^{\mu\nu} X^{4}, \qquad \mu,\nu = 
0,\ldots,3 \,,
 \label{P-P-CR}
 \end{aligned}
\end{align}
which justifies to consider $T^\mu$ as translation generators, 
and 
\begin{align}
\Box_T T^b =  [T_a,[T^a,T^b]] = +4 T^b 
\label{P-Box}
\end{align}
Note the different signs in \eqref{P-Box} and \eqref{X-Box}, which arise from  
of $\eta^{55}=-1 = - \eta^{44}$.

\subsection{\texorpdfstring{$SO(3,1)$}{SO(3,1)}-covariant fuzzy spaces}
\label{sec:poincare}
In analogy to section \ref{sec:so31-proj-class}, we consider the 
$SO(3,1)$-covariant fuzzy generators
\begin{align}
 \tilde X^\mu = \cM^{\mu a}\a_a, 
 \qquad \tilde T_\mu = \cM^{\mu a}\b_a
 \label{XP-def-general}
\end{align}
where $\a,\b$ are $SO(3,1)$-invariant.
 They satisfy
\begin{align}
 [\tilde X^\mu,\tilde X^\nu] &= (\a\cdot\a) \cM^{\mu\nu}, \qquad
 [\tilde T^\mu,\tilde T^\nu] = (\b\cdot\b) \cM^{\mu\nu} \nn\\ 
 [\tilde X^\mu,\tilde T_\nu] &= \im \big(\d^\mu_\nu\,\cM^{ab}\a_a\b_b + 
\a\cdot\b \cM^{\mu\nu}\big)\nn\\
   &= \im  \big(\d^\mu_\nu\, \a\wedge\b D + \a\cdot\b \cM^{\mu\nu}\big)
 \label{X-P-CR-general}
 \end{align}
 where $\a\wedge\b  = \a^4\b^5-\a^5 \b^4$ and $D= \cM^{45}$.
 For $\a\cdot\a \approx 0$ and $\a\cdot\b\approx 1\approx \a\wedge\b$, the 
$\tilde X^\mu$ become almost commutative and the commutation relations are not 
far from the Poincare algebra:
\paragraph{Poincare algebra.}
In particular for light-like $\a=\frac 1{\sqrt{2}}(1,-1)$ and $\b=\frac 1{\sqrt{2}}(1,1)$, we obtain 
\begin{align}
 K_\mu \coloneqq  \frac 1{\sqrt{2}}(\cM_{\mu 5} - \cM_{\mu 4}), \qquad 
  \tilde T^\mu = \frac 1{\sqrt{2}}(\cM^{\mu 5} + \cM^{\mu 4})
 \label{P-def}
\end{align}
 which satisfy
\begin{align}
\begin{aligned}
 [ \tilde T^\mu, \tilde T^\nu] &= 0 =  [K^\mu,K^\nu]  \,,\\ 
 [ \tilde T^\mu,K_\nu] &= \im (\d^\mu_\nu\, D + \cM^{\mu\nu}) \, .
 \end{aligned}
 \end{align}
Hence the $\tilde T^\mu$ together with $\cM^{\mu\nu}$ generate the Poincare algebra $ISO(3,1)$ as
sub-algebra of $\mso(4,2)$, with special conformal generators $K^\mu$ 
and the dilatation operator $D$
\begin{align}
 [D,\tilde T_\mu] &= \im \tilde T_\mu, \quad [D,K_\mu] = - \im K_\mu .
\end{align}
\subsection{Oscillator realization, minireps and coherent states}
\label{sec:oscillator-realization}
The Hilbert space $\cH_n$ is a highest-weight 
unitary representation of $SU(2,2)$, which can be obtained by quantizing the
spinorial construction of $\C P^{2,1}$ in \eqref{Hopf}.
For the quantization one replaces the classical 4-component spinor $\psi_\a$ by 
4 operators, which satisfy 
\begin{align}
 [{\psi}_\a, \bar{\psi}^\b] = \d_\a^\b \; .
 \label{osc-alg-NC-general}
\end{align}
The associated bilinears
\begin{align}
\cM^{ab} \coloneqq \bar{\psi}\Sigma^{ab}\psi
\label{Mab-osc-def}
\end{align}
realize the Lie algebra \eqref{M-M-relations-noncompact} of $SO(4,2)$, due to
\begin{align}
\left[ \bar{\psi}\Sigma^{ab}\psi,\bar\psi\Sigma^{cd}\psi\right]=
\bar{\psi}\left[ \Sigma^{ab},\Sigma^{cd}\right]\psi .
\end{align}
The $\cM_{ab}$ are self-adjoint operators, since 
\begin{align}
 {\Sigma^{ab}}^\dagger = \g^0 \Sigma^{ab} {\g^0}^{-1} \; .
 \label{spinor-hermitian-so42}
\end{align}
As a consequence, they implement unitary representations of $SU(2,2)$ on the 
Fock space $\cF = {\rm span}\{\bar\psi \ldots \bar\psi |0\rangle\}$ 
of the bosonic oscillators, which decomposes into an infinite number of 
irreducible positive energy unitary representations $\cH_\L$.

The oscillator algebra \eqref{osc-alg-NC-general} can be realized explicitly 
as follows (cf.\ \cite{Chiodaroli:2011pp,Govil:2013uta}):
Consider bosonic creation and annihilation operators $a_{i}, b_j$ which satisfy
\begin{align}
 [a_i, a_j^\dagger] = \delta_{i}^{j} \, , \qquad
[b_i, b^\dagger_j] = \delta_{i}^{j}  \qquad \text{for }i,j=1,2 \; .
\end{align}
Using the $a_{i}, b_j$ we form spinorial operators  
\begin{align}
\psi \coloneqq 
\begin{pmatrix}
a^\dagger_{1} \\  a^\dagger_{2} \\ b_1 \\ b_2
\end{pmatrix}
\label{spinor-oscillator}
\end{align}
with Dirac conjugates 
\begin{align}
\bar{\psi}\equiv {\psi}^{\dagger}\gamma^{0}
=
\left(-a_{1},-a_{2}, b^\dagger_{1}, b^\dagger_{2}\right).
\end{align}
Then 
\begin{align}
 [\psi^\a,\bar\psi_\b] = \d^\a_\b
\end{align}
as required, and the $SO(4,2)$ generators are 
\begin{align}
 \cM^{ab} = \bar{\psi}\Sigma^{ab}\psi = 
 \left(-a_{1},-a_{2}, b^\dagger_{1}, b^\dagger_{2}\right) \Sigma^{ab} 
\begin{pmatrix}
a^\dagger_{1} \\  a^\dagger_{2} \\ b_1 \\ b_2
\end{pmatrix} \; .
\end{align}
The  generators of $SU(2)_{L}$ and $SU(2)_{R}$ are defined by
\begin{align}
\begin{aligned}
L^{k}_{i} &\coloneqq {a}^\dagger_{k} {a}_{i}
-\frac{1}{2} \delta^{k}_{i}N_{a} \nn\\
R^{i}_{j} &\coloneqq {b}^\dagger_{i} {b}_{j}
-\frac{1}{2} \delta^{i}_{j} N_{b}
\end{aligned}
\end{align}
and the time-like generator $X^0$  (or the ``conformal Hamiltonian'' $E$) is 
given by
\begin{align}
 r^{-1}\, X^0 =  E = M^{05}  = \bar{\psi}\Sigma^{05}\psi = \frac 12 \psi^\dagger \psi = \frac 12 (N_a + N_b + 2),
\label{conf-energy}
\end{align}
where $N_{a} \equiv {a}^\dagger_{i} {a}_{i}$,
$N_{b} \equiv {b}^\dagger_{j} {b}_{j}$ are the bosonic number operators, and 
\begin{align}
 \hat N =  \bar{\psi}\psi = -N_a + N_b - 2 
\end{align}
is  invariant.
The non-compact generators are
given by linear combinations of creation and annihilation
operators of the form ${a}_{i}^\dagger {b}_{j}^\dagger$ and
${a}_{i} {b}_{j}$.

\paragraph{Minireps.}
The simplest class of unitary representation has lowest weight space given by the Fock vacuum 
$a_i \left|0\right\rangle = 0 = b_i \left|0 \right\rangle$, which defines 
\cite{Chiodaroli:2011pp}
\begin{align}
 \left|\Omega \right\rangle \coloneqq \left|1,0,0\right\rangle \eqqcolon 
\left|0\right\rangle, \qquad E = 1, j_L = j_R = 0 \; .
\end{align}
This gives the \emph{doubleton} minireps  built on the lowest weight vectors
\begin{align}
\begin{aligned}
  \left|\Omega \right\rangle \coloneqq 
  \left|E,\frac{n}{2},0 \right\rangle &\coloneqq 
a_{i_1}^\dagger \ldots a_{i_n}^\dagger \left|0 \right\rangle, 
 \qquad E = 1+\frac{n}{2}, j_L = \frac n2, \  j_R = 0 \\
  \left|\Omega \right\rangle \coloneqq  
  \left|E,0,\frac{n}{2} \right\rangle &\coloneqq 
b_{i_1}^\dagger \ldots b_{i_n}^\dagger \left|0 \right\rangle, 
 \qquad E = 1+\frac n2, j_L = 0, j_R = \frac{n}{2}
 \end{aligned}
 \label{minireps-Fock}
\end{align}
which are annihilated by all $L^-$ operators, i.e. of the form  $a_i b_j$,
\begin{align}
  a_i b_j |\Omega\rangle \equiv 0
\end{align}
and
\begin{align}
 n^2 \coloneqq \left(\hat{N} +2 \right)^2 = \left(N_a-N_b \right)^2, \qquad 
n=0,1,2,\ldots \; .
\end{align}
Acting with all operators 
of the form $a^\dagger_i b^\dagger_j$ of $L^+$ on $|\Omega\rangle $, one obtains positive energy discrete series UIR's 
$\cH_{\L}$ of $U(2,2)$ with lowest weight $\L=\left(E,\frac{n}{2},0\right)$ and 
$\L=\left(E,0,\frac{n}{2}\right)$. We will largely ignore the distinction 
and denote both as $\cH_n$.
These are known as \emph{minireps} of $\mso(4,2)$, 
because they are free of multiplicities in weight space\footnote{This can be 
seen e.g.\ from the characters given in \cite{Heidenreich:1980xi}}. 
They correspond to fields living on the boundary of $AdS^5$.
The minireps remain irreducible under $SO(4,1)$ as well as $SO(3,2)$,
and they can be interpreted as massless fields on $AdS^4$, or 
as conformal fields on Minkowski space.
The lowest weight state $\left|E,0,\frac{n}{2}\right\rangle$ of $\cH_n$ 
generates a $(n+1)$-dimensional 
irreducible representation of either $SU(2)_L$ or $SU(2)_R$ with degenerate 
$X^0$, naturally interpreted as fuzzy $S^2_n$.

Comparing the above oscillator construction \eqref{Mab-osc-def}  with 
\eqref{Mab-trace-Z}, it is manifest 
that for each $\cH_n$, with $n>0$, the $\cM_{ab}$ generators can be 
interpreted as quantized embedding functions 
\begin{align}
 \cM_{ab} \sim m_{ab}: \quad  \C P^{1,2} \to \mso(4,2) \cong \R^{15} \; .
\end{align}
This provides the quantization of the coadjoint orbits \eqref{SU22-orbit}, which 
defines fuzzy $\C P^{1,2}_n$.
Since $X^0 \geq 1$,
they should be viewed as quantized bundles with base space $H^4_n$ described by 
$X^a$, 
and fiber $S^2_n$, for $n=1,2,3,\ldots$.
The implicit constraints defining these varieties will be elaborated below.
For $n>0$, these spaces have been briefly discussed in \cite{Hasebe:2012mz,Valenzuela:2015gia}, and we will 
mostly focus on that case. The minimal $n=0$ case
is different, but also very interesting, and we discuss it in some detail in 
appendix \ref{sec:app-minimal}.

\paragraph{Coherent states and quantization.}
The above discrete series irreps $\cH_n$ provide a natural definition of coherent states $|m\rangle = g\cdot |\Omega\rangle \in \cH_n$, 
which are given by the 
$SO(4,1)$ 
orbit\footnote{It turns out that the $SO(4,2)$ orbit of these states defines an 8-dimensional bundle over $H^4$; this was missed in a previous version of this paper.}
through the 
lowest weight state $\left|\Omega \right\rangle$. The set of coherent states 
forms a $U(1)$-bundle over $\C P^{1,2}$, and allow to
recover the semi-classical geometry of $\C P^2$ as $S^2$-bundle over $H^4$ via 
$m^{ab} = \left\langle m\right|\cM^{ab}\left|m\right\rangle$.
In particular, the lowest weight state is located at the reference point 
$\left\langle\Omega \right|X^a\left|\Omega\right\rangle =x^a_\xi=(R,0,0,0,0)$, 
see \eqref{eq:reference_point}. The local $SO(4)$ generators $\cM^{ij}$  act on 
the coherent states over $\xi$ in a spin $\frac{n}{2}$ irrep.

These coherent states $\left|m\right\rangle$ also provide a 
$SO(4,1)$-equivariant
quantization map from the classical space of functions on $\C P^{1,2}$
to the fuzzy functions $\End(\cH_n)$:
\begin{align}
\begin{aligned}
 \cQ: \quad \cC(\C P^{1,2}) &\to \End(\cH_n)  \\
 f(m)  &\mapsto \int\limits_{\C P^{1,2}} d\mu\, f(m) 
\left|m\right\rangle \left\langle m\right|
 \end{aligned}
 \label{quantization-map}
\end{align}
where $\left|m\right\rangle$ is a coherent state\footnote{Observe that the 
phase ambiguity of the coherent states drops out here.}, 
and $d \mu$ is the $SO(4,2)$-invariant measure.
For polynomial functions, this 
corresponds to Weyl quantization, mapping irreducible polynomials $P(m^{ab})$ to the corresponding totally symmetrized 
polynomials $P(\cM^{ab})$; in particular $\cQ(m^{ab}) = \cM^{ab}$.
Likewise, square-integrable functions on $\C P^{1,2}$ are mapped to Hilbert-Schmidt 
operators in $\End(\cH_n)$. 
However, the map $\cQ$ is not surjective;
its image only comprises a finite tower of higher-spin modes truncated at $s=n$, as shown in appendix \ref{sec:funct-spin-coherent}.
These modes can be understood as semi-classical sector of the full space of functions.


\subsection{Algebraic properties of fuzzy \texorpdfstring{$H^4_n$}{H4n}}
\label{sec:alg-id-H4n}
Using the aforementioned oscillator realization, one can derive a number of 
useful identities for the above operators on $\cH_n$; we refer the reader to 
appendix \ref{sec:fuzzyH4-derivations} for the details.
To begin with, consider the $SO(4,1)$-invariant radius operator
\begin{align}
 \cR^2 \coloneqq \sum_{a,b=0,1,2,3,4} \eta_{ab} X^a X^b \, .
\end{align}
Since $\cH_\L$ is irreducible under $\mso(4,1)$, it must follow that $\cR^2 
\sim \one$. Indeed, one finds
\begin{align}
 X_a X^a =  -\frac {r^2}4 \hat N  (\hat N+4) =  - \frac {r^2}4 (n^2-4) 
\eqqcolon - R^2
 \label{R2-fuzzy-explicit}
\end{align}
where $n= |\hat N +2| = 0,1,2,\ldots$.
Note that $\cR^2$ is positive for $n=0,1$, which seems strange because $X^0$ is 
positive.  
However, this is a quantum artifact, and the expectation values $\left\langle 
X^a \right\rangle$ under coherent states still sweep out the usual $H^4$. 
Additionally we compute the quadratic $SO(4,1)$ and $SO(4,2)$ Casimir 
operators 
\begin{align}
 C^2[\mso(4,1)] &= \sum_{a<b\leq 4} \cM_{ab} \cM^{ab} 
   = \frac{1}{2} (n^2-4) \; ,
   \label{C2-minirep} \\
 C^2[\mso(4,2)] &= \sum_{a<b\leq 5} \cM_{ab} \cM^{ab}  
    = \frac{3}{4} (n^2-4) \ .
\label{C2-SO42-minirep} 
\end{align}
We note that \eqref{C2-minirep} agrees with  \cite{Hasebe:2012mz}.
Further identities can be obtained from the $\mso(6)_\C$ identity 
\eqref{Sigma-Sigma-contract}, which
entails
\begin{align}
  \eta_{cc'}\cM^{ac} \cM^{bc'} + (a \leftrightarrow b) 
   = \frac{1}{2} (n^2-4)  \eta_{ab} \ .
   \label{so6-MM-rel}
\end{align}
This implies the $\mso(4,1)$ relation
 \begin{align} 
\eta_{cc'}\Theta^{ac} \Theta^{bc'} + (a \leftrightarrow b)
  &= r^2 \left( 2R^2\eta_{ab} + \left(X^a X^b + X^b X^a \right) \right) \ .
 \label{Theta-constraint}
\end{align} 
These correspond to  \eqref{so42-MM-rel-class}, \eqref{so41-MM-rel-class}. 
Moreover, one finds
\begin{align}
 X_b \cM^{ab} +  \cM^{ab} X_b = 0 \; ,
 \label{M-tangential}
\end{align}
which means that the $SO(4,1)$ generators $\cM^{ab}$ are tangential to $H_n^4$. 
 Another interesting identity is 
\begin{align}
\begin{aligned}
\epsilon_{abcdef}\cM^{ab} \cM^{cd} &=  4 n \cM_{ef} \\
\epsilon_{abcde}\cM^{ab} \cM^{cd} &=  4 nr^{-1} X_e
  \end{aligned}
  \label{epsilonMM-id}
\end{align}
cf.\ \eqref{epsilonMM-id-1}, \eqref{epsilonMM-id-2}.
Finally, the self-duality relation \eqref{selfdual-class} becomes
\begin{align}
 \epsilon_{abcde}\cM^{ab} X^{c} 
  = n r \cM_{de} \ .
 \label{selfdual-fuzzy}
\end{align}
To summarize, we have found counterparts for all relation of the 
classical geometry in section \ref{sec:CP12-H4}, 
which vindicates the choice of representation $\cH_n$.

\subsection{Wave-functions and spin Casimir}
\label{sec:wavefunct-spin-modes-H}
Given a  representation $\cH_n$ of $SO(4,2)$,
the most general ``function''  in $\End(\cH_n)$  can always be expanded as follows
\begin{align}
 \phi = \phi(X) + \phi_{ab}(X) \cM^{ab} + \ldots  \qquad \in \End(\cH_n) \ 
\eqqcolon \cC \;,
 \label{phi-M-expand}
\end{align}
which transform in the adjoint representation $\cM^{ab} \mapsto 
[\cM^{ab},\cdot]$ of $\mso(4,2)$.
The $\phi_{ab}(X)$ will be interpreted as quantized tensor fields on $H^4$, 
which transform under $SO(4,1)$.
We define an $SO(4,2)$-invariant inner product on $\cC$ via
\begin{align}
 \left\langle\phi,\psi\right\rangle = \tr_\cH\left(\phi^\dagger \psi\right) \ .
 \label{inner-prod-End}
\end{align}
For polynomials generated by the $X^a$, this trace diverges. 
However this is only an IR-divergence, and we are mainly interested in 
normalizable fluctuations corresponding to physical scalar fields.
Technically speaking, we will be working with Hilbert-Schmidt operators in 
$\End(\cH)$.
These can be expanded into modes obtained by 
decomposing $\End(\cH)$ into unitary representations of the isometry group 
$SO(4,1)$ of the background. 

\paragraph{Spin Casimir.}
To proceed, we require a characterization of the above $SO(4,1)$ modes 
in terms of a Casimir operator which measures spin. One can achieve this by the 
$SO(4,1)$-invariant 
\begin{align}
\cS^2 \coloneqq C^2[\mso(4,1)] +  r^{-2}\Box = \sum_{a<b\leq 4} 
[\cM^{ab},[\cM_{ab},\cdot]] + r^{-2} [X_a,[X^a,\cdot]] \; ,
\label{Spin-casimir}
 \end{align}
which measures the spin along the $S^2$ fiber. To understand this,
we locally decompose $\mso(4,1)$ , for example at the reference point 
\eqref{eq:reference_point}, into $\mso(4)$ generators $\cM^{\mu\nu}$ and 
translation generators $P^\mu = \frac{1}{R}\cM^{\mu 0}$.
Then $C^2[\mso(4,1)] = - R^2 P_\mu P^\mu + C^2[\mso(4)]$, and 
 $R^2 P_\mu P^\mu \sim - r^{-2} [X_\mu, [X^\mu,\cdot]]$ if acting on functions 
$\phi(x)$, cf.\ \eqref{Box-delsquare}.
 Therefore $\cS^2 \sim C^2[\mso(4)]$ should vanish on scalar functions $\phi(X)$ 
on $H^4$, but not on higher-spin functions involving $\theta^{ab}$. 
 We will see that this is indeed the case, and
$\End(\cH_n)$ contains modes up to spin $n$ as measured by $\cS^2$ 
\eqref{End-decomposition-Cn}.
We also observe
\begin{align}
\begin{aligned}
 C^2[\mso(4,2)] &=  C^2[\mso(4,1)] -  r^{-2}\Box = \cS^2 - 2  r^{-2}\Box \;, \\
  C^2[\mso(4,2)] &= 2 C^2[\mso(4,1)] - \cS^2  \ .
  \end{aligned}
\end{align}
Note that $\cS^2, \ \Box$, and $C^2[\mso(4,2)]$ commute and can be 
diagonalized simultaneously.

\paragraph{Higher-spin modes on $H^4_n$.}
 To determine the spectrum of $\cS^2$ for the modes  in \eqref{phi-M-expand}
we first prove the following identity for any $f\in\cC$:
\begin{align}
 \cS^2(\{f, X_a\}_+) = \{\cS^2 f, X_a\}_+   \; ,
 \label{Spin-X-id}
\end{align}
where $\{\cdot,\cdot\}_+$ denote the anti-commutator.
To see this, consider  
\begin{align}
 \cS^2(f X_a) &= (\cS^2 f) X_a +  [\cM^{cd},f][\cM_{cd}, X_a] + 2r^{-2}[X^{c},f][X_{c},X_a]  \nn\\
   &= (\cS^2 f) X_a + 2 i [\cM^{ad},f] X_d  - 2i[X^{c},f] \cM_{ca}  \nn\\
  &= (\cS^2 f) X_a + 2 i [\cM^{ad}X_d,f]   - 2 i \cM^{ad}[X_d,f]   - 2i[X^{c},f] \cM_{ca} \ .
\end{align}
Similarly,
\begin{align}
 \cS^2(X_a f) 
  &= X_a(\cS^2 f)  + 2 i [X_d\cM^{ad},f]   - 2 i [X_d,f]\cM^{ad}   - 2i\cM_{ca} [X^{c},f] 
\end{align}
and adding them yields \eqref{Spin-X-id}.
Next, starting from 
\begin{align}
 \Box_X X^a &= -4 X^a = - C^2[\mso(5)] X^a 
\end{align}
this identity immediately implies
\begin{align}
 \cS^2 P_n(X) &= 0
\end{align}
for totally symmetrized polynomials $P_n(X)$ in $X$.
More generally, we show in appendix \ref{sec:funct-spin-coherent} that this 
holds for any scalar field $\phi$ on $H^4$ quantized via coherent states, i.e. 
\begin{align}
   \cS^2 \phi = 0 \qquad \text{for any} \quad \phi = \int_{\C P^{1,2}} \phi(x) 
|x\rangle\langle x| \ .
   \label{spin-scalar-field}
\end{align}
As a next step, we consider the higher-spin fields. Using
\begin{align}
 2 C^2[\mso(4,1)] \cM_{ab} &= [\cM^{cd},[\cM_{cd}, \cM_{ab}]] 
  = 12 \cM_{ab}  \nn\\
\Box_X  \cM_{ab} &= [X^c,[X_c, \cM_{ab}]] 
  = -2\cM_{ab} \nn\\
 \cS^2 \cM_{ab} &= 4 \cM_{ab}
\end{align}
we find
\begin{align}
 \cS^2 \phi^{(1)} = 4 \phi^{(1)} \qquad \text{for any} \quad \phi^{(1)} = 
\phi_{ab}(x) \cM^{ab}
\end{align}
with quantized functions $\phi_{ab}(X)$ on $H^4$ in \eqref{spin-scalar-field}, 
etc.
We can similarly compute $\cS^2$ for any irreducible polynomial function in 
$\cM^{ab}$, and obtain
\begin{align}
 \cS^2(\Xi^s_{\und{\a}}) &= 2s(s+1) \Xi^s_{\und{\a}},  \qquad \quad
 \Xi^s_{\und{\a}} = 
 (\cP_{\und{\a}})_{a_1b_1 \ldots a_sb_s}\, \cM^{a_1b_1} \ldots  \cM^{a_sb_s}
\end{align}
where $\cP_{\und{\a}} \sim \tiny \Yvcentermath1 \yng(3,3)$ is a 2-row rectangular Young projector.
The restriction to these Young diagrams follows from the commutation 
relations \eqref{M-M-relations-noncompact} and the self-duality relation 
\eqref{epsilonMM-id}.
This leads to the  decomposition
\begin{align}
\boxed{
 \ \  \cC \coloneqq \End(\cH_n) = \bigoplus\limits_{s=0}^\infty \ \cC^s,  \qquad 
\cS^2 |_{\cC^s} =  2s(s+1) \ 
}
 \label{End-decomposition-Cn}
\end{align} 
where $\cC^s$ is the eigenspace of  $\cS^2 = 2s(s+1)$.
We refer to appendix \ref{sec:funct-spin-coherent} for the details.
Of course the $\cC^s$ contain also forms of the type
$\phi_{\und{\a}}(X)\, \Xi^s_{\und{\a}}$. However since
the multiplication does not respect the grading, we can only say that 
$\cC^s$ is the \emph{quantization} of tensor fields $\phi_{\und{\a}}(x)$
 taking values in the vector space spanned by $\Xi^{\und{\a}}$, 
\begin{align}
 \cC^s  \ \ni \ \cQ(\phi_{a_1\ldots a_s;b_1 \ldots b_s}(x) m^{a_1b_1} \ldots 
m^{a_sb_s})
   \ \equiv \cQ(\phi_{\und{\a}}(x)\, \Xi^{\und{\a}}) \  ,  \quad \ s \leq n \ .
\label{spin-n-expand}
 \end{align}
 where $\Xi^{\und{\a}}$ denotes both the polynomials in $\cM^{ab}$ and 
$m^{ab}$.
The truncation\footnote{We do not claim that for example the algebra of 
functions generated by $P^a = \cM^{a4}$ is truncated at order $n$; this is not 
the case.} 
at $n$ only holds for the semi-classical regime, since the  classical expressions \eqref{spin-n-expand} with $s>n$ are 
annihilated by $\cQ$.
In fact they correspond to spin $s>n$ irreps of the local $SO(4)$, 
which are not supported by the local fiber spanned by the coherent states, 
which is a fuzzy 2-sphere $S^2_n$. See appendix \ref{sec:funct-spin-coherent} 
for more details.

This is a very remarkable structure, which leads to higher-spin fields on $H^4$.
For small $n$, the uncertainty scale  $L_{\rm NC}^2 \approx R^2$, see 
\eqref{LNC-def}, is set by the curvature scale  of $H^4 \subset \R^{1,4}$, so  
that the space is far from classical.
Nevertheless, the case of small $n$ may  be interesting after projection to the 
cosmological space-time $\cM^{3,1}$ as discussed in section 
\ref{sec:projection-Lorentzian}.
\paragraph{The $n=2$ case.}
The case $\hat{N} = 0 = n-2$ is special, because then $C^2[\mso(4,1)] = 0 =R^2$.
To avoid this we will assume $n\neq 2$ in this paper.
\paragraph{The $n=0$ case.}
In that case, \eqref{epsilonMM-id} gives 
\begin{align}
\epsilon_{abcdef}\cM^{ab} \cM^{cd} = 0 \; ,
  \label{n=0-epsilon}
\end{align}
which is a relation in the Joseph ideal \cite{Joung:2014qya}.
Then the $\cM^{ab}, \ a,b=0,\ldots,5$
generate Vasiliev's higher-spin algebra associated to $\mso(4,2)$. 
However here we will not aim for a higher-spin theory on $AdS^5$, but
reduce $\cH_n$ for $n\neq 0$ to the $\mso(4,1)$ generators  $\cM^{ab}$, 
$a,b=0,\ldots,4$, and the remaining $X^a$ generators.
Then the  $\cM^{ab}$, $a,b=0,\ldots,4$  satisfy relations which are locally  
similar to the $\hs$ algebra of $\mso(4,1)$, while the 
$X^a$ generate the underlying space.

\subsection{Semi-classical limit and Poisson calculus}
\label{sec:wavefunct-semiclassical-H}
Now consider the semi-classical limit of fuzzy $H_n^4$, which is obtained 
for large $n$, and is indicated by $\sim$. Then $X^a \sim x^a$ and $\Theta^{ab} \sim \theta^{ab}$, and
the above relations on $\cH^4_n$ reduce to
\begin{subequations}
\begin{align}
  x_a x^a &= -R^2, \label{eq:xx=R} \; ,\\
 \theta^{ab} x_b &= 0  \; ,\\
 \epsilon_{abcde} \theta^{ab} x^{c} &= n r\theta_{de} \sim 2 R   \theta_{de}     
 \; ,  \label{SD-H-class}\\
  \g^{bb'} \coloneqq \eta_{aa'}\theta^{ab} \theta^{a'b'} &= \frac{L^4_{NC}}{4}  
P^{bb'}\; ,
 \label{theta-constraint}
 \end{align}
 \end{subequations}
  where  the scale of non-commutativity is
\begin{align}
 L^4_{NC} &\coloneqq \theta^{ab}\theta_{ab} = 4 r^2 R^2 \ .
 \label{LNC-def}
\end{align}
Here
 \begin{align}
 P^{ab} = \eta^{ab} + \frac{1}{R^2}x^a x^b  
 \qquad \text{with} \quad 
 P^{ab} x_b = 0  
 \quad \text{and} \quad 
 P^{ab}P^{bc}=P^{ac}
\end{align}
is the Euclidean projector on $H^4$
(recall that $H^4$ is a Euclidean space).
Hence the algebra of functions on fuzzy $H^4_n$  reduces 
for large $n$  to the algebra of functions 
\begin{align}
 \End(\cH_n) \ \sim \  \cC(\C P^{1,2})\ = \ \oplus_s\ \cC^s
\end{align}
on the classical Poisson manifold $\C P^{1,2}$,
as described in section \ref{sec:CP12-H4}. We denote the eigenspaces of $\cS^2$ again with $\cC^s$, which are now
modules over  the algebra $\cC^0 = \cC(H^4)$ of functions on $H^4$, thus 
encoding the structure of a bundle over $H^4$.
From now on we work in the semi-classical limit.
The bundle structure can be made more explicit by writing
\begin{align} 
 \theta^{ab} = \eta^{ab}_i J^i 
 \label{tHooftsymbol-H4}
\end{align}
as in \eqref{tHooft-classical},
where $\eta^{ab}_i$ are the tangential self-dual t'Hooft symbols;
 ``tangential'' follows from $x_a \theta^{ab} = 0$.
The $J^i$ transform as vectors of the local $SU(2)_L \subset SO(4)$, and describe the 
internal $S^2$ fiber.
\paragraph{Derivatives.}
It is useful to define the following derivations (cf.\ \cite{Sperling:2017gmy})
\begin{align}
  \eth^a \phi \coloneqq -\frac{1}{r^2 R^2}\theta^{ab}\{x_b,\phi\} 
  = \frac{1}{r^2 R^2}x_b \{\theta^{ab} ,\phi\}, \qquad 
\phi\in\cC \; ,
\end{align}
which are tangential  $x^a\eth_a = 0$, satisfy the Leibniz rule, and  
are $SO(4,1)$-covariant.
Equivalently,
\begin{align}\boxed{
 \{x^a,\cdot\} = \theta^{ab}\eth_b \, . \ 
 }
 \label{x-nabla-id}
\end{align}
In particular, the following holds:
\begin{align}
 \eth^a x^c &= -\frac{1}{r^2 R^2}\theta^{ab}\{x_b,x^c\} = P_T^{ac} \, \nn\\
 [\eth_a,\eth_b]\phi &=  -\frac 1{r^2 R^2} \{\theta_{ab},\phi\} 
 \label{del-del-CR}
\end{align}
as shown in appendix \ref{sec:semiclass-id-appendix}.
The first line shows that $\eth$ act as isometries on functions, 
such that the $\eth_a$ can be viewed as a set of five Killing vector fields on 
$H^4$ with Lie bracket  given by \eqref{del-del-CR}.
Furthermore,
\begin{align}
 \eth^a \theta^{cd} &= \frac{1}{r^2R^2}\theta^{ab}\{x_b,\theta^{cd}\} 
   = -\frac{1}{R^2}\theta^{ab}(\eta^{bc}x^d - \eta^{bd}x^c) \nn\\
   &= \frac{1}{R^2}(-\theta^{ac}x^d + \theta^{ad}x^c) .
   \label{del-M}
\end{align}
We also note that the $SO(4,1)$ rotations of scalar functions are generated by
$\{\cM^{ab},\cdot\}$, which can be written as 
\begin{align}
 \{\cM^{ab},\phi\} &= - (x^a\eth_b - x^b \eth_a)\phi , \qquad \phi \in \cC^0 \ . 
  \label{M-ad-explicit}
 \end{align}
To see this, it suffices to verify the action on the $x^c$ generators,
 \begin{align}
 \{\cM^{ab},x^c\} &= - (x^a\eth_b - x^b \eth_a)x^c = - (x^a P^{bc} - x^b P^{ac}) = - (x^a \eta^{bc} - x^b \eta^{ac})
 \end{align}
  since both sides are derivations. 
 Finally, the semi-classical limit of the $\Box$ operator \eq{X-Box} can be expressed in terms of the derivatives as 
follows: 
\begin{align}
 \Box \phi &= - \{x_a,\{x^a,\phi\}\} 
  = - \{x_a,\theta^{ab} \eth_b \phi \} \nn\\
  &=  - \{x_a,\theta^{ab}\} \eth_b \phi - \theta^{ab} \{x_a,\eth_b\phi \}  \nn\\
  &= r^2 \{\cM^{ab}, x_a\} \eth_b\phi - \theta^{ab} \theta^{ac} \eth_b\del_c\phi\nn\\
  &= - r^2 R^2 P^{ab} \eth_a\eth_b \phi
     \label{Box-delsquare}
\end{align}
for any $\phi \in \cC$.

\paragraph{Connection.}
We define an $SO(4,1)$-covariant connection on the module $\cC$ 
\eqref{C-module-semiclass} by \cite{Sperling:2017gmy}
\begin{align}
 \nabla = P_T \circ \eth
\end{align}
so that for $\nabla_a \equiv \nabla_{\eth_a}$ 
\begin{align}
 \nabla_a \phi_b &= \del_a \phi_b - \frac{1}{R^2}x_b \phi_a , \nn\\
 \nabla_a \phi_{bc} &= \del_a \phi_{bc} - \frac{1}{R^2}(x_b \phi_{ac} + x_c \phi_{ba})
 \label{nabla-a-phi-b}
\end{align}
etc.\ if $\phi_a$, $\phi_{ab}$ are tangential. 
Comparing with \eqref{del-M} and using  \eqref{M-ad-explicit}
it follows that the connection is  compatible with $\theta^{ab}$, i.e.\
\begin{align}
 \nabla\theta^{ab}  &= 0, \qquad 
 \nabla\{f,g\}  = \{\nabla f,g\} + \{f,\nabla g\}
 \label{nabla-poisson}
 \end{align}
 and  $\nabla_a P_{bc} =  0$.
 The associated curvature
 \begin{align}
  \cR_{ab} \coloneqq \cR[\eth_a,\eth_b] =  [\nabla_a,\nabla_b] - 
\nabla_{[\eth_a,\eth_b]} \ 
 \label{nabla-relations}
\end{align}
is computed in appendix \ref{curvature-general}, and reduces to the Levi--Civita 
connection on tensor fields.
Thus $H^4_n$ is a quantum space which is fully $SO(4,1)$-covariant, and 
we have found a calculus which is defined solely in terms of the Poisson 
bracket, i.e.\ the semi-classical limit of matrix commutators. This is very 
important for the present non-commutative framework.

\paragraph{Averaging over the fiber.}
There exist a canonical map 
\begin{align}
\begin{aligned}
 [\cdot]_0: \cC(\C P^{1,2}) &\to \cC(H^4)  \\
   f(\xi) &\mapsto f(x) = \int_{S^2} f(\xi)
   \end{aligned}
\end{align}
defined by integrating over the fiber at each $x\in H^4$.
This projects the functions on the total space to functions on the base space. 
On fuzzy $S^4_N$, this averaging can be defined in terms of a  $SO(5)$-invariant projection to some 
sub-space of $\End(\cH)$. For $H^4_n$, $[\cdot]_0$ is nothing but the 
projection 
to $\cS^2 = 0$ i.e.\ to $\cC^0$, as discussed below.

Explicitly, the averaging $[\cdot]_0$  over the internal $S^2$ is given  by 
\begin{align}
  \left[\theta^{ab}  \theta^{cd}\right]_{0} 
  &=\frac{1}{12}  L_{NC}^4(P^{ac} P^{bd} - P^{bc} P^{ad} + \varepsilon^{abcde} \frac 1{R} x^e) \nn\\
   &= \frac{r^2 R^2}{3}  (P^{ac} P^{bd} - P^{bc} P^{ad} + \varepsilon^{abcde} 
\frac 1{R} x^e) \; .
\label{average-H}
\end{align}
One can generalize the averaging to higher powers of $\theta^{ab}$, e.g.\ 
\cite{Sperling:2017gmy}
\begin{align} 
 [\theta^{ab} \theta^{cd } \theta^{ef} \theta^{gh}]_0 &= \frac{3}{5} \left(
 [\theta^{ab} \theta^{cd}]_0 [\theta^{ef} \theta^{gh}]_0
 +[\theta^{ab} \theta^{ef}]_0 [\theta^{cd} \theta^{gh}]_0 
 +[\theta^{ab} \theta^{gh}]_0 [\theta^{cd} \theta^{ef}]_0
\right) \, .
\label{averaging-theta4}
\end{align}
Alternatively, one could proceed to  define a star product for functions on 
$H^4$, which is presumably commutative, but not associative, 
in analogy to the case of $S^4_N$ \cite{Ramgoolam:2001zx}. 
On the other hand, for $n=0$ there is nothing to project, and 
the full algebra of functions on $H^4$ is non-commutative and associative without extra generators.

\paragraph{Integration.}
As for any quantized coadjoint orbit, the trace on $\End(\cH)$ corresponds to 
the integral over the underlying symplectic space, defined by the symplectic 
volume form. Explicitly,
\begin{align}
 \Tr \cQ(\phi) =  \int d\mu\, \phi \ = \int_{H^4} \rho [\phi]_0 \ , 
\quad\qquad \rho \ \mathrel{\hat{=}} \ \frac{\dim(\cH)}{\vol(H^4)} 
 \label{integral-trace}
\end{align}
replacing the ill-defined fraction $\frac{\dim(\cH)}{\vol(H^4)}$  
with the symplectic volume form $d\mu$, which reduces to $\rho$ on $H^4$.
This is best seen via coherent states \eqref{quantization-map}. 
We will often drop  $d\mu$ and $\cQ$ in the semi-classical limit. Finally,
  note that the $\eth^a$  are \emph{not} self-adjoint  under the integral, but
\begin{align}
 \int \eth_a f\, g = - \int  f\, \eth_a g + \frac{1}{\theta R^2}\int f \{x_b,\theta^{ab}\} g 
  = - \int  f\, \eth_a g - \frac{4}{R^2}\int x^a f  g 
  \label{partint}
\end{align}
using $\{x_b,\theta^{ba}\}= 4 r^2 x^a$.

\section{Functions, tensors and higher-spin modes}  
\label{sec:irred-tensors-in-C}
We have seen that the  algebra $\End(\cH_n)$ of fuzzy $H^4_n$ reduces in the 
semi-classical limit to the algebra of functions on $\C P^{1,2}$.
The results of section \ref{sec:wavefunct-spin-modes-H} 
provide  a more detailed  decomposition of $\cC$ into modules 
\eqref{End-decomposition-Cn}
\begin{align}
\boxed{
 \ \ \cC = \bigoplus\limits_{s=0}^\infty \cC^s  \quad \ni \ 
 \phi^s_{a_1\ldots a_s;b_1 \ldots b_s}(x) \ m^{a_1b_1} \ldots m^{a_sb_s} 
 \ \equiv \phi^s_{\und{\b}}(x)\, \Xi^{\und{\b}},
}
 \label{C-module-semiclass}
\end{align}
over the algebra of functions $\cC^0$ on $H^4$, due to \eqref{Spin-X-id}. 
This means that $\cC$ is a bundle over $H^4$, whose structure is determined by the constraints 
\eqref{so42-MM-rel-class}, \eqref{epsilonMM-id-1} and \eqref{selfdual-class}.
An explicit description is given by the
one-to-one map \footnote{Note that $\Gamma^{(s)} H^4$ 
is not a module over $\cC^0$, hence this is not a module isomorphism.
In \cite{Sperling:2017gmy}, a different convention was used 
for the map $\phi_{a_1\ldots a_s}(x) \leftrightarrow \phi^{(s)}$. The present 
convention avoids the appearance of square-roots of Casimirs in this map.}

\begin{align}
 \boxed{ \
\begin{aligned}
    \Gamma^{(s)} H^4 \  &\rightarrow  \    
\cC^s  \\
  \phi^{(s)}_{a_1 \ldots  a_s}(x)   &\mapsto  \ 
 \phi^{(s)} =  \{x^{a_1},\ldots\{x^{a_s},\phi^{(s)}_{a_1\ldots a_s}\}\ldots\} 
\ .
  \end{aligned} \
  }
 \label{psi-iso}
\end{align}
Here $\Gamma^{(s)}H^4$ denotes the space of totally symmetric,
traceless, divergence-free rank $s$ tensor fields on $H^4$, which are identified  
with (symmetric tangential divergence-free traceless)
tensor fields $\phi^{(s)}_{a_1 \ldots  a_s}$ with  $SO(4,1)$ indices,
as discussed in 
section \ref{sec:local_decomp} and in \cite{Sperling:2017gmy}.
The inverse map of \eqref{psi-iso} (up to normalization) can be given by 
\begin{align}
 \cC^s \ni \ \phi^{(s)} \mapsto \{x^{a_1}, \ldots 
\{x^{a_s},\phi^{(s)}\}\ldots \}_0\ \in \Gamma^{(s)} H^4
\label{eq:inverse_phi}
\end{align}
which is  symmetric due to $[\cdot]_0$, as well as traceless, divergence-free 
and tangential.
These statements are analogous to the results in \cite{Sperling:2017gmy}.

Some comments on the map \eqref{psi-iso} are in order. We show in sections 
\ref{subsec:spin-1}--\ref{subsec:spin-s} that pure divergence modes would 
be mapped to zero by \eqref{psi-iso}. 
Injectivity will be shown below by establishing \eqref{eq:inverse_phi}.
To see surjectivity,  it suffices to consider
the vicinity of a chosen reference point, for 
instance \eqref{eq:reference_point}. Then polynomial functions suffice to approximate 
any element in $\cC^{s}$. Then the $\mso(4,2)$ representation theory allows to  characterize all polynomials 
in $\cC^{s}$ uniquely by Young diagrams, as explained in detail in 
\cite[section 3]{Sperling:2017gmy}. These in turn are captured by the map \eqref{psi-iso},
and an alternative inverse map can be used \cite{Sperling:2017gmy}
\begin{align}
 \cC^s \ni \ \phi^{(s)}_{a_1\ldots a_s;b_1 \ldots b_s}(x)\, m^{a_1 b_1} \ldots m^{a_s b_s} 
\mapsto \phi^{(s)}_{a_1\ldots a_s;b_1 \ldots b_s}(x)\, x^{b_1} \ldots x^{b_s}\ 
\in \Gamma^{(s)} H^4 \ ,
\end{align}
which is equivalent to \eqref{eq:inverse_phi} up to normalization. 

Hence $\cC^s$ encodes one and only one irreducible spin $s$ field on $H^4$, 
given by square-integrable tensor fields on $H^4$. 
The generators $\Xi^{\und\b}$ form a basis of irreducible totally symmetric 
polynomials in $m^{ab}$,  i.e.\ of Young tableaux
\begin{align}
 \oplus\ {\tiny \Yvcentermath1 \yng(3,3)} \  \cong  \ \hs \coloneqq 
\oplus_{s=1}^\infty\,  \Xi^{s,\und{\b}} \; . 
 \label{HS-def}
\end{align}
As in \cite{Sperling:2017gmy}, $\hs$ is closely related to the higher-spin 
algebra of Vasiliev theory\footnote{Note that $\cH_0$ is a minirep
of $SO(4,2)$ but not of $SO(4,1)$. 
This explains why we get an extension of Vasiliev's $\hs$ algebra by functions 
of $X$.}.
Hence $\cC$ can be viewed as functions on $H^4$ taking values in $\hs$.

\subsubsection{Spin 1 modes}
\label{subsec:spin-1}
The unique spin 1 field is encoded in $\phi_{ab} m^{ab}$. 
According to the above statements, it can be  expressed in terms of a
tangential, divergence-free  tensor field $\phi_a \in \cC^0$ on $H^4$, i.e.
\begin{align}
 x^a \phi_a = 0 = \eth^a\phi_a \ .
\end{align}
Given such a $\phi_a$, we define 
\begin{align}
\begin{aligned}
 \phi^{(1)} &\coloneqq \{x^a,\phi_a\}  = \theta^{ab}\eth_b \phi_a 
    = \frac 12\theta^{ab} \cF_{ab} \quad \in \cC^1 , \\
  \cF_{ab} &= \eth_b \phi_a - \eth_a\phi_b
  \end{aligned}
  \label{phi-1-def}
\end{align}
which encodes the field strength of the vector field.
This is not tangential, but
\begin{align}
 x^a \cF_{ab} = x^a\eth_b \phi_a - x^a \eth_a\phi_b =  x^a\eth_b \phi_a  = - 
\phi_b
\end{align}
using \eqref{x-del-phi-id}.
Conversely, the ``potential'' $\phi_a(x)$ is recovered from $\phi^{(1)} $ via a 
projection 
\begin{align} 
\boxed{
 -\{x_a,\phi^{(1)}\}_0  = \a_1 (\Box - 2 r^2) \phi_a,
 \qquad \a_1 = \frac{1}{3}
 }
 \label{x-phi-comm-spin1}
\end{align}
where $\phi^{(1)}=\{x^c,\phi_c\}$ for a tangential, divergence-free $\phi_a \in 
\cC^0$.
The derivation of \eqref{x-phi-comm-spin1} is detailed in 
\eqref{eq:aux_calc_spin1_1}--\eqref{xdeldelphi}.
The generalization of this formula for higher-spin is discussed below.
If $\phi_a$ is an irrep of $SO(4,1)$, we may abbreviate this as
\begin{align}
 -\{x_a,\phi^{(1)}\}_0  \eqqcolon \hat\a_1 \phi_a 
 \label{x-phi-comm-spin1-short}
\end{align}
where $\hat\a_1$ is the value of $\a_1(\Box-2r^2)$ on $\phi_a$.
\paragraph{Pure gauge modes.}
Finally, one can verify that for $\tilde\phi_a = \eth_a\phi$, 
the associated ``field strength'' tensor is $\cF_{ab} \propto 
\{\theta^{ab},\phi\}$, but the field strength form $\phi^{(1)}$ vanishes 
identically:
\begin{align}
 \phi^{(1)} &= \{x^a,\eth_a\phi\} = 0
\end{align}
using \eqref{x-del-id}.
This expresses the gauge invariance (or irreducibility) of $\phi^{(1)}$.
\subsubsection{Spin 2 modes}
\label{subsec:spin-2}
Similarly, spin 2 modes  can be realized in terms of a tangential, 
divergence-free, traceless, symmetric rank 2 tensor 
$\phi_{ab}(x) = \phi_{ba}(x) \in \cC^0$, i.e.
\begin{align}
 x^a \phi_{ab} = 0 = \eth^a\phi_{ab} = \eta^{ab}\phi_{ab}\ .
\end{align}
We define the associated ``potential form''
\begin{align}
 \phi^{(2)}_a = \{x^b,\phi_{ab}\} = \theta^{bc}\eth_c \phi_{ab} 
  = -\omega_{a;cb}\theta^{cb} \quad \in \cC^1 \; ,
  \label{spin-2-connection}
\end{align}
which can be viewed as $\mso(4,1)$-valued one-form with 
\begin{align}
 \omega_{a;cb} = \frac{1}{2} (\eth_c \phi_{ab} -\eth_b \phi_{ac} ) \,.
 \label{omega-connection-def}
\end{align}
Note that $\phi_{c}^{(2)}$ is indeed tangential,
\begin{align}
 x^c \phi_{c}^{(2)} &= x^c \{x^{a},\phi_{ca}\}
  = -  \{x^{a},x^c \} \phi_{ca} = 0 \; .
  \label{phi2-a-tangential}
\end{align}
The $\mso(4)$-valued components of $\phi^{(2)}_a$
correspond to the spin connection, while its translational components
\begin{align}
 x^c\omega_{a;cb} = - \frac 12 x^c\eth_b \phi_{ac} = \frac 12 \phi_{ab}
\end{align}
reduce to  $\phi_{ab}$, as on fuzzy $S^4_N$  \cite{Sperling:2017gmy}.
The  ``field strength form'' corresponding to $\phi^{(2)}_a$ is
\begin{align}
 \phi^{(2)} &= \{x^a,\phi^{(2)}_a\} \eqqcolon 
\frac{1}{2} \theta^{ad}\cR_{ad}[\phi]  \nn\\
 &=  -\theta^{ad}\eth_d(\omega_{a;cb}\theta^{cb}) \
   = -\theta^{cb}\theta^{ad}\eth_d\omega_{a;cb} - \theta^{ad}\omega_{a;cb} 
\eth_d\theta^{cb}  \nn\\
   &= \frac 12 \theta^{ad}\theta^{cb}(\eth_a\omega_{d;cb} 
-\eth_d\omega_{a;cb})\nn\\
   &\eqqcolon \frac 12\theta^{ad}\theta^{bc}\cR_{ad;bc}[\phi]   \qquad \in \cC^2
 \label{spin-2-field-R}
\end{align}
noting that the $\eth_d\theta^{cb} $ terms drop out for
for traceless, tangential $\phi_{ab}$, using \eqref{del-M}.
This encodes the linearized Riemann  curvature tensor associated to $\phi_{ab}$,  
\begin{align}
  \cR_{ad}[\phi] &\coloneqq -\eth_a\phi_{d}^{(2)} + \eth_d\phi_{a} ^{(2)} =  
\cR_{ad;bc} \theta^{bc} \quad \in \cC^1,  \nn\\
 \cR_{ad;bc} &= \eth_a\omega_{d;bc} -\eth_d\omega_{a;bc} \nn\\
 &= \frac 12( \eth_d\eth_c \phi_{ab} -\eth_a\eth_c \phi_{db} - \eth_d\eth_b 
\phi_{ac} +\eth_a\eth_b \phi_{dc}) \; .
 \label{Riemann-tensor-def}
\end{align}
Although the $\eth_e$ do not commute among another, their commutator is  
radial due to \eqref{M-ad-explicit}, i.e.
\begin{align}
 P \cR_{ad;bc} - P\cR_{bc;ad} = 0 \; .
 \label{R-symm-tang}
\end{align}
Hence the tangential components of $P\cR_{ae;bc}[\phi]$ coincide with the usual linearized Riemann tensor.
The connection form  $\phi^{(2)}_c$ (i.e.\ $\omega$) is recovered by a 
projection
\begin{align} 
\boxed{
 - \{x_a,\phi^{(2)}\}_{1} = \a_2(\Box - 2 r^2) \phi^{(2)}_a \quad \in \cC^1,
 \qquad \a_2=\frac 25
 }
 \label{x-phi-comm-spin2}
\end{align}
generalizing \eqref{x-phi-comm-spin1}. Here, we defined 
$\phi^{(2)}=\{x^b,\{x^c,\phi_{bc}\}\}$ 
for a tangential, divergence-free, traceless $\phi_{ab} \in \cC^0$.
Similarly to the spin 1 case, \eqref{x-phi-comm-spin2} could be 
obtained via formula \eqref{averaging-theta4};
however, we provide a more transparent derivation by means of an inner product 
below.
If the underlying tensor $\phi_{ab}$ is an irrep of $SO(4,1)$, we may abbreviate this as
\begin{align}
 -\{x_a,\phi^{(2)}\}_0  \eqqcolon \hat\a_2 \phi_a ^{(2)}
 \label{x-phi-comm-spin2-short}
\end{align}
where $\hat\a_2$ is the value of $\a_2(\Box-2r^2)$ on $\phi_a^{(2)}$.
\paragraph{Spin 2 pure gauge modes.}
Again, consider a pure gauge rank 2 tensor
 \begin{align}
 \tilde\phi_{ab}^{(1)} = \nabla_a \phi_b + \nabla_b \phi_a 
 \label{spin2-puregauge}
\end{align}
which is tangential and traceless (provided $\eth^a \phi_a = 0$), but no longer 
divergence-free.
Then 
\begin{align}
 \tilde\phi_a^{(1)} &\coloneqq \{x^b,\tilde\phi_{ab}^{(1)} \}
  =  \{x^b,\eth_a \phi_b + \eth_b \phi_a - \frac{1}{R^2} (x^a \phi_b + x^b \phi_a)\}  \nn\\
  &= \{x^b,\eth_a \phi_b\}  - \frac{1}{R^2} \{x^b,x^a \phi_b\} \nn\\
  &= \eth_a \phi^{(1)}  + \frac{2}{R^2} \theta^{ac}\phi_c 
 \label{spin2-puregauge-id}
\end{align}
using \eqref{x-eth-phi-rel} and  $\phi^{(1)} = \{x^a,\phi_a\}$.
This satisfies 
\begin{align}
 \{x^a, \tilde\phi_a^{(1)}\} = \{x^a, \eth_a \phi^{(1)} + \frac{2}{R^2} 
\theta^{ac}\phi_c \} 
  = \frac{2}{R^2}\{x^a, \theta^{ac}\phi_c \} = 0
  \label{fieldstrengthform-spin1-gauge}
\end{align}
using \eqref{gaugefix-modes},
which expresses the gauge invariance of $\phi^{(2)}$.
\subsubsection{Spin $s$ modes and Young diagrams}
\label{subsec:spin-s}
As observed above, elements in $\cC^s$ 
can be identified with totally symmetric, traceless, divergence-free rank $s$ 
tensor fields $\phi_{a_1\ldots a_s}$ on $H^4$ via
\begin{align}
  \phi^{(s)} = \{x^{a_1},\ldots\{x^{a_{s}},\phi_{a_1\ldots a_{s}}\}\ldots \} \ 
\in \cC^{s} \ .
\end{align}
It is useful to define also the mixed spin $s$ objects, such as
the ''connection $(2s{-}1)$-form''
\begin{align}
 \phi_a^{(s)} = \{x^{a_1},\ldots ,\{x^{a_{s-1}},\phi_{a_1\ldots a_{s-1} 
a}\}\ldots \} \ \in \cC^{s-1}
 \label{phi-a-s}
\end{align}
which are all tangential and
associated to the underlying irreducible rank $s$ tensor field. 
Then the ``field strength'' form can be written as 
\begin{align}
 \phi^{(s)} &= \{x^a,\phi^{(s)}_a\} \eqqcolon \frac 12 \cR_{ad}[\phi] 
\theta^{ad} \nn\\
 &= \theta^{a_1 b_1}\eth_{b_1} \ldots 
\theta^{a_{s}b_s}\eth_{b_s}\phi_{a_1\ldots a_{s}}
  = \theta^{a_1 b_1} \ldots  \theta^{a_{s}b_s}\eth_{b_1} \ldots  
\eth_{b_s}\phi_{a_1\ldots a_{s}} \nn\\
  &\eqqcolon\cR_{a_1\ldots a_s;b_1\ldots b_{s}}(x)\, \theta^{a_1 b_1} \ldots  
\theta^{a_{s}b_s} 
   \quad \equiv \ \cR_{\und{\a}}(x)\, \Xi^{\und{\a}}   \qquad \in \cC^s
\end{align}
noting that the $\eth\theta^{\ldots } $ terms drop out 
for traceless tangential $\phi_{a_1\ldots  a_s}$, using \eqref{del-M}.
Here
\begin{align}
 \cR_{ad}[\phi] &\coloneqq-\eth_a\phi_{d}^{(s)} + \eth_d\phi_{a} ^{(s)} \nn\\
\cR_{b_1\ldots b_s;a_1\ldots a_{s}}(x) &= \cP\eth_{a_1} \ldots  \eth_{a_s} 
\phi_{b_1\ldots b_{s}}
 \label{Riemann-tensor-def-s}
\end{align}
is some antisymmetrized derivatives corresponding to some two row rectangular 
Young projector $\cP \sim \tiny \Yvcentermath1 \yng(3,3)$,
which can be regarded as linearized higher-spin curvature.
We will show below that the potential $\phi^{(s)}_a $ is then recovered from the following projection
\begin{align} 
\boxed{
 - \{x_a,\phi^{(s)}\}_{s-1} = \a_s (\Box - 2 r^2) \phi^{(s)}_a  \qquad \in 
\cC^{s-1}.
 }
 \label{x-phi-comm-spins}
\end{align}
If the underlying  $\phi_{a_1\ldots a_s} \in \cC^0$
is an irrep of $SO(4,1)$, we may abbreviate this as
\begin{align}
 -\{x_a,\phi^{(s)}\}_{s-1}  \eqqcolon\hat\a_s \phi_a ^{(s)}
 \label{x-phi-comm-spins-short}
\end{align}
where $\hat\a_s$ is the value of $\a_s(\Box-2r^2)$ on $\phi_a^{(s)}$.
\paragraph{Pure gauge modes.}
Finally, one can verify that a pure gauge rank $s$ tensor 
\begin{align}
 \tilde\phi^{(s-1)}_{a_1\ldots  a_s} = \nabla_{(a_s}\phi_{a_1\ldots a_{s-1})}
 \label{spins-puregauge}
\end{align}
drops out from the field strength form,
\begin{align}
 \{x^{a_1},\ldots ,\{x^{a_{s}}, \tilde\phi^{(s-1)}_{a_1\ldots  a_s}\}\ldots \}  
= 0 \ .
\end{align}
As before, this is a manifestation of the gauge invariance of $\phi^{(s)}$.
One way to see this is to move $\nabla$ out of the brackets using 
\eqref{nabla-poisson} and, finally, use \eqref{del-Q-id}.

One may wonder about the meaning of the infinitesimal transformations 
\begin{align}
  \phi^{(s)} \mapsto \{\L, \phi^{(s)}\} \ .
\end{align}
These correspond to symplectomorphisms on $\C P^{1,2}$ generated by 
the Hamiltonian 
vector field $\{\L,\cdot \}$, which mix the different spin modes in a 
non-trivial way. They 
do not correspond to the above pure gauge modes \eqref{spins-puregauge}, 
but see section \ref{sec:local-gauge}.

\subsection{Inner product and quadratic action}
It is interesting and useful to compute the inner product \eqref{inner-prod-End}
of the above spin $s$ fields $\phi^{(s)}$
defined by the trace in $\End(\cH)$.
In the spin 1 case, consider the quadratic form 
\begin{align}
  \int \phi^{(1)}\phi^{(1)} &=  \frac 14\int [\theta^{ab} \theta^{cd} ]_0  \cF_{ab} \cF_{cd}  \nn\\
   &=  \frac {r^2R^2}{12}\int (2P^{ac}P^{bd} + \frac{x^f}{R}\varepsilon^{abcdf})  \cF_{ab} \cF_{cd}  
\end{align}
which looks like the action for self-dual (abelian) Yang--Mills. 
In the spin 2 case, consider the analogous quadratic form
\begin{align}
 \int \phi^{(2)}\phi^{(2)} &=  \frac 14\int [\theta^{ae}\theta^{bc}\theta^{a'e'}\theta^{b'c'}]_0 \cR_{ae;bc}[\phi]\cR_{a'e';b'c'}[\phi]  \nn\\
  &= \frac{3}{10} \int [\theta^{ae}\theta^{a'e'}]_0[\theta^{bc}\theta^{b'c'}]_0 \cR_{ae;bc}[\phi]\cR_{a'e';b'c'}[\phi]  \nn\\
  &= \frac{1}{30} \int (2P^{aa'}P^{ee'} + \frac{x^f}{R}\varepsilon^{aea'e'f})
   (2P^{bb'}P^{cc'} + \frac{x^f}{R}\varepsilon^{bcb'c'f}) \cR_{ae;bc}[\phi]\cR_{a'e';b'c'}[\phi]  \nn\\
  &= \frac{2}{15} \int P^{aa'}P^{ee'} P^{bb'}P^{cc'} 
\cR_{ae;bc}[\phi]\cR_{a'e';b'c'}[\phi] \quad + \text {topological terms}
\label{spin2-action-phi_aux}
\end{align}
because $[\phi^{(2)}]_0 =0$. Note that we used the symmetries 
\eqref{R-symm-tang} of $\cR_{ae;bc}$ or rather of its tangential part 
$P\cR_{ae;bc}$, as the radial contributions drop out anyway.
We observe that \eqref{spin2-action-phi_aux} is a (self-dual) linearized 
quadratic gravity action\footnote{The topological terms are the linearized 
Pontrjagin and Euler class (i.e.\ Gauss--Bonnet term).}, which can be written 
in 
terms of the $\cR_{ab}$ ``forms`` as follows:
\begin{align}
 \int \phi^{(2)}\phi^{(2)} &=  \frac 14\int [\theta^{ae}\theta^{bc}\theta^{a'e'}\theta^{b'c'}]_0 \cR_{ae;bc}\cR_{a'e';b'c'}  \nn\\
  &= \frac{3}{10} \int [\theta^{ae}\theta^{a'e'}]_0 \cR_{ae;bc}\theta^{bc}\cR_{a'e';b'c'}\theta^{b'c'}   \nn\\ 
  &= \frac{3}{10} \int [\theta^{ae}\theta^{a'e'}]_0 \cR_{ae}\cR_{a'e'}   \nn\\
 &=  \frac{2r^2 R^2}{5}\int (P^{ac}P^{bd} - P^{ad}P^{bc}  + \frac{x^e}{R} \varepsilon^{abcde}) 
      \eth_b \phi_a^{(2)} \eth_d \phi_c^{(2)} \ .
 \label{spin2-action-phi}
 \end{align} 
Similarly for spin $s$, we have 
\begin{align}
 \int \phi^{(s)}\phi^{(s)}  &= \int [\theta^{ae}\theta^{a'e'}\cR_{ae}\cR_{a'e'}]_0  \nn\\
 &=  \a_s r^2 R^2\int (P^{ac}P^{bd} - P^{ad}P^{bc}  + \frac{x^e}{R} \varepsilon^{abcde}) 
      \eth_b \phi_a^{(s)} \eth_d \phi_c^{(s)} 
 \label{spins-action-phi}
 \end{align} 
which is again some self-dual quadratic Fronsdal-type higher-spin action 
\cite{Fronsdal:1978rb}.
The factor $\a_s$ will be determined below.
This suggests that a matrix model based on a single $\phi \in \End(\cH)$ should 
define some higher-spin theory, which is however expected to be more or less 
trivial. 
Nevertheless it would be interesting to study the action defined by higher-order 
polynomials, and to understand its relation with Vasiliev's theory \cite{Vasiliev:1990en}.
In the remainder of this paper, we will show how a non-trivial higher-spin 
gauge 
theory arises from multi-matrix models.
\subsubsection{Projections, positivity and determination of 
\texorpdfstring{$\a_s$}{alpha-s}}
Now consider the spin $s$ modes $\phi^{(s)} \in \cC^s$ as above, determined by 
some irreducible rank $s$ tensor 
field on $H^4$. 
We have seen that this in one-to-one correspondence to a spin $s$ potential $\phi_a^{(s)} \in \cC^{s-1}$ as above.
Then 
\begin{align}
  -\int \phi_a^{(s)} \{x_a,\phi^{(s)}\}_{s-1} = -\int \phi_a^{(s)} \{x_a,\phi^{(s)}\}  = \int \{x^a,\phi_a^{(s)}\} \phi^{(s)} 
 = \int  \phi^{(s)}  \phi^{(s)} \ .
\label{a-s-normalization}
 \end{align}
 This provides the following relations:

\paragraph{Spin 1 case.}
For spin $s=1$, the projection $\{x_a,\phi^{(s)}\}_0 $ in 
\eqref{a-s-normalization} was computed in \eqref{x-phi-comm-spin1}, which gives
\begin{align}
 \int \phi^{(1)}\phi^{(1)} &=  \a_1\int \phi_a^{(1)}( \Box  - 2 r^2 )\phi_a^{(1)} \ \geq 0 .
  \label{phi-spin1-normalization}
\end{align}
for Hermitian $\phi^{(1)}$
Therefore
\begin{align}
 \a_1 =\frac{1}{3} \ .
 \label{a1-spin1-explicit}
\end{align}
and in particular $ \Box  - 2 r^2 $ is \emph{positive} on $\cC^0$.

\paragraph{Spin 2 case.}
 We can evaluate the right-hand side of \eqref{a-s-normalization} using 
\eqref{spin2-action-phi}  as
 \begin{align} 
 \int \phi^{(2)}\phi^{(2)} &= \frac{2r^2 R^2}{5}\int P^{ac}\eth_b \phi_a^{(2)} \eth^b \phi_c^{(2)}
   - \eth^d \phi_a^{(2)}  \eth^a\phi_d^{(2)} 
    + \frac{x^e}{R} \varepsilon^{abcde} \eth_b \phi_a^{(2)} \eth_d \phi_c^{(2)} \nn\\
   &= \frac{2r^2 R^2}{5}\int - \phi_a^{(2)} \eth_b\eth^b \phi_a ^{(2)}
   + \frac 1{R^2} \phi_b^{(2)} \phi_b^{(2)} - \frac{4}{R^2} \phi_a^{(2)}  \phi_a^{(2)} \nn\\
   &\qquad \quad  + \frac{1}{r^2 R^2} \phi_a^{(2)} \Big( \{\theta^{ad},\phi_d^{(2)}\} - \frac{1}{2R} \varepsilon^{abdce} x^e \{\theta_{bd}, \phi_c^{(2)}\}  \Big) \nn\\
   &= \a_2\int \phi_a^{(2)}( \Box  - 2 r^2 )\phi_a^{(2)} 
 \label{phi-spin2-normalization}
\end{align}
using \eqref{partint}, \eqref{int-tensor-id}, the self-duality relation 
\eqref{theta-del-SD} and \eqref{SD-id-del-phi}.
Therefore
\begin{align}
 \a_2=\frac{2}{5} \ .
  \label{a1-spin2-explicit}
\end{align}
This holds  in fact for any tangential divergence-free $\phi_c \in \cC^1$.
Together with  \eqref{a-s-normalization}, this establishes the formula 
\eqref{x-phi-comm-spin2}.
On the other hand, \eqref{a1-spin1-explicit} and \eqref{phi-spin1-normalization}
implies also e.g.\footnote{since $ \phi^{(2)}_a = \{x^c,\phi_{cb}\}$ for tangential traceless divergence-free $\phi_{ab} \in \cC^0$;
the  index $a$ is irrelevant here.}
\begin{align}
 \int \phi^{(2)}_a\phi^{(2)}_a
  = \frac{1}{3} \int \phi_{ab}(\Box - 2r^2)\phi_{ab} 
  \label{normalization-rank1-2}
\end{align}
if $\phi_{ab}\in\cC^0$ is divergence-free, traceless and tangential (by fixing one index).

\paragraph{Generic spin $s$ case.}
In the generic case, we obtain similarly
\begin{align}
  \int \phi^{(s)}\phi^{(s)} &=  \int \phi_a^{(s)}\hat\a_s \phi_a^{(s)} , \qquad
 \hat\a_s \phi^{(s)} = \a_s(\Box- 2r^2) \phi_a ^{(s)}  \; ,
 \label{phi-int-alpha-general} \\
\int   \phi^{(s)}   \phi^{(s)} &= \int \phi^{(s)}_{a_1\ldots a_s} 
\hat\a_s \ldots  \hat\a_2\hat\a_1   \phi^{(s)}_{a_1\ldots  a_s} \ .
  \label{phi-spins-normalization}
\end{align}
Explicit expressions for $\a_s$ for $s\geq 3$ could be computed similarly but 
are not required for our purposes.
\subsection{Local decomposition}
\label{sec:local_decomp}
Finally consider any point on $H^4$, for instance the reference point 
\eqref{eq:reference_point}.
We denote the four tangential coordinates with $x^\mu$, and the time-like 
coordinate 
on $\R^{4,1}$ with $x^0$.
Then the $\mso(4,1)$ generators decompose (locally) into $\mso(4)$ generators $m^{\mu\nu}$, and the 
remaining translation generator by $p^\mu = m^{\mu 0}$.
We can then decompose e.g.\ the spin $s = 1$ modes locally as
\begin{align}
 \phi_{ab}(x) m^{ab} =  \phi_\mu(x)\, p^\mu + \phi_{\mu\nu}(x) m^{\mu\nu}   \qquad \in \cC^1 \ 
 \label{phi-M-expand-local}
\end{align}
and similar for higher-spin. From this point of view, the main lesson of the 
above results
is that the $\phi_\mu(x)$ and $\phi_{\mu\nu}(x)$  are \emph{not} independent 
fields, 
but determined by the same irreducible spin 1 field $\phi_a(x)$, and similarly 
for higher-spin fields. For generalized fuzzy spaces these constraints may 
disappear, as considered in \cite{Steinacker:2016vgf}.
For the basic spaces $H^4_n$ and for $S^4_N$ \cite{Sperling:2017gmy}, 
the  formalism developed above takes these constraints properly into account.
%
%
\section{Matrix model realization and  fluctuations}
\label{sec:matrix_model}
Now consider the IKKT matrix models with mass term,
\begin{align}
 S[Y] &= \frac 1{g^2}\Tr \Big([Y^a,Y^b][Y^{a'},Y^{b'}] \eta_{aa'} \eta_{bb'} \, - \mu^2 Y_a Y^a  \Big) \ . 
 \label{bosonic-action}
\end{align}
Here $\eta_{ab} = \diag(-1,1,\ldots ,1)$ 
is interpreted as Minkowski metric of the target space $\R^{1,D-1}$.
The positive mass $\mu^2 > 0$ should ensure stability.
The above model leads to the classical equations of motion 
\begin{align}
 \Box_Y Y^a + \mu^2 Y^a = 0 
 \label{eom-lorentzian-M}
\end{align}
where 
\begin{align}
  \Box_Y = \left[Y^a,\left[Y_a,\cdot\right]\right] \ \sim - 
\{y^a,\{y_a,\cdot\}\}
  \label{Box-Y}
\end{align}
plays the role of the Laplacian. 
Note that \eqref{eom-lorentzian-M} are precisely the equation of motions for 
the IKKT model put forward in \cite{Kim:2011cr} after taking an IR 
cutoff into account.
\subsection{Fuzzy \texorpdfstring{$H^4_n$}{H4n} solution and tangential 
fluctuation modes}
\label{sec:H4-tang}
Consider the solution $Y^a = X^a$ of \eqref{X-Box} corresponding to fuzzy 
$H^4_n$, and add fluctuations 
\begin{align}
 Y^a = X^a + \cA^a
\end{align}
on $H^4_n$. They naturally separate into tangential modes $x_a\cA^a{=}0$ and 
radial modes $x_a\cA^a{\neq}0$.
The $SO(4,1)$-invariant inner product
\begin{align}
 \left\langle \cA^{(i)}, \cA^{(j)} \right\rangle  \coloneqq \int \cA_a^{(i)} 
\cA_b^{(j)} \eta^{ab}
 \label{innerproduct-A}
\end{align}
is positive definite for (Hermitian) tangential $\cA_a$ on $H^4$, and negative  
for the radial modes.
Since $\cA^a \in \End(\cH_n) \otimes \R^5$, 
we expect four tangential fluctuation modes and one radial mode for each spin 
(except for spin 0),  
as for  $S^4_N$ \cite{Sperling:2017gmy}.
Our strategy will be to remove the radial modes, 
and to find a useful basis of tangential modes in the semi-classical limit.

\paragraph{Intertwiners.}
Define the $SO(4,1)$ intertwiners
\begin{align}
 \cI(\cA^a) &\coloneqq \{\theta^{ab},\cA_b\}   \nn\\
 \tilde\cI(\cA_a) &\coloneqq P_{aa'}\{\theta^{a'b},\cA_b\}  \nn\\ 
 \cG(\cA^a) &\coloneqq\{x^a,\{x^b,\cA_b\}\} \ .
 \label{gaugefix-intertwiner}
\end{align}
They are Hermitian w.r.t.\ the 
inner product \eqref{innerproduct-A}, and tangential 
except for $\cI$, noting that  
\begin{align}
 x^a\, \cI(\cA_a) &= x^a\,\{\theta^{ac},\cA_c\} =  - r^2 R^2\,\eth^a \cA_a \  \nn\\
\eth^a\cI(\cA_a) &=
 \eth_a\{\theta^{ad},\cA_d\} = \frac{1}{r^2 R^2} x_b\{\theta^{ab},\{\theta^{ad},\cA_d\}\} \nn\\
  &= -\frac{1}{r^2 R^2}\, x^b\, \cI^2(\cA_b)  \ .
 \label{I-radial-id}
\end{align}
The $SO(4,1)$ Casimir for the fluctuation modes can be expressed using $\cI$ as 
follows:
\begin{align}
 C^2[\mso(4,1)]^{\rm (full)} \cA^a &= \frac 1 2([\cM_{cd},\cdot] + 
M^{(5)}_{cd})^2\cA^a \nn\\
  &=  C^2[\mso(4,1)]^{(\ad)} \cA^a - 2 r^{-2} \cI(\cA^a) +4  \nn\\
  &= (-r^{-2}\Box  - 2 r^{-2}\cI + \cS^2 + 4) \cA^a\nn\\
  &= (R^2 \eth\cdot\eth  - 2r^{-2} \cI + \cS^2 +4) \cA^a
\label{C2-total-id}
\end{align}
using \eqref{Spin-casimir}, and $C^2[\mso(4,1)] = 4$ for the vector 
representation $\C^{5}$.
 This can be seen by expressing $\cI$ as follows:
\begin{align}
 -\theta (M_{cd}^{(\ad)} \otimes M_{cd}^{(5)}\cA)^a \sim
 -\big(M^{(5)}_{cd}\big)^a_b \im \{\theta^{cd},\cdot \}  \cA^b 
 = 2\{\theta_{ab},\cA^b\} = 2\, \cI(\cA)^a \, .
 \label{I-MM-id}
 \end{align}
Here 
\begin{align}
(M_{ab}^{(5)})^c_d &= \im(\d^c_b \eta_{ad} - \d^c_a \eta_{bd})\, 
\label{M-rep-5}
\end{align}
is the  vector generator of $\mso(4,1)$, 
and $M_{bc}^{(\ad)} = \im \{\cM_{bc},\cdot \}$ denotes the representation of 
$\mso(4,1)$ induced by the Poisson structure on $\cS^4$.
As a check, we note that $C^{\rm (full)}(x^a) = 0$, since $\cI(x^a) = 4 x^a$. 
This reflects the full $SO(4,1)$-invariance of the background $x^a$.
\subsubsection{Spin 0 modes}
Let $\phi \in \cC^0$ be a spin 0 scalar field.
There are two tangential spin 0 mode, which read
\begin{align}
\begin{aligned}
 \cA_a^{(1)} &= \eth_a\phi \ \in \cC^0 , \qquad \phi \in \cC^0 \; ,\\
 \cA_a^{(2)} &= \theta^{ab}\eth_b\phi \ = \{x^a,\phi\} \ \in \cC^1 \; .
 \end{aligned}
 \label{A-H4-spin0}
\end{align}
These modes satisfy 
\begin{align}
\begin{aligned}
 \{x^a,\cA_a^{(1)} \} &=  \{x^a,\eth_a \phi\}  = 0 \; , \\
 \{x^a,\cA_a^{(2)} \} &= - \Box \phi \; ,
 \end{aligned}
 \label{gaugefix-spin0}
\end{align}
using \eqref{x-del-id}. Clearly only $\cA_a^{(1)}$ is physical, while 
$\cA_a^{(2)}$ is a pure gauge field.
Let us compute the action of the $\cI$ intertwiner; to start with
\begin{align}
 \cI(\cA_a^{(2)}) &\coloneqq \{\theta^{ab},\{x^b,\phi\} \} 
  =  \{\{\theta^{ab},x^b\},\phi\} \} +  \{x^b,\{\theta^{ab},\phi\} \} \nn\\
  &= 4 r^2 \{x^a,\phi\} + r^2\{x^b,(x^a \eth^b - x^b \eth^a) \phi \} \nn\\
  &= 4 r^2 \{x^a,\phi\} + r^2\theta^{ba} \eth^b \phi  \nn\\
  &= 3 r^2 \cA^{(2)} \; .
  \label{I-A2-spin0-direct}
\end{align}
Similarly, one finds
\begin{align}
 \cI(\cA_a^{(1)}) &\coloneqq \{\theta^{ab},\eth_b\phi\} = r^2\cA_a^{(1)}
\end{align}
Now we can use the identities
\begin{align}
\cI(\cA_a^{(2)})&=  \{\theta^{ab},\theta^{bb'}\cA_{b'}^{(1)}\}   \nn\\
  &= \{\theta^{ab},\theta^{bb'}\} \cA_{b'}^{(1)} + \theta^{bb'}\{\theta^{ab},\cA_{b'}^{(1)}\}   \nn\\
   &= 3 r^2 \theta^{ab'} \cA_{b'}^{(1)} 
   + \{\theta^{bb'}\theta^{ab},\cA_{b'}^{(1)}\} - \theta^{ab}\{\theta^{bb'},\cA_{b'}^{(1)}\} \nn\\
    &=  3 r^2 \theta^{ab'} \cA_{b'}^{(1)} 
   - r^2 R^2 \{P^{ab'},\cA_{b'}^{(1)}\} - \theta^{ab}\cI(\cA_{b}^{(1)})  \nn\\
    &= 3 r^2 \theta^{ab'} \cA_{b'}^{(1)} 
  - r^2 (x^a  \{x^c,\cA_{c}^{(1)}\} - \theta^{ac}\cA_{c}^{(1)}) - \theta^{ab}\cI(\cA_{b}^{(1)})  \nn\\
   &=  4 r^2 \theta^{ac} \cA_{c}^{(1)} - \theta^{ab}\cI(\cA_{b}^{(1)})  \; ,
 \label{cI-A2-Id}
\end{align}
wherein we used
\begin{align}
 R^2\{P^{ab},\phi_{b}\} &=  \{ x^a x^c,\phi_c\} =  x^a  \{x^c,\phi_c\} +  x^c \{ x^a ,\phi_c\} \nn\\
   &= x^a  \{x^c,\phi_c\} - \theta^{ac} \phi_c  \;,
   \label{xx-phi-CR}
\end{align}
for any tangential $\phi_c$, and the gauge fixing relations 
\eqref{gaugefix-modes}. 
Therefore
\begin{align}
\theta^{ab}\cI(\cA_{b}^{(1)})  &= 4 r^2 \cA_{a}^{(2)} - \cI(\cA_a^{(2)})   \nn\\
\tilde\cI(\cA_{a}^{(1)})  &= 4 r^2 \cA_{a}^{(1)} + \frac 1{r^2 
R^2}\theta^{ad}\cI(\cA_d^{(2)})  \;.
\label{I-A1-general}
 \end{align}
For $s=0$, this gives
\begin{align}
 \cI(\cA_a^{(2)}) &=  3r^2 \cA^{(2)}
 \label{I-A2-spin0}
\end{align}
since $\cS^2 \cA^{(2)}= 4\cA^{(2)}$, in agreement with 
\eqref{I-A2-spin0-direct}. 
Then \eqref{I-A1-general} gives
\begin{align}
 \tilde\cI(\cA_{a}^{(1)})  &= r^2  \cA_{a}^{(1)} 
\end{align}
because $\cA_{a}^{(1)}$ is tangential.
To summarize,
\begin{align}
\tilde\cI
\begin{pmatrix}
 \cA_a^{(1)} \\\cA_a^{(2)}
\end{pmatrix}
 &= r^2 \begin{pmatrix}
     1  & 0   \\
     0 &  3   \\
   \end{pmatrix}
   \begin{pmatrix}
 \cA_a^{(1)} \\\cA_a^{(2)}
\end{pmatrix}
   \label{cI-modes-explicit-spin1}
\end{align}
\subsubsection{Spin 1 modes}
Now let $\phi_a \in \cC^0$ be a tangential, divergence-free spin 1 field.
Then there are four tangential spin 1 modes, given by
\begin{align}
\begin{aligned}
 \cA_a^{(1)} &= \eth_a \phi^{(1)}  \quad \in \cC^1 , \qquad  \phi^{(1)} = 
\{x^a,\phi_a\} \ \in \cC^1   \\
 \cA_a^{(2)} &= \theta^{ab}\eth_b \phi^{(1)}  \ 
  = \{x^a,\phi^{(1)}\}  \quad \in \cC^2 \oplus \cC^0 \,,  \\
 \cA_a^{(3)} &= \phi_a  \quad \in \cC^0 \, , \\
 \cA_a^{(4)} &= \theta^{ab}\phi_b  \quad \in \cC^1 \,.
 \end{aligned}
 \label{A-H4-spin1}
\end{align}
Here $\phi^{(1)}$ is the unique spin 1 mode in $\End(\cH)$.
$\cI$ can be computed on the $\cA_a^{(3)}$ and $\cA_a^{(4)}$ modes
using
\begin{align}
 \cI(\phi_a) &\coloneqq\{\theta^{ab},\phi_b\} = r^2 \phi_a 
 \label{I-phi-spin-1}
\end{align}
due to \eqref{theta-phi-id}, which gives
\begin{align}
 \cI(\cA_a^{(3)}) &\coloneqq\{\theta^{ab},\phi_{b}\}  = r^2 \phi_a =  
r^2\cA_a^{(3)} .
 \end{align} 
Furthermore,
 \begin{align}
 \tilde\cI(\cA_a^{(4)}) &\coloneqq P^{aa'} \{\theta^{a'b},\theta^{bc}\phi_c \} 
\nn\\
 &= P^{aa'} \theta^{bc}\{\theta^{a'b},\phi_c\} + P^{aa'} \{\theta^{a'b},\theta^{bc} \}\phi_c \nn\\
 &= P^{aa'}\{\theta^{bc}\theta^{a'b},\phi_c\} - \theta^{ab} \{\theta^{bc},\phi_c\} 
  + 3 r^2 \theta^{ac} \phi_c  \nn\\
 &= - r^2R^2 P^{aa'}\{P^{a'c},\phi_c \}  + 2 r^2 \theta^{ac} \phi_c\nn\\
 &= r^2 P^{aa'}\theta^{a'c} \phi_c +2 r^2 \cA^{(4)}_a\nn\\
 &=  3r^2 \cA^{(4)}_a 
\end{align} 
using \eqref{xx-phi-CR}.
$\cI(\cA_a^{(2)})$ and $\cI(\cA_a^{(1)})$  will be computed
for the general case below.

\subsubsection{Spin 2 modes}
Now let $\phi_{ab} = \phi_{ba} \in \cC^0$ be a tangential, divergence-free, 
traceless spin 2 field,
and let $\phi_a^{(2)} = \{x^b,\phi_{ab}\} \in \cC^1$.
Then there are four tangential spin 2 modes, given by
\begin{align}
\begin{aligned}
 \cA_a^{(1)} &= \eth_a \phi^{(2)} \ \in \cC^2 , \qquad  \phi^{(2)} = 
\{x^a,\phi_a^{(2)}\} \in \cC^2 \; ,  \\
 \cA_a^{(2)} &= \theta^{ab}\eth_b \phi^{(2)}  \ = \{x^a,\phi^{(2)}\} \quad \in 
\cC^3 \oplus \cC^1   \; ,\\
 \cA_a^{(3)} &= \phi_{a}^{(2)} \ \in \cC^1 \; ,  \\
 \cA_a^{(4)} &= \theta^{ab}\phi_b^{(2)} \ \in \cC^2 \; .  
 \end{aligned}
 \label{A-H4-spin2}
\end{align}
Here $\phi^{(2)}$  is the unique spin 2 mode in $\End(\cH)$, 
which involves the linearized  Riemann tensor.
They satisfy the gauge fixing relations derived below, see 
\eqref{gaugefix-modes}. Also recall from \eqref{phi2-a-tangential} that 
$\phi_{c}^{(2)}$ is indeed tangential.
Furthermore,
\begin{align}
\cI(\cA_a^{(3)}) &\coloneqq\{\theta^{ab},\phi_{b}^{(2)}\}  
 =  \{\theta^{ab},\{x^{c},\phi_{bc}\}\} \nn\\
 &= - \{\phi_{bc},\{\theta^{ab},x^{c}\}\} - \{x^c,\{\phi_{bc},\theta^{ab}\}\} \nn\\
 &=  r^2\{\phi_{bc},\eta^{ac}x^b - \eta^{bc} x^a\}  + r^2\{x^c,\phi_{ac}\} \nn\\
 &= 0
\end{align}
using \eqref{I-phi-spin-1} for the last term.
Adapting \eqref{cI-A4}, we obtain
 \begin{align}
 \tilde\cI(\cA_a^{(4)}) 
 &= P^{aa'}\{\theta^{bc}\theta^{a'b},\phi_c\} - \theta^{ab} \{\theta^{bc},\phi_c\} 
  + 3 r^2 \theta^{ac} \phi_c  \nn\\
 &= - r^2 R^2 P^{aa'}\{P^{a'c},\phi_c \}  + 3 r^2 \theta^{ac} \phi_c\nn\\
 &= r^2 P^{aa'}\theta^{a'c} \phi_c +3 r^2 \cA^{(4)}\nn\\
 &=  4r^2 \cA^{(4)}  \ .
  \label{cI-A4}
\end{align} 
It is illuminating to display the explicit tensor content of the spin 2 modes, recalling that 
$\phi_b^{(2)}$ is the spin connection \eqref{spin-2-connection} and
$\phi^{(2)}$ is the curvature tensor. Using \eqref{spin-2-field-R}, this is
\begin{align}
\begin{aligned}
 \cA_a^{(1)} &= \eth_a \phi^{(2)} 
   = \frac 12\eth_a (\theta^{ed}\theta^{bc}\cR_{ed;bc}[\phi^{(2)}]) \; ,\\
 \cA_a^{(2)} &= \frac 12\theta^{aa'} \eth_{a'} 
(\theta^{ed}\theta^{bc}\cR_{ed;bc}[\phi^{(2)}]) \; ,  \\
 \cA_a^{(3)} &= - \omega_{a;de}\theta^{de}  \; ,  \\
 \cA_a^{(4)} &= - \theta^{ab}\omega_{b;de}\theta^{de} \; .
 \end{aligned}
\end{align}
In particular, $\cA_a^{(4)}=\theta^{ab}A_b$ encodes a $\mso(4,1)$-valued 
 gauge field $A_b = - \omega_{b;de}\theta^{de}$ given by the linearized
 spin connection of $\phi_{ab}$.

\subsubsection{Spin \texorpdfstring{$s\geq 1$}{s>=1} modes.}
Now consider the generic case.
Let $\phi_{a_1\ldots  a_s} \in \cC^0$ be a tangential, divergence-free, 
traceless, symmetric spin $s$ field,
and let $\phi_a^{(s)} = \{x^{a_1},\ldots \{x^{a_{s-1}},\phi_{a_1\ldots a_{s-1} 
a}\}\ldots \} \ \in \cC^{s-1}$.
Then there are four tangential spin s modes, given by
\begin{align}
\boxed{
\begin{aligned}
 \cA_a^{(1)} &= \eth_a \phi^{(s)} \ \in \cC^s  , \qquad  \phi^{(s)} = \{x^a,\phi_a^{(s)}\} \ \in \cC^s   \\
 \cA_a^{(2)} &= \theta^{ab}\eth_b \phi^{(s)}  \ = \{x^a,\phi^{(s)}\} 
                              \quad \in \cC^{s+1} \oplus \cC^{s-1}   \\
 \cA_a^{(3)} &= \phi_{a}^{(s)}  \ \in \cC^{s-1} \,,  \\
 \cA_a^{(4)} &= \theta^{ab}\phi_b^{(s)}  \ \in \cC^s  \,.
 \end{aligned}
 }
 \label{A-H4-spins}
\end{align}
Here $\phi^{(s)}$ is the unique spin s mode in $\End(\cH)$.
The modes \eqref{A-H4-spins} satisfy the gauge-fixing relations
\begin{align}
\begin{aligned}
 \{x^a,\cA_a^{(1)} \} &=  \{x^a,\eth_a \phi^{(1)}\}  = 0 \,, \\
 \{x^a,\cA_a^{(4)} \} &=  \{x^a,\theta^{ab}\phi_b\}
  = \theta^{ab}\{x^a,\phi_b\} = r^2 R^2 \eth^a\phi_a = 0 \,, \\
 \{x^a,\cA_a^{(3)} \} &= \{x^a,\phi_a \} = \phi^{(1)} \,, \\
 \{x^a,\cA_a^{(2)} \} &= - \Box \phi^{(1)} \,,
 \end{aligned}
  \label{gaugefix-modes}
\end{align}
using \eqref{x-del-id}, and 
\begin{align}
\begin{aligned}
 \eth^a \cA_a^{(1)} &= 
     -\frac{1}{r^2 R^2} \Box \phi^{(1)}  \; ,\\
 \eth^a \cA_a^{(2)} &= 0 = \eth^a \cA_a^{(3)} \; ,\\
 \eth^a \cA_a^{(4)} &= \eth_a(\theta^{ab}\phi_b) = \{x^a,\phi_a\} = \phi^{(1)} 
\;.
\end{aligned}
 \label{del-A-modes}
\end{align}
These relations hold for any spin.
Together with \eqref{I-radial-id}, it follows that $\cI(\cA^{(2)})$ and 
$\cI(\cA^{(3)})$ are tangential, while
$\cI(\cA^{(1)})$ and $\cI(\cA^{(4)})$ are not.
Let us proceed to $\tilde{\cI}$; we first show that 
\begin{align}
 \tilde\cI(\cA_a^{(3)}) &\coloneqq\{\theta^{ab},\phi_{b}^{(s)} \}  
 = (2-s) r^2 \cA_a^{(3)}
 \label{I-A3-generic}
\end{align}
This can be proven inductively as follows:
\begin{align}
 \{\theta^{ab},\phi_{b}^{(s)}\} &=  \{\theta^{ab},\{x^{c},\phi_{bc}^{(s)}\}\} \nn\\
 &= - \{\phi_{bc}^{(s)},\{\theta^{ab},x^{c}\}\} - 
\{x^c,\{\phi_{bc}^{(s)},\theta^{ab}\}\} \nn\\
 &= r^2\{\phi_{ba}^{(s)},x^b \}  + (3-s) r^2\{x^c,\phi_{ac}^{(s)}\}\nn\\
 &= (2-s) r^2 \phi_{a}^{(s)}
\end{align}
using \eqref{I-phi-spin-1}, where $\phi_{ab}^{(s)} =  
\{x^{a_1},\ldots ,\{x^{a_{s-2}},\phi_{a_1\ldots a_{s-2} ab}\}\ldots \} \in 
\cC^{s-2}$.
Note that we employed the relation $\{\theta^{ab},\phi_{bc}^{(s)}\} = (3-s) 
r^2 \phi_{bc}^{(s)}$ for $\phi_{bc}^{(s)} \in \cC^{s-2}$, which can be 
derived via induction, too.
Adapting \eqref{cI-A4}, this yields
 \begin{align}
 \tilde\cI(\cA_a^{(4)}) 
 &= P^{aa'}\{\theta^{bc}\theta^{a'b},\phi_c\} - \theta^{ab} \{\theta^{bc},\phi_c\} 
  + 3 r^2 \theta^{ac} \phi_c  \nn\\
 &= - r^2 R^2 P^{aa'}\{P^{a'c},\phi_c \} - (2-s) r^2 \theta^{ab} \phi_{b}^{(s)} + 3 r^2 \theta^{ac} \phi_c \nn\\
 &= r^2 P^{aa'}\theta^{a'c} \phi_c + (s+1) \phi_{a}^{(s)} r^2 \cA^{(4)}\nn\\
 &=  (s+2)r^2 \cA^{(4)}  .
  \label{cI-A4-s}
\end{align} 
To compute $\cI(\cA_a^{(2)})$, consider 
\begin{align}
 \cI(\cA_a^{(2)}) &\coloneqq \{\theta^{ab},\cA_b^{(2)}\} =  
\{\{x^a,x^b\},\cA_b^{(2)}\} \nn\\
  &= - \{\{x^b,\cA_b^{(2)}\},x_a\} - \{\{\cA_b^{(2)},x_a\},x_b\} \;,
\end{align}
where the second term can be rewritten as 
\begin{align}
 - \{\{\cA_b^{(2)},x_a\},x_b\} &= -\{\{\{x_b,\phi\},x_a\},x_b\} \nn\\
   &= \{\{\{\phi,x_a\},x_b\},x_b\} + \{\{\theta^{ab},\phi\},x_b\} \nn\\
   &= \Box \cA_a^{(2)} - \{\{\phi,x_b\},\theta^{ab}\} - 
\{\{x_b,\theta^{ab}\},\phi\} \nn\\
   &= \Box \cA_a^{(2)} - \cI(\cA_b^{(2)}) + 4 r^2\cA_a^{(2)} \;.
\end{align}
So that we obtain 
\begin{align}
 2\cI(\cA_a^{(2)}) &= - \{\{x^b,\cA_b^{(2)}\},x_a\} + \Box \cA_a^{(2)}  + 4 r^2\cA_a^{(2)}  \nn\\
 &=  \{ \Box \phi ,x_a\} +  \big(\Box  + 4 r^2\big)\cA_a^{(2)} 
 \label{cI-A2-general-id}
 \end{align}
 for any spin $s \geq 1$. Therefore
 \begin{align}
  (\Box - 2 \cI)(\cA_a^{(2)}) &= -\{ \Box \phi ,x_a\} - 4 r^2\cA_a^{(2)} \,.
  \label{D2-A2-id}
 \end{align}
 On the other hand, for a spin $s$ field $\phi$ we have
\begin{align}
\{x_a,\Box\phi\} &=
\cA^{(2)}[\Box\phi] = r^2\cA^{(2)}[(-C^2 + \cS^2) \phi] =  r^2(2s(s+1)- C^2_{\rm full}) \cA^{(2)}[\phi] \nn\\
 &= ((\Box +2\cI) - r^2\cS^2 +2r^2 s(s+1) - 4r^2) \cA^{(2)}[\phi] \,,
\end{align}
using the intertwiner property and \eqref{C2-total-id}, hence
\begin{align}
 \{x_a,\Box\phi\} - \Box \{x_a,\phi\} &= (2\cI - r^2\cS^2 +2r^2 s(s+1) - 4r^2) 
\cA^{(2)}[\phi] \,.
\end{align}
Comparing with \eqref{cI-A2-general-id}, this gives 
\begin{align}
 2\cI(\cA_a^{(2)}) &= - \{x^a, \Box \phi\} +  \big(\Box  + 4 r^2\big)\cA_a^{(2)} 
  = (4r^2 - 2\cI + r^2\cS^2 -2r^2 s(s+1) + 4r^2) \cA^{(2)} \nn
 \end{align} 
  such that
\begin{align} \boxed{
 2\cI(\cA_a^{(2)}) =  r^2\Big(\frac 12 \cS^2 - s(s+1) + 4\Big) \cA^{(2)}
 }
\label{I-A2-general}
\end{align}
for $s\geq 1$, which is  tangential. 
Hence if $\cS^2$ is diagonal then $\cI$ is also diagonal, 
and the Casimir $C^2[SO(4,1)]$ \eqref{C2-total-id} can be diagonalized  
simultaneously.
To evaluate \eqref{I-A2-general},we decompose $\cA_a^{(2)}$ into its components 
in $\cC^{s-1} \oplus \cC^{s+1}$ as follows:
\begin{align}
 \cA_a^{(2)} &= -\a_s (\Box-2r^2) \cA_a^{(3)} + \cA_a^{(2')} \qquad \in 
\cC^{s-1} \oplus \cC^{s+1} \; ,\nn\\[1ex]
  \cA_a^{(2')} &\coloneqq\cA_a^{(2)} + \a_s(\Box-2r^2) \cA_a^{(3)}\qquad \in  
\cC^{s+1} \; ,
  \label{A2prime-def}
\end{align}
using \eqref{x-phi-comm-spins}; recall that $\cA_a^{(3)}\equiv \phi_a^{(s)}$.
Note that \eqref{A2prime-def} is simultaneously a decomposition into
eigenvectors of $\cI$, 
\begin{align}
 2\cI(\cA_a^{(2')} ) 
 &= 2(s+3) r^2 \cA_a^{(2')},  \nn\\
 2\cI(\cA_a^{(3)} ) 
  &= 2(2-s)r^2 \cA_a^{(3)}
\end{align}
consistent with \eqref{I-A3-generic}.
Then we arrive at
\begin{align}
  2\cI(\cA_a^{(2)}) &= -2(2-s)\a_s r^2 (\Box-2r^2)\cA_a^{(3)} + 2(s+3) r^2  
\cA_a^{(2')} \nn\\
   &= 2(s+3) r^2  \cA_a^{(2)}  + 2(2s+1)\a_s r^2(\Box-2r^2) \cA_a^{(3)} 
\end{align}
and $\cI(\cA_a^{(1)})$ is obtained from \eqref{I-A1-general}, 
\begin{align}
\tilde\cI(\cA_{a}^{(1)}) 
&= 4 r^2 \cA_{a}^{(1)} + \frac 1{R^2}\theta^{ad}((s +3) \cA^{(2)}_d + (2s +1)\a_s(\Box-2r^2) \cA_d^{(3)})  \nn\\
 &=  r^2 (1 - s) \cA^{(1)}_a 
       + \frac{2s+1}{R^2} \a_s \theta^{ab}(\Box-2r^2)\cA_a^{(3)} \nn\\
 &=  r^2 (1 - s) \cA^{(1)}_a + \frac{2s+1}{R^2} \hat\a_s \cA_a^{(4)} 
\end{align}
where the last line is only a short-hand notation which applies to irreps,
cf. \eqref{x-phi-comm-spin1-short}.
Hence $\tilde\cI$ is diagonalized as follows:
\begin{align}
 \tilde\cI(\cA_{a}^{(1')}) &= r^2 (1 - s)\cA_{a}^{(1')},  \nn\\
  \cA_{a}^{(1')} &=  \cA_{a}^{(1)} - \frac{\a_s}{R^2 r^2} \tilde\theta^{ab}(\Box-2r^2)\cA_{b}^{(3)}
  \equiv \cA_{a}^{(1)} - \frac{\hat\a_s}{R^2 r^2} \cA_{a}^{(4)} \, ,
 \label{A1prime-def}
\end{align}
using \eqref{cI-A4-s}.
Accordingly, we define the eigenmodes
\begin{align}
 (\cB_a^{(1)},\cB_a^{(2)},\cB_a^{(3)},\cB_a^{(4)}) &\coloneqq
(\cA_a^{(1')},\cA_a^{(2')},\cA_a^{(3)},\cA_a^{(4)}) \;,
 \label{B-modes-def}
\end{align}
 which satisfy
 \begin{subequations}
\begin{align}
 \tilde\cI 
 \begin{pmatrix} 
\cB_a^{(1)} \\  \cB_a^{(2)}\\  \cB_a^{(3)}\\  \cB_a^{(4)}\\
\end{pmatrix}
&= r^2\begin{pmatrix}
                               1-s &0&0&0\\
                               0&s+3 &0&0\\
                               0&0&2-s &0\\
                               0&0&0&2+s
                              \end{pmatrix}
                              \begin{pmatrix} 
\cB_a^{(1)} \\  \cB_a^{(2)}\\  \cB_a^{(3)}\\  \cB_a^{(4)}\\
\end{pmatrix} \; ,  \\
 \cS^2 
 \begin{pmatrix} 
\cB_a^{(1)} \\  \cB_a^{(2)}\\  \cB_a^{(3)}\\  \cB_a^{(4)}\\
\end{pmatrix}
 &= 2\begin{pmatrix}
                         s(s+1)&0&0&0\\
                         0&(s+1)(s+2)&0&0\\
                         0&0&(s-1)s&0\\
                         0&0&0&s(s+1)
                        \end{pmatrix}  
\begin{pmatrix} 
\cB_a^{(1)} \\  \cB_a^{(2)}\\  \cB_a^{(3)}\\  \cB_a^{(4)}\\
\end{pmatrix} \, .
\end{align}
\end{subequations}
This shows that all these modes are distinct, and it
will allow to diagonalize and evaluate explicitly the quadratic action.
It also implies that we did not miss any modes, since there can be only 5 modes
for each spin (including the radial one, see below).
\paragraph{Gauge fixing term.}
The intertwiner $\cG$ of \eqref{gaugefix-intertwiner} takes the values 
\begin{align}
  \cG(\cA_a^{(2)}) &= -\{x^a,\Box\phi^{(s)}\}   \nn\\
  \cG(\cA_a^{(3)}) &= \{x^a,\phi^{(s)}\} = \cA_a^{(2)} \nn\\
   \cG(\cA_a^{(1)}) &=  \cG(\cA_a^{(4)}) = 0  \ .
  \label{gaugefix-G-modes}
\end{align}
\subsection{Recombination, \texorpdfstring{$\hs$}{hs}-valued gauge fields and 
Young diagrams}
\label{sec:reducible}
The distinct modes $\cA^{(i)}$ are useful to disentangle the different degrees of freedom. On the other hand
we can relax the requirements that the underlying tensor fields $\phi_{a_1\ldots 
a_s}$
are irreducible, so that the modes can be captured in a simpler way.
\paragraph{Trace contributions.} These arise from
\begin{align}
 \phi_{a_1\ldots a_s} = \eta_{a_1 a_2}\phi_{a_3\ldots a_s} \ .
\end{align}
Then
\begin{align}
  \tilde\phi^{(s)}_a &=  \{x^{a_1},\ldots \{x^{a_{s-1}},\eta_{a_1 
a_2}\phi_{a_3\ldots a}\}\ldots \} \} 
   = -\Box  \phi^{(s-2)}_a \ \in \cC^{s-2} ,\nn\\
  \tilde\phi^{(s)} &= \{x^{a}, \tilde \phi^{(s)}_a \} \ = -\{x^{a}, \Box \phi^{(s)}_a \}
\end{align}
which enters the four modes as follows
\begin{align}
 \cA_a^{(1)} &= \eth_a \tilde\phi^{(s-2)} \ \in \cC^{s-2}  , \qquad   \nn\\
 \cA_a^{(2)} &= \theta^{ab}\eth_b \tilde\phi^{(s-2)}  \ = \{x^a,\phi^{(s-2)}\} 
                              \quad \in \cC^{s-1} \oplus \cC^{s-3}   \nn\\
 \cA_a^{(3)} &= \tilde\phi_{a}^{(s-2)}  \ \in \cC^{s-3} ,  \nn\\
 \cA_a^{(4)} &= \theta^{ab}\tilde\phi_b^{(s-2)}  \ \in \cC^{s-2}  .
 \label{A-H4-spins-red}
\end{align}
Hence the trace components reproduce the four modes with spin $s-2$,
as long as $\Box \phi^{(s)}_a \neq 0$.
\paragraph{Divergence modes.}
Now we  drop the requirement that $\phi_{ab}$ is divergence-free.
Consider the case of rank 2 tensors, expressed in terms of spin 1 modes as in 
\eqref{spin2-puregauge}
\begin{align}
 \tilde\phi_{ab}^{(1)} = \nabla_a \phi_b + \nabla_b \phi_a \ .
\end{align}
Then according to \eqref{fieldstrengthform-spin1-gauge},  these contribution to
the would-be spin 2 modes $\cA_a^{(1)}, \cA_a^{(2)}$ 
vanish identically.
The  contribution to $\cA_a^{(3)}$ reduces to a combinations of the
spin 1 modes of $\cA_a^{(1)}$ and $\cA_a^{(4)}$, and 
the contribution to $\cA_a^{(4)}$ reduces to a combinations of the
spin 1 modes of $\cA_a^{(2)}$ and $\cA_a^{(3)}$.
Hence if we drop the divergence-free condition, it 
would suffice to keep the $\cA_a^{(3)}$ and $\cA_a^{(4)}$ 
modes\footnote{However, the spin 0 modes cannot be recovered from divergence modes:
for $\phi_a = \eth_a\phi$ we get $\tilde\phi^{(1)} = \{x^a,\eth_a \phi\} = 0$ 
due to \eqref{x-del-id}. }.
In particular, we  need not  worry about these constraints  
upon projecting $H^4$ to $\cM^{3,1}$.
It will suffice to impose the appropriate divergence- 
and trace- conditions for $\cM^{3,1}$.

Finally as for $S^4_N$ \cite{Sperling:2017gmy},
we can collect \emph{all} tangential fluctuation modes as $\hs$-valued 
tangential gauge fields 
\begin{align}
 \cA^a &= \theta^{ac}{\bf A}_c, \qquad {\bf A}_c = A_{c,\und{\a}}(x) \, \Xi^{\und{\a}} \ 
 \label{A-theta-recombined}
\end{align}
where $A_{c,\und{\a}}(x)$ are double-traceless tensor fields corresponding to 
2-row Young diagrams of the type ${\tiny \Yvcentermath1 \yng(3,2)}$.
The external leg is associated to the extra box in the Young diagram. However 
the $A_{c,\und{\a}}(x)$ are 
in fact higher curvatures of the underlying symmetric tensor fields 
$\phi_{a_1\ldots  a_s}$ as in \eqref{Riemann-tensor-def-s},
which characterize the irreducible physical degrees of freedom  $\cA_a^{(i)}$.

\subsubsection{Inner products}
\label{sec:inner-products}

The inner products \eqref{innerproduct-A} of the tangential fluctuations are 
given by
\begin{align}
 \int \cA_b^{(1)} \cA_b^{(1)} &= \int \eth_b \phi^{(s)} \eth_b \phi^{(s)} 
  =  \frac{\a_s}{r^2R^2}\int \phi^{(s)}_a (\Box +2r^2 s)(\Box-2r^2) \phi^{(s)}_a 
\,, \nn\\
  \int \cA_b^{(1)} \cA_b^{(4)} &= \int \eth_a \phi^{(s)} 
\theta^{ab}\phi^{(s)}_b 
    = \int  \phi^{(s)} \{x_b,\phi^{(s)}_b\}
    = \int  \phi^{(s,1)} \phi^{(s,4)} 
   = \a_s\int \phi^{(s)}_a (\Box - 2r^2) \phi^{(s)}_a \,, \nn\\
 \int \cA_b^{(3)} \cA_b^{(2)} &= \int \phi^{(s)}_b \{x^b, \phi^{(s)}\} 
  = -\int \{x^b,\phi^{(s)}_b \} \phi^{(s)} 
   = -\a_s\int \phi^{(s)}_a (\Box -2r^2) \phi^{(s)}_a \,, \nn\\
 \int \cA_b^{(2)} \cA_b^{(2)} &= \int \{x_b, \phi^{(s)}\} \{x^b, \phi^{(s)}\}
  = \a_s\int \phi^{(s)}_a (\Box +2r^2 s)(\Box-2r^2) \phi^{(s)}_a \,, \nn\\
 \int\cA_a^{(3)} \cA_a^{(3)} &= \int \phi^{(s)}_b  \phi^{(s)}_b  \,, \nn\\
  \int \cA_a^{(4)} \cA_a^{(4)} &= \int \theta^{ab}\phi^{(s)}_b  
\theta^{ac}\phi^{(s)}_c    = r^2 R^2 \int \phi^{(s)}_b  \phi^{(s)}_b \,,  \nn\\
  \int \cA_b^{(1)} \cA_b^{(2)} &=  \int \cA_b^{(1)} \cA_b^{(3)} =  \int 
\cA_a^{(4)} \cA_a^{(2)} = \int \cA_a^{(4)} \cA_a^{(3)} = 0 \,,
  \label{A-phi-normalization}
\end{align}
using \eqref{Box-delsquare}, \eqref{x-del-id}, \eqref{Box-Phi-intertwiner}
and  $[\theta^{ab}\phi^{(s,4)}_b  \phi^{(s,3)}_a]_0 = 0$;
we drop the labels $\phi^{(s,4)}_b \equiv \phi^{(s)}_b$ if no confusion can arise.

Now consider the eigenstates \eqref{B-modes-def} of $\cI$.
We verify that $\cB_a^{(2)}$ and $\cB_a^{(1)}$  satisfy the 
orthogonality relations 
\begin{align}
 \int \cB_a^{(2)} \cB_a^{(3)} &= \int (\cA_a^{(2)} + \hat\a_s \cA_a^{(3)}) 
\cA_a^{(3)}  = 0  \,, \nn\\
 \int \cB_{a}^{(1)} \cB_a^{(4)}  &= 0\,, 
\end{align}
using the definitions \eqref{A2prime-def},  
\eqref{A1prime-def} as well as \eqref{phi-int-alpha-general}. 
Therefore $\{\cB_{a}^{(i)}\}_{i=1}^4$ form an orthogonal basis of eigenmodes.
The normalization can be computed as
\begin{align}
 \int \cB_a^{(2)} \cB_a^{(2)} &= 
 \int \left(\cA_a^{(2)}+ \hat\a_s \cA_a^{(3)} \right) 
\left(\cA_a^{(2)}+ \hat\a_s \cA_a^{(3)}\right) \nn\\
  &= \a_s\int\phi^{(2')}_a  \Big((\Box +2r^2 s)
  -  \a_s(\Box - 2r^2)\Big)(\Box - 2r^2) \phi^{(2')}_a \nn\\
  &= \int\phi^{(2')}  \Big((1-\a_s)\Box + 2r^2(s+1)\Big) \phi^{(2')} 
  \label{A2prime-normalization}
\end{align}
and
\begin{align}
 \int \cB_{a}^{(1)} \cB_a^{(1)} &= 
 \int \left( \cA_a^{(1)} - \frac{\hat\a_s}{R^2 r^2} \cA_a^{(4)} 
\right)
\left(\cA_a^{(1)} - \frac{\hat\a_s}{R^2 r^2} \cA_a^{(4)}\right) \nn\\
 &= \frac{\a_s}{r^2 R^2}\int\phi^{(1')}_a  \Big((\Box +2r^2 s)
 - \a_s(\Box-2r^2) \Big)(\Box-2r^2) \phi^{(1')}_a \nn\\
 &= \frac{1}{r^2 R^2}\int\phi^{(1')}  \Big((1-\a_s)\Box + 2r^2 \a_s(s+1) \Big) \phi^{(1')} \ .
 \label{inner-A1prime}
\end{align}
Note that all $\cA_a^{(2)}$ modes are pure gauge modes, and they will drop out in the action.

\subsection{Radial modes}
\label{sec:radial-modes}

Finally consider the radial fluctuation modes.
These are  given by
\begin{align}
 \cA^{(r)}_a[\phi^{(s)}] = x_a \phi^{(s)}, \qquad \phi^{(s)} \in \cC^s \ .
 \label{A-modes-radial}
\end{align}
They are dangerous because the radial metric in $\R^{1,4}$ is negative,
\begin{align}
 \int \cA_b^{(r)} \cA_b^{(r)} &= \int x_a \phi^{(s)} x^a \phi^{(s)} = - R^2 \int  \phi^{(s)}  \phi^{(s)}
\end{align}
recalling that $x_a x^a = - R^2 < 0$. 
However, they disappear after the projection to $\cM^{3,1}$. 
If we include these radial fluctuations, we should first diagonalize $\cI$. We have
\begin{align}
 \cI(\cA^{(r)}_a) &= \{\theta^{ab},x_b\phi^{(s)}\} 
 = \{\theta^{ab},x_b\} \phi^{(s)} + x_b \{\theta^{ab},\phi^{(s)}\} \nn\\
 &=  4 x^a \phi^{(s)} - \theta^{ab} \{x_b,\phi^{(s)}\} \nn\\
  &= 4 \cA^{(r)}_a + r^2 R^2 \cA^{(1)}_a  \ .
  \label{I-Ar}
\end{align}
Recall that $\cI(\cA^{(2',3)})$ is tangential, but  $\cI(\cA^{(1,4)})$ is not, with 
\begin{align}
 x^a \cI(\cA^{(1)}[\phi]) &= \Box \phi \,,  \nn\\
 x^a \cI(\cA^{(4)}[\phi]) &= - r^2 R^2 \phi \,, \nn\\
 x^a \cI(\cA^{(1')}[\phi]) &= (\Box + \hat\a_s) \phi \,,
\end{align}
using \eqref{I-radial-id} and \eqref{del-A-modes}. 
Hence the radial modes may couple to the $\cA^{(1,2)}$ or the $\cB^{(1,2)}$ modes, and the $\cI$ eigenmodes
seem to mix completely all 3 components $\cA^{(1')},\cA^{(4)},\cA^{(r)}$. 
However since the radial modes are
negative definite, we will  focus on the tangential modes,
and on its projection to $\cM^{4,1}$ in the next stage.
%
%
\subsection{\texorpdfstring{$SO(4,1)$}{SO(4,1)}-invariant quadratic action on 
\texorpdfstring{$H^4$}{H4}}
\label{sec:quad-action-H4}
The quadratic fluctuations for the fluctuation modes $y^a = x^a + \cA^a$
are governed by the action 
\begin{align}
 S[y] &= S[x]  +  S_2[\cA] + O(\cA^3),  
 \end{align}
 where 
 \begin{align}
S_2[\cA] &= \frac{2}{g^2}\int\limits_{} d\mu \Big( \cA_a (\cD^2 \cA)^a + 
\{x^a,\cA_a\}^2  \Big)
  = \frac{2}{g^2}\int\limits_{} d\mu \, \cA_a (\cD^2  + \cG)\cA^a \ .
\label{eff-S-expand}
\end{align}
Here 
\begin{align}
(\cD^2 \cA) &\coloneqq \left(\Box - 2\cI+ \frac{1}{2} \mu^2\right)\cA 
\label{vector-Laplacian}
\end{align}
is the  ``vector'' (matrix) Laplacian, and 
$\cG(\cA)$  \eqref{gaugefix-intertwiner} ensures gauge invariance.
The mass term determines $r^2$ via the on-shell condition for $H^4_n$,
\begin{align}
 0 = \left(\Box + \frac{1}{2} \mu^2\right) x^a, \qquad \frac{1}{2} \mu^2 = 4 
r^2 \ .
\label{on-shell-mu}
\end{align}
\paragraph{Gauge-invariant action.}
Consider first the gauge-invariant kinetic term 
\begin{align}
 S_2[\cA] &=  \frac{2}{g^2}\int\limits_{} d\mu \, \cA_a (\cD^2  +\cG)\cA^a \ .
\end{align}
We verify that the pure gauge modes $\cA_a^{(2)}$ are null modes using 
\eqref{gaugefix-G-modes} and \eqref{D2-A2-id}:
\begin{align}
 (\cD^2 + \cG)\cA_a^{(2)} = -\{ \Box \phi^{(s)} ,x_a\} +\left(\frac{1}{2}\mu^2 
- 4 r^2\right)\cA_a^{(2)} -\{x^a,\Box\phi^{(s)}\} = 0
\end{align}
for any spin, taking into account the on-shell condition $\frac 12 \mu^2 = 4 r^2$.
Hence the pure gauge modes $\cA_a^{(2)}$ indeed decouple.

For spin 0, we determine the action explicitly for the $\cB^{(1)}$ and 
$\cB^{(2)}$ modes
\begin{align}
 (\cD^2 + \cG)
 \begin{pmatrix}
 \cB_a^{(1)} \\ \cB_a^{(2)} 
 \end{pmatrix}
  = 
 \begin{pmatrix}
  \Box + 2r^2  & 0 \\
   0 & 0
 \end{pmatrix}
   \begin{pmatrix}
 \cB_a^{(1)} \\ \cB_a^{(2)} 
 \end{pmatrix} \;.
   \label{kinetic-term-spin0}
\end{align}
The inner product  is diagonal for spin 0, and
the quadratic action is given by
\begin{align}
 S_2[\cA] = \int \cB_a^{(i)} 
   \begin{pmatrix}
    \Box + 2 r^2  & 0  \\
     0 & 0  \\
   \end{pmatrix}
   \cB_a^{(i)} \ .
\end{align}
Since $\cB_a^{(1)} \in \cC^0$, this is indeed positive define (except for the 
pure gauge mode)
due to \eqref{A-phi-normalization},
recalling that $\Box \propto -\eth\cdot\eth$ for spin 0 \eqref{Box-delsquare}.
\paragraph{Gauge-fixed action and positivity.}
\label{sec:gaugefixed-action}
Now we consider a gauge-fixed action, which is obtained by canceling $\cG$ with 
a suitable Faddeev--Popov (or BRST) term:
\begin{align}
 S_{2, (\mathrm{fix})}[\cA] &=  \frac{2}{g^2}\int\limits_{} d\mu\, \cA \cD^2 \ 
\cA \ . 
 \label{eq:gauge_fixed_action}
\end{align}
We work in the basis $\{\cB^{(i)}\}$ \eqref{B-modes-def} where $\cI$ is 
diagonal. 
Then the eigenvalues of the kinetic operator $\cD^2$ are elaborated in the 
appendix \ref{sec:positivity}. 
Together with the inner products in section \ref{sec:inner-products},
we obtain the following diagonalized quadratic action
\begin{subequations}
\label{kinetic-gaugefixed-general}
\begin{align}
 \int \cB^{(1)}_a \cD^2 \cB^{(1)}_a 
   &= \frac{\a_s}{r^2 R^2}\int\phi^{(s)}_a  \Big((\Box +2r^2 s) - 
\a_s(\Box-2r^2) \Big)(\Box-2r^2) (\Box+ 2r^2(3s+2))\phi^{(s)}_a \nn\\
 &= \frac{1}{r^2 R^2}\int\phi^{(s)} \Big((1-\a_s)\Box + 2r^2 \a_s(s+1)\Big) 
(\Box+ 4r^2(s+1))\phi^{(s)} \,,\\
\int \cB^{(2)}_a \cD^2 \cB^{(2)}_a
  &= \a_s\int\phi^{(s)}_a  \Big((\Box +2r^2 s) -  \a_s(\Box - 2r^2)\Big)(\Box - 
2r^2) (\Box + 2r^2s) \phi^{(s)}_a \nn\\
   &= \int\phi^{(s)} \Big((1-\a_s)\Box + 2r^2 \a_s(s+1)\Big) \Box \phi^{(s)} \,,
\\
\int \cB^{(3)}_a \cD^2 \cB^{(3)}_a
  &= \int \phi^{(s)}_b (\Box +2r^2 s) \phi^{(s)}_b  \,,\\
\int \cB^{(4)}_a \cD^2 \cB^{(4)}_a
  &=   r^2 R^2 \int \phi^{(s)}_b (\Box - 2r^2  s) \phi^{(s)}_b \,.
 \end{align}
 \end{subequations}
All  these  terms are non-negative, because
\begin{align}
 (\Box +2r^2 s) - \a_s(\Box-2r^2)  &= (1-\a_s)\Box + 2r^2 (s +\a_s) > 0  .  \nn\\
 \cA[(\Box - 2r^2  s) \phi^{(s)}_b] &\propto \cA[(\Box + r^2 
s(s-3))\phi_{a_1\ldots a_s}]
\end{align}
for any intertwiner $\cA$, using \eqref{BoxA4-2}. The first line is positive 
because $1>\a_s$, and the second line is positive since $\Box + r^2 s(s-3)$
is manifestly positive for $s\geq 3$, while for $s=1,2$ it coincides with $\Box-2r^2$
which is also positive 
on divergence-free tensor fields as shown in \eqref{phi-spin1-normalization}.
As usual, the unphysical modes will be canceled  by  Faddeev--Popov ghosts.

We consider explicitly the case of spin 1 and spin 2. 
For spin 1, we have 
\begin{subequations}
\label{kinetic-gaugefixed-spin1}
\begin{align}
 \int \cB^{(1)}_a \cD^2 \cB^{(1)}_a 
 &= \frac{\a_1}{r^2 R^2}\int\phi_a  \Big((\Box +2r^2) - \a_1(\Box-2r^2) 
\Big) (\Box-2r^2) (\Box+ 10 r^2)\phi_a \nn\\ 
     &= \frac{1}{r^2 R^2}\int\phi^{(1)} \Big((1-\a_1)\Box + 4r^2 \a_1\Big)  
(\Box+ 8 r^2)\phi^{(1)} \,,\\  
 \int \cB^{(2)}_a \cD^2 \cB^{(2)}_a 
   &= \a_1\int\phi_a  \Big((\Box +2r^2)
  -  \a_1(\Box - 2r^2)\Big)(\Box - 2r^2) (\Box + 2r^2) \phi_a  \nn\\
  &= \int\phi^{(1)} \Big((1-\a_1)\Box + 4r^2 \a_1\Big) \Box \phi^{(1)} \,, \\
  \int \cB^{(3)}_a \cD^2 \cB^{(3)}_a  &= \int \phi_a(\Box +2r^2 ) \phi_a \,,\\
  \int \cB^{(4)}_a \cD^2 \cB^{(4)}_a 
  &= r^2 R^2 \int \phi_{a}  (\Box -2 r^2 )\phi_{a} \,,
\end{align}
\end{subequations}
and for spin 2
\begin{subequations}
\label{kinetic-gaugefixed-spin2}
\begin{align}
 \int \cB^{(1)}_a \cD^2 \cB^{(1)}_a 
  &= \frac{\a_2}{r^2 R^2}\int\phi^{(2)}_a  \Big((\Box +4r^2) - \a_2(\Box-2r^2) 
\Big)(\Box-2r^2) (\Box+ 16r^2)\phi^{(2)}_a \nn\\
  &= \frac{1}{r^2 R^2}\int\phi^{(2)} \Big((1-\a_2)\Box + 6r^2 \a_2\Big) (\Box+ 
12r^2)\phi^{(2)} \nn\\
    &= \frac{\a_1\a_2}{r^2 R^2}\int \phi_{ab} (\Box + 6 r^2 - \a_2\Box)(\Box 
+18r^2)(\Box-2r^2)\Box \phi_{ab} \,, \\
 \int \cB^{(2)}_a \cD^2 \cB^{(2)}_a  
    &= \a_2\int\phi^{(2)}_a  \Big((\Box +4r^2)
  -  \a_2(\Box - 2r^2)\Big)(\Box - 2r^2) (\Box + 4r^2) \phi^{(2)}_a \nn\\
   &= \int\phi^{(2)} \Big((1-\a_2)\Box + 6r^2 \a_2\Big) \Box \phi^{(2)} \nn\\
  &= \a_2\a_1\int\phi_{ab}  \Big(\Box +6r^2
  -  \a_2\Box \Big) (\Box + 6r^2)(\Box-2r^2) \Box\phi_{ab} \,, \\
 \int \cB^{(3)}_a \cD^2 \cB^{(3)}_a  &= \int \phi^{(2)}_a(\Box +4r^2 ) 
\phi^{(2)}_a\nn\\
    &= \a_1\int \phi_{ab}(\Box + 6r^2 )(\Box-2r^2) \phi_{ab}  \,,\\
 \int \cB^{(4)}_a \cD^2 \cB^{(4)}_a 
  &=   r^2 R^2 \int \phi^{(2)}_b (\Box - 4r^2) \phi^{(2)}_b \nn\\
  &=  \a_1 r^2 R^2 \int \phi_{ab}  (\Box -2 r^2 )^2\phi_{ab} \,,
\end{align}
\end{subequations}
using \eqref{normalization-rank1-2}.
Note that  we only include tangential fluctuation modes here. 
If we would also include the radial fluctuations as in section \ref{sec:radial-modes}, they 
would be negative definite or ghost modes, because the metric in the radial direction is time-like.
However this is resolved upon projecting to $\cM^{3,1}$, as discussed below.

\subsection{Yang--Mills gauge theory}
We can write the full action \eqref{bosonic-action} in a  conventional 
(higher-spin) Yang--Mills form
for the recombined higher-spin gauge fields \eqref{A-theta-recombined}
$\cA^a = \theta^{ab}{\bf A}_b$. Then the field strength is
\begin{align}
 \cF^{ab} &= [X^a+\cA^a,X^b+\cA^b] 
 \ \sim \ \theta^{ab}+ \theta^{aa'}\theta^{bb'}F_{a'b'}  ,  \nn\\
 F_{ab} &= \nabla_{a} {\bf A}_{b} - \nabla_{b} {\bf A}_{a'} + [{\bf A}_{a},{\bf A}_{b}]
\end{align}
recalling that $\nabla \theta^{ab} = 0$.
Hence the action \eqref{bosonic-action}
\begin{align}
  S[Y] &\sim \frac 1{g_{YM}^2}\int_{H^4} \big( F_{ab}F_{a'b'} \eta^{aa'} \eta^{bb'} \, 
  -  \frac{2}{R^2} {\bf A}_a {\bf A}_{a'} \eta^{aa'} \big)
 \label{bosonic-action-YM}
\end{align}
is basically a $\hs$-valued
Yang--Mills action\footnote{We used $x_a \cA^a=0$; 
the apparent ``mass'' term  is at
the cosmological curvature scale, and would presumably disappear upon imposing
the non-linear constraint $Y_a Y^a = -R^2$.} (dropping surface terms and using  $\mu^2 = 8r^2$), where
\begin{align}
 \frac{1}{g_{\rm YM}^2} = \rho\ \frac{L_{NC}^8}{4 g^2} 
 \label{YM-coupling}
\end{align}
is the a dimensionless Yang--Mills coupling constant.
For nonabelian spin 1 modes $\cA^{(4)}_a$  on stacks of $H^4_n$ branes, 
the usual Yang--Mills action is  recovered.
For spin 2, one would expect this to describe some type of quadratic gravity action 
\cite{Stelle:1977ry,Alvarez-Gaume:2015rwa,Salvio:2018crh}. However  this does not happen as shown below,
since the graviton is obtained by a
field redefinition \eqref{graviton-a4-explicit} and does not propagate at the 
classical level.
However the Yang--Mills framework suggests that no ghost modes appear also for 
higher-spin
(as opposed to quadratic gravity), hence gravity might emerge at the quantum level.

\subsection{Metric and gravitons on \texorpdfstring{$H^4$}{H4}}
\label{sec:graviton}
Now we take some of the leading (cubic) interactions of these modes into 
account, focusing on the contributions of the spin 2 (and spin 1) modes to the 
kinetic term on $H^4$. These contributions are expected to give rise 
to linearized gravity. 
The kinetic term for all fluctuations on a given background $Y^a \sim y^a$ 
 arises in the matrix model from\footnote{One might worry about the 
contributions from  $\{y^a,\cdot\}$ on the generators $\theta^{bc}$ for 
higher-spin modes. However the metric is  always defined by the two derivative 
terms 
acting on the tensor fields.}
\begin{align}
  S[\phi] &= -  \Tr [Y^a,\phi][Y_a,\phi] 
  \ \sim \ \int \rho\,\{y^a,\phi\}\{y_a,\phi\}  \nn\\
 &= \int \rho \g^{ab}\eth_a \phi \eth_b \phi 
  \ \stackrel{\xi}{=}  \ \int \, \rho \, 
  \g^{\mu\nu}\del_\mu \phi \del_\nu \phi \ \nn\\
   &= \ \int_{H^4} d^4 x\, \sqrt{|G_{\mu\nu}|}\,
  G^{\mu\nu}\del_\mu \varphi \del_\nu \varphi \ 
  \label{eff-G-action}
\end{align} 
using \eqref{integral-trace}; some dimensionful constants are 
absorbed in $\varphi$, and Greek indices indicate local coordinates.
Here $\g^{ab}$ is a symmetric  tensor in $SO(4,1)$ notation
\begin{align}
 \g^{ab} &= \eta_{cc'}e^{ca}e^{c'b}, \qquad e^{ca} = \{y^c,x^a\} \ 
\end{align}
which in local coordinates near some reference point $\xi$ reduces to 
$\g^{\mu\nu}$, cf. \eqref{theta-constraint}. Hence the effective metric is given by 
\cite{Steinacker:2017vqw,Steinacker:2016vgf,Steinacker:2010rh}
\begin{align}
   G^{\mu\nu} = \frac{4\a}{L_{NC}^4}\, \g^{\mu\nu} \ , \qquad \a = 
\sqrt{\frac{L_{NC}^4}{4|\g^{\mu\nu}|}} \ 
   \label{eff-metric-G}
\end{align}
and $e^{ca}$ can be interpreted as vielbein.
For a deformation of the $H^4$ background of the form 
\begin{align}
 y^a = x^a + \cA^a \; ,
 \label{eq:fluctuation_background}
\end{align}
the metric is perturbed due to $\g^{ab} =  \obar\g^{ab} + \d_\cA \g^{ab} + 
O(\cA^2)$ with
\begin{align}
 \d_\cA \g^{ab} \eqqcolon H^{ab}[\cA]
  &= \{x^c,x^a\}\{\cA_c,x^b\} + (a \leftrightarrow b)  \nn\\
  &= \theta^{c a}\{\cA_c,x^b\} + (a \leftrightarrow b)  \nn\\
   &= \{\theta^{ca}\cA_c,x^b\} +  \{\theta^{cb}\cA_c,x^a\}
 + r^2 \left(\cA^b x^a + \cA^a x^b  - 2 \eta^{ab} \left(\cA_c x^c\right) 
\right)  \ .
  \label{gravitons-H1}
\end{align}
Here $H^{ab}[\cA]$ is an $SO(4,1)$ intertwiner and tangential,
\begin{align}
  H^{ab} x_a = 0 \ , \qquad H \coloneqq\eta_{ab}H^{ab} = \frac 12 L_{NC}^4 
\eth^a \cA_a \ .
\end{align}
Then the  linearized effective metric 
\eqref{eff-metric-G} becomes  in $SO(4,1)$-covariant notation
\begin{align}
 G^{ab} = P^{ab} + \widetilde{h}^{ab}  \; , \qquad 
\text{with} \qquad 
\widetilde{h}^{ab} \coloneqq \left(h^{ab} - \frac{1}{2} P^{ab} h \right) \; ,
\label{eq:def_phys_graviton}
\end{align}
thus defining the \emph{physical graviton} $\widetilde{h}^{ab}$, where
 \begin{align}
h^{ab} = \frac{4}{L_{NC}^4}[H^{ab}]_0, \qquad h = \eta_{ab} h^{ab} \; 
\end{align}
is dimensionless.
We study the graviton modes \eqref{eq:def_phys_graviton} for the spin $s=0,1,2$ 
fluctuations of \eqref{A-H4-spins} in more detail below. 
\subsubsection{Spin 0 gravitons}
To begin with, consider the perturbation \eqref{gravitons-H1} of the metric for 
the two spin 0 modes of \eqref{A-H4-spin0}. One finds
\begin{align}
 H_{ab}[\cA^{(1)}] &= \theta^{ac}\theta^{bd}\left( \eth_c \eth_d \phi^{(0)} + 
\eth_d \eth_c \phi^{(0)} \right) \,,\nn \\
 H_{ab}[\cA^{(2)}] &=r^2 (x_a \theta^{bd} \eth_d \phi^{(0)} - R^2 \theta^{ad} 
 \eth_d \eth_b \phi^{(0)}) + (a\leftrightarrow b)  \, .
\end{align}
Upon averaging, one obtains
\begin{align}
 h_{ab}[\cB^{(1)}] &= \a_1 \left(2P_{ab} \eth \cdot \eth \phi^{(0)} - (\nabla_a 
\nabla_b\phi^{(0)} + \nabla_b \nabla_a\phi^{(0)}) \right) \,,  \\
h_{ab}[\cB^{(2)}] &=0 \,, \nn
\end{align}
and the expressions satisfy
\begin{align}
 h[\cB^{(1)}] = 6\Box\phi , \qquad  \nabla^a h_{ab}[\cB^{(1)}] = 0 \,,
\end{align}
using $\nabla_a h^{ab} =   \eth_a h^{ab} - \frac{1}{R^2} x_b h$.
Then the physical graviton of \eqref{eq:def_phys_graviton} satisfies the de 
Donder gauge,
\begin{align}
  \nabla^a\tilde h_{ab}[\cB^{(1)}]  -\frac 12  \nabla_b\tilde h = 0 
  \qquad \text{with}  \qquad
  \widetilde{h}_{ab}[\cB^{(1)}] = h_{ab}[\cB^{(1)}] -\frac{1}{2} P_{ab} \tilde 
h \;.
\end{align}
The spin 0 contribution to the metric is interesting because its off-shell modes 
have the wrong (ghost-like) sign in GR. 
This does not happen in the present Yang--Mills model, which is important for 
quantization.
\subsubsection{Spin 1 gravitons}
\label{sec:spin-1-grav}
Next, we compute the spin one contributions to the gravitons on $H^4$.
Taking into account the $\cC^s$ gradation, the averaged metric perturbation 
\eqref{gravitons-H1} is non-vanishing only for the modes $\cA_a^{(3)}$ and  
$\cA_a^{(2')}$.
\paragraph{Spin 1 graviton $\cA_a^{(2)}$.}
Here, we observe
\begin{align}
 H_{ab}[\cA^{(2)}] &=-r^2 R^2 \{P_{ab},\phi^{(1)}\}
 -r^2 R^2 \left(\nabla_a \cA^{(2)}_b+\nabla_b \cA^{(2)}_a \right) 
 +r^2 \left( x_a \cA^{(2)}_b + x_b \cA^{(2)}_a \right) \nn \\
 &=-r^2 R^2 \left(\nabla_a \cA^{(2)}_b+\nabla_b \cA^{(2)}_a \right) 
 \label{spin-1-grav-A2}
\end{align}
such that the averaging yields
\begin{align}
 h_{ab}[\cA^{(2)}]=\a_1\left( \nabla_a (\Box -2 r^2) \phi_b + \nabla_b 
(\Box -2 r^2) \phi_a \right) \;.
\end{align}
This    has the form of pure gauge (diffeomorphism) contributions.
Since the $\cA^{(2)}$ modes are pure gauge, they are not  physical 
in the present model.
\paragraph{Spin 1 graviton $\cA_a^{(3)}$.}
Similarly, we have
\begin{align}
 H_{ab}[\cA^{(3)}] = \theta^{ad}  \theta^{bf} \left( \eth_f \phi_d
+\eth_d \phi_f\right) 
\end{align}
such that averaging yields
\begin{align}
 h_{ab}[\cA^{(3)}]= -\a_1 \left( \nabla_a \phi_b + \nabla_b \phi_a 
\right) \; .
\end{align}
\paragraph{Physical spin 1 gravitons.}
For the spin 1 eigenmodes $\cB^{(i)}$ of \eqref{B-modes-def}, we therefore obtain the 
following physical gravitons:
\begin{align}
\begin{aligned}
\widetilde{h}_{ab}[\cB^{(1)}] &= \widetilde{h}_{ab}[\cB^{(4)}] =0 
 \, ,  \\
 \widetilde{h}_{ab}[\cB^{(2)}] &=\a_1(1-\a_1) \left( \nabla_a (\Box -2 r^2) 
\phi_b + 
\nabla_b (\Box -2 r^2) \phi_a \right)
 \, , \\
 \widetilde{h}_{ab}[\cB^{(3)}] &=-\a_1 \left( \nabla_a \phi_b + \nabla_b \phi_a
\right)  \, .
\end{aligned}
\end{align} 
Hence there is indeed a physical spin 1 mode
 $\widetilde{h}_{ab}[\cB^{(3)}]$  contributing to the metric fluctuations. 
Nevertheless,
since it has the form of pure gauge (diffeomorphism) contributions,  
it will decouple from a conserved energy-momentum tensor $T_{\mu\nu}$. 
\subsubsection{Spin 2 gravitons}
Finally, we consider the spin 2 fluctuations of the background and evaluate 
their associated gravition modes.
\paragraph{Spin 2 graviton $\cA_a^{(1)}$.}

Since $\cA_a^{(1)} = \eth_a \phi^{(2)}$ with  $\phi^{(2)} = 
\{x^a,\{x^b,\phi_{ab}\}\} \in \cC^2$, we have
\begin{align}
 H_{ab}
 &= \theta^{da}\{\eth_d \phi^{(2)},x^b\} + (a \leftrightarrow b)   \nn\\
 &= \{x^b,\{\theta^{ad}\eth_d \phi^{(2)}\} - \{\theta^{da},x^b\}\eth_d \phi^{(2)} + (a \leftrightarrow b)   \nn\\
 &= \{x^b,\{x^a, \phi^{(2)}\}\} - \{\theta^{da},x^b\}\eth_d \phi^{(2)} + (a 
\leftrightarrow b)  \; .
\end{align}
The second term drops out in the projection to $\cC^0$, and using 
\eqref{x-phi-comm-spins} twice one finds
\begin{align}
 h_{ab}[\cA^{(1)}] = \frac{4}{L_{NC}^4}[\{x^b,\{x^a, \phi^{(2)}\}\}]_0 
 =  \frac{2}{R^2 r^2}\hat\a_1\hat\a_2\phi_{ab} \; .
\end{align}
\paragraph{Spin 2 graviton $\cA_a^{(2)}$.}
For $\cA_a^{(2)} = \{x^c,\phi_{bc}\}  \in \cC^1 \oplus \cC^3$, it follows that
$H_{ab} \in \cC^{\odd}$ and therefore 
\begin{align}
 h_{ab} \propto [H_{ab}]_0 = 0 \; .
\end{align}
In fact this is a pure gauge mode in the model.
\paragraph{Spin 2 graviton $\cA_a^{(3)}$.}
Next, consider $\cA_a^{(3)} = \{x^c,\phi_{ac}\} \ \in \cC^1$. Then 
$H_{ab} \in \cC^{\odd}$, and again
\begin{align}
  h_{ab} \propto [H_{ab}]_0 = 0 \; .
\end{align}
\paragraph{Spin 2 graviton $\cA_a^{(4)}$.}
Finally, consider the mode  $\cA_a^{(4)} = \theta^{ae}\{x^c,\phi_{ec}\} \ \in 
\cC^2$. Then 
\begin{align}
 H_{ab}
 &= \theta^{da}\{\theta^{de}\{x^c,\phi_{ec}\},x^b\} + (a \leftrightarrow b) 
\nn\\
 &= \theta^{da}\theta^{de}\{\{x^c,\phi_{ec}\},x^b\} 
  +  \theta^{da}\{\theta^{de},x^b\}\{x^c,\phi_{ec}\}  + (a \leftrightarrow 
b)\nn\\
 &= r^2 R^2 \{\{x^c,\phi_{ac}\},x^b\} - r^2 \theta^{ba} \{x^c,\phi_{ec}\}x^e + 
(a \leftrightarrow b)\nn\\
 &= -r^2 R^2 \{x^b,\{x^c,\phi_{ac}\}\} + (a \leftrightarrow b)
\end{align}
using \eqref{phi2-a-tangential}.
Recall that \eqref{x-phi-comm-spins} implies $[\{x^b,\{x^c,\phi_{ac}\}]_0 = - 
\hat\a_1 \phi_{ab}$, and  therefore
\begin{align}
 h_{ab}[\cA^{(4)}] = 2\a_1 (\Box-2r^2) \phi_{ab} \; .
\label{graviton-a4-explicit}
 \end{align}
\paragraph{Physical gravitons.}
Computing the gravitons for the eigenmodes $\cB^{(i)}$, we find
\begin{align}
 \widetilde{h}_{ab}[\cB^{(i)}] &=0 \qquad \text{for} \quad i=1,2,3 \nn\\
  \widetilde{h}_{ab}[\cB^{(4)}]&= 2  \a_1 (\Box -2r^2) \phi_{ab} , \qquad \eth^a 
\widetilde{h}_{ab}[\cB^{(4)}] = 0 = \nabla^a \widetilde{h}_{ab}[\cB^{(4)}]
\end{align}
using \eqref{del-Hab-id}.
The trivial result for $\cB^{(i)}$, $i=2,3$, is obvious, as the individual 
contributions for $\cA^{(i)}$, $i=2,3$, vanish. However, the vanishing contribution  
of $\cB^{(1)}$ is the result of a non-trivial cancellation of the contributions from $\cA^{(1)}$ and 
$\cA^{(4)}$. 

In summary, the physical fields contributing to the metric fluctuations  
are a spin 2 field, a spin 1 field, and a spin 0 field.
This is somewhat reminiscent of scalar-vector-tensor gravity.
The spin 0 and spin 2 modes both satisfy the de Donder gauge.

To understand the present organization into spin modes, recall that 
the linearized metric fluctuations $h_{ab}$ decompose in general as
\begin{align}
 h_{ab} = h_{ab}^{(2)} + \nabla_a \xi_b + \nabla_b \xi_a
  + \frac{1}{4} \eta_{ab} h
\end{align}
where $h_{ab}^{(2)}$ is a divergence-free, traceless spin 2 tensor.
This  corresponds to our spin 2, spin 1 and spin 0 contribution to the graviton;
note that $\xi_a$ contains another spin 0 (divergence) mode.
While the $\xi_a$ fields are unphysical pure gauge modes,
the spin 0 part $h$ is a physical field which 
is in general sourced by the trace of the energy-momentum tensor. 
In the Einstein-Hilbert action,  this spin 0 field enters with the 
``wrong'' sign, cf. \cite{Antoniadis:1986sb}.
This does not happen here, which is certainly welcome for the quantization of 
the model.
%
%
\subsection{Classical action for metric fluctuations}
Having defined the notion of physical graviton in 
\eqref{eq:def_phys_graviton}, an effective 4-dimensional action for 
$\widetilde{h}^{ab}$ is desirable. 
By writing the trace as an integral as in \eqref{integral-trace},
one can express the (gauge-fixed) kinetic term for $\cB^{(4)}$ in terms of 
$\widetilde{h}_{ab} \equiv \widetilde{h}_{ab}[\cB^{(4)}]$ as follows:
\begin{align}
 S_2 
 &= \frac{1}{g^2}  \int \rho \ \cB^{(4)} \cD^2 \cB^{(4)}  \
= \frac{1}{g^2} \a_1 r^2 R^2 \int  \rho \
\phi_{ab}^{(2)}[\cB^{(4)}] (\Box -2r^2 )^2 \phi_{ab}^{(2)}[\cB^{(4)}] \nn \\
  &= \frac{1}{4\a_1 g_{\rm YM}^2 L_{NC}^4} \int  \
 \widetilde{h}_{ab}[\cB^{(4)}] \widetilde{h}_{ab}[\cB^{(4)}] 
 \label{bare-action-massterm}
\end{align}
where $g_{\rm YM}$ is the dimensionless Yang--Mills/Maxwell coupling constant 
\eqref{YM-coupling}.
Superficially, this looks like a mass term for the graviton; 
however this is only the spin two mode, which is by definition invariant under 
diffeomorphisms.
Hence \eqref{bare-action-massterm} could also be viewed as
the quadratic contribution to the cosmological constant in GR\footnote{Hence 
a large positive  mass would not imply large curvature but rather a short range 
of these modes. See e.g.\ \cite{Gabadadze:2003jq} for a related 
discussion.}.

Taking into account a coupling to matter of the form $\d_h S = \frac{1}{2} \int 
\widetilde{h}_{ab} T^{ab}$, the equations of motion for $\widetilde{h}_{ab}$ 
become
\begin{align}
 \widetilde{h}_{ab}[\cA^{(4)}] 
 =  - \frac{4}{3} \, \frac{ g^2\, }{ \r L^4_{NC} \,}  T_{ab} 
 =  - \frac{1}{3} \, g_{\rm YM}^2 L^4_{NC}  T_{ab}  \; .
\end{align}
Clearly $\widetilde{h}_{ab}$ is not propagating, but acts like an auxiliary 
field which tracks $T_{ab}$. As a consequence, the pure matrix model action 
\eqref{bosonic-action} does not lead to gravity on $H^4$, similar to the case of 
$S^4_N$ \cite{Sperling:2017gmy}. 
Nevertheless, the action \eqref{bosonic-action} does define a non-trivial, and 
apparently not pathological, spin 2 theory in 4 dimensions with a propagating 
spin 2 field $\phi_{ab}$, which should be suitable for quantization. 
Gravity may then arise upon quantization, as discussed next.
%
%
\subsection{Induced gravity}
\label{sec:induced_gravity}
At first sight it may seem disappointing that gravity does not arise from the classical action. 
On the other hand, since classical GR  is not renormalizable, it should 
presumably be viewed as a low-energy effective theory. 
Adopting this point of view, it is reasonable that the starting point of 
an underlying quantum theory can be very different at the classical level, as 
for instance in the approach advocated here. 
This train of though is exactly the idea of \emph{emergent 
gravity}\footnote{In fact, it is known that the Type IIB bulk gravity in the 
IKKT model arises 
only at one loop \cite{Ishibashi:1996xs}. However, this is a different issue, 
since the present degrees of freedom are only 4-dimensional.}.

As soon as quantum effects in the matrix model are taken into account, the 
effective metric $\widetilde{h}_{ab}$ will unavoidably acquire a kinetic term, 
and therefore propagate.
More specifically, it is well-known that induced gravity terms arise at one 
loop, 
upon integrating out fields that couple to the effective metric 
\cite{Sakharov:1967pk,Visser:2002ew,Donoghue:2017vvl}.
The induced terms include the cosmological constant and Einstein-Hilbert terms.
The maximal supersymmetry of the underlying model\footnote{This really requires the maximal supersymmetry of the IKKT model, 
otherwise UV/IR mixing effects will render the model strongly non-local and probably pathological, 
cf. \cite{Minwalla:1999px,Steinacker:2016nsc}.} 
along with the finite density of states of the solution
strongly suggests that the model is UV finite and ``almost-local''.
Moreover, the usual large contribution to the cosmological constant  $\int 
\sqrt{g} \L^4$ is avoided here, cf.\ the one-loop computation in 
\cite{Steinacker:2016vgf}.
Canceling also the induced Einstein-Hilbert term 
is more subtle\footnote{On Moyal--Weyl backgrounds, $\cN=1$ SUSY is sufficient 
to 
cancel the induced ``would-be'' cosmological constant term, while the induced 
Einstein--Hilbert term is only canceled in the $\cN=4$ 
case \cite{Klammer:2008df,Blaschke:2010rr}. This is reflected by the absence of UV/IR mixing. 
Here the background and the explicit mass term induce a spontaneous and soft 
breaking of $\cN=4$ SUSY. Nevertheless, the suggested scenario seems 
reasonable.}, 
and it is plausible that the supersymmetry breaking $H^4$ background does lead 
to an induced Einstein-Hilbert term with scale $\tilde\L = 
O\left(\frac{1}{r}\right)$. 

Motivated by these considerations, one may add a term  $\int \sigma \tilde\L^2 
\widetilde{h}_{ab} \eth\cdot\eth \widetilde{h}^{ab}$ to the action 
\eqref{bare-action-massterm}, 
with $\sigma =\pm1$, such that the total action coupled to matter reads
\begin{align}
 S  &= \int \sigma \tilde\L^2 \widetilde{h}_{ab} \eth\cdot\eth  
\widetilde{h}^{ab} 
+ \frac{4}{3g^2_{\rm YM}L_{NC}^4} \int   \ \widetilde{h}_{ab} 
\widetilde{h}_{ab}
  + \frac{1}{2} \int \widetilde{h}_{ab} T^{ab}  \; .
\end{align}
The equation of motion for $\widetilde{h}_{ab}$ are then
\begin{align}
 \left( \eth\cdot\eth + \frac{4}{3\s g_{\mathrm{YM}}^2 
L^4_{NC}\tilde{\L}^2}\right) \widetilde{h}_{ab} 
 &= - \frac{1}{4\s \tilde{\L}^2}  T_{ab}
\end{align}
where $\tilde{\L}$ is the effective cutoff scale set by induced gravity.
For $\s=-1$, this is indeed a reasonable equation for linearized gravity, with the 
effective Newton constant 
\begin{align}
 8\pi G_N = \frac{1}{8\tilde{\L}^2} 
\end{align}
and mass scale 
\begin{align}
 m^2 = O\left( \frac{1}{g_{\rm YM}^2 L^4_{NC}\tilde{\L}^2}\right) \,.
\end{align}
The mass scale can become very small $m^2 = O\left(\frac{1}{R^2}\right)$  if 
$\tilde{\L} = O\left(\frac{1}{r}\right)$ and $n$ is large, or upon 
projection to the Minkowski space-time $\cM^{3,1}$,
where the universe  grows in time.
Of course, the mass term will acquire quantum corrections too,
which will be suppressed by supersymmetry. It would be desirable to study 
this in more detail elsewhere.

Even though such a mass term might be interpreted in terms of a cosmological 
constant in linearized GR, its meaning here is somewhat different. 
As in GR, a proper interpretation  requires the full non-linear 
theory.
However, it is plausible that a positive mass term may simply imply an IR cutoff for gravity here,
while the large-scale structure of the background solution might not be affected.
Therefore a small, but non-zero mass term is quite welcome in the presented 
setting to ensure stability, while the large-scale cosmology would be 
determined by the background solution,
as illustrated in section \ref{sec:projection-Lorentzian}.
%
%
\subsection{Local gauge transformations}
\label{sec:local-gauge}
Among the higher-spin gauge transformations $\d_\L(x^a+\cA^a) \coloneqq 
\{x^a+\cA^a,\L\}$  generated by $\L \in \cC$, consider the spin 1 gauge transformations
 generated by
\begin{align}
 \L^{(1)} = \{x^a,v_a\} \ =   \ \theta^{ab}\eth_b v_a \quad \in \cC^1
\end{align}
with $v_a(x)$ a divergence-free vector field. 
These  correspond to (volume-preserving) diffeomorphisms on $H^4$. 
The action on scalar functions $\phi(x)$ reads
\begin{align}
 \d_\L \phi &= \{\phi(x),\L^{(1)}\} =   \{\phi(x), \theta^{ab}\eth_b v_a\}
\end{align}
so that the action on vector fluctuations is
\begin{align}
 \d_\L \cA_a = \d_\L x_a  + \{\cA_a,\L^{(1)}\} 
\end{align}
with\footnote{Note that $\{\cA_a,\L^{(1)}\}$ is not necessarily tangential. 
However that term vanishes in the 
semi-classical limit, and is significant only for nonabelian gauge fields which 
we do not consider.
The proper treatment is of course to impose the non-linear constraint $Y_a Y^a = 
-R^2$, which would restore gauge invariance.}
\begin{align}
 \d_\L x_a  &= \{x_a,\L^{(1)}\} = \{x_a,\L\}_0 + \{x_a,\L^{(1)}\}_2 \nn\\
  &= \a_1(\Box-2 r^2) v_a + \cA_a^{(2')}[\L^{(1)}]
\end{align}
using \eqref{x-phi-comm-spin1}.
The first term describes a diffeomorphism corresponding to the 
vector field $\tilde v_a =  \a_1(\Box-2 r^2) v_a$.
The second term accounts for the spin 1 pure gauge mode $\cA_a^{(2')}$ 
as discussed in section \ref{sec:H4-tang}, whose contribution to the 
graviton $\widetilde{h}_{ab}$ was computed in \eqref{spin-1-grav-A2}.
The higher-spin gauge transformations could be worked out similarly.

Since there is only one such gauge invariance, but several fields for each spin,
one may worry about the consistency of the model.
However, recall that the gauge-fixed action \eqref{eq:gauge_fixed_action} has 
been proven to be well-defined and non-degenerate in section 
\ref{sec:gaugefixed-action}. 
Hence there is no problem at least in the  Euclidean setting.
This is due to the special origin of the fluctuation modes in $\End(\cH_n)$, see
\eqref{End-decomposition-Cn}.
%
%
\section{Lorentzian quantum space-times from fuzzy 
\texorpdfstring{$H^4_n$}{H4n}}
\label{sec:projection-Lorentzian}
Having disentangled the fluctuations on $H^4_n$, we would like to apply these 
tools to the more interesting cosmological space-time solutions $\cM^{3,1}$.
Since the latter is obtained by a projection considered in section 
\ref{sec:poincare}, many considerations remain valid. 
Most importantly, the fluctuation modes originate from the \emph{same} 
$\End(\cH_n)$ such that we can rely on the same spin operator $\cS^2$, and
our classification can be carried over. 
Moreover, the tangential fluctuations on $H^4_n$ are in 
one-to-one correspondence to the full set of fluctuation 
modes on $\cM^{3,1}$, as will be shown below.
The symmetry group is reduced to 
$SO(3,1)$ instead of $SO(4,1)$, which is weaker, but should still be very 
useful.
\subsection{Cosmological space-time solutions}
By projecting fuzzy $H^4_n$ onto the $0123$ plane via $\Pi$ of 
\eqref{proj-class} 
i.e.\ by keeping the 
$Y^\mu = \cM^{\mu a}\a_a$ for $\mu=0,1,2,3$ and dropping $Y^4$,
we obtain $(3+1)$-dimensional fuzzy space-time solutions.
Since the embedding metric $\eta^{\mu\nu}$ is compatible with $SO(3,1)$, we have 
\begin{align}
  [Y_\r,[Y^\r,Y^\mu]] &= i(\a\cdot\a)  [Y_\r,\cM^{\r\mu}]  = -i (\a\cdot\a) [\cM^{\r\mu},Y_\r] \nn\\
    &= (\a\cdot\a)
    \begin{cases}
          Y^\mu, & \mu\neq\r \\
               0, & \mu = \r
    \end{cases}
\qquad\text{(no sum)} 
\end{align}
such that
\begin{align}
 \Box_{Y} Y^\mu =  [Y^\rho,[Y_\rho,Y^\mu]] &= 3 (\a\cdot\a) \, Y^\mu \; .
 \label{Y-Box}
\end{align}
Depending on $\a\cdot \a$ we obtain three different types of 
quantized space-time solutions with Minkowski signature in the IKKT model with 
mass term. These are:
\begin{align}
\begin{aligned}
 \Box_X X^\mu &= -3 r^2\, X^\mu  \; ,\\
  \Box_T T^\mu &= 3 r^2\,  T^\mu \; , \\
  \Box_Z Z^\mu &= 0 \; .
  \end{aligned}
\end{align}
Choosing a positive mass term to ensure stability, we focus on the 
solution 
\begin{align}
 Y^\mu &= X^\mu, \ \qquad  r^2 = \frac 13 m^2  \ .
 \label{X-solution-4}
\end{align}
This is the
homogeneous and isotropic quantized FLRW cosmological space-time $\cM_n^{3,1}$  with $k=-1$
introduced\footnote{We change notation from \cite{Steinacker:2017bhb}, 
where $Y^1$ was dropped instead of $Y^4$.} in \cite{Steinacker:2017bhb}.
Here $m^2$ sets the scale $r^2$, while $n$ remains 
undetermined.
These backgrounds are $SO(3,1)$-covariant, which is the symmetry respected by 
$\Box_X$. 
\subsection{Semi-classical geometry}
We first recall the  semi-classical limit of this space 
\cite{Steinacker:2017bhb}, with $x^\mu$ for $\mu=0,1,2,3$ as coordinates on  
$\cM$. By $SO(3,1)$-invariance, we can 
always consider the local reference point $\xi$ on $H^4$ resp. $\cM$
\begin{align}
 \xi = (x^0,0,0,0,x^4) \stackrel{\Pi}{\to} \ (x^0,0,0,0) , \qquad x^0 = R\cosh(\eta),  \ \ x^4 = R\sinh(\eta) \ .
\end{align}
Globally, we have the following constraints
\begin{align}
 x_\mu x^\mu &= -R^2 - x_4^2 = -R^2 \cosh^2(\eta) \, ,  \nn\\
 t_{\mu} t^{\mu}  &=  r^{-2}\, \cosh^2(\eta) \,, \nn\\
 t_\mu x^\mu &= 0, \qquad \mu,\nu = 0,\ldots ,3 
\end{align}
where $\eta$ will be a global ''cosmic`` time coordinate.
From the radial constraint $x_a x^a = -R^2$ on $H^4$  one deduces $\{x_a x^a, 
x^\mu\} = 0$, which further implies  
\begin{align}
 0 &= x_a m^{a\mu} = x_\nu m^{\nu \mu} + x_4 m^{4\mu} \ .
 \label{radial-constraint}
\end{align}
This establishes a relation between the momenta and the  $t^{\mu}$, 
\begin{align}
 t^\mu &=  \frac 1R\, m^{\mu4} = \frac{1}{R r^2 x^4}\, x_\nu\theta^{\nu\mu} 
  \ \stackrel{\xi}{=} \ \frac{1}{R r^2}\, \frac{1}{\tanh(\eta)}\theta^{0\mu} \ .
\end{align}
Furthermore, the self-duality constraint \eqref{SD-H-class} reduces 
to\footnote{Note that 
this form only applies in the special $\mso(3,1)$ adapted frame, and it is not generally covariant; 
of course on Minkowski manifolds, there is no notion of self-duality. However there can be a $SO(3,1)$-invariant relation as above 
which holds in the preferred cosmological frames, and this is what happens here. 
This is one reason why it is important 
to \emph{not} have full Poincare covariance in the Minkowski case.}  
\begin{align}
t^i &= \frac 1R m_{i 4} = \frac 1{nRr^2} \epsilon_{abc i 4} \theta^{ab} x^{c} 
\ \stackrel{\xi}{=} \ \frac 1{nr^2} \cosh(\eta)\epsilon^{i jk} \theta^{jk}  , \nn\\
t^0  &\stackrel{\xi}{=} \  0 \,, 
\end{align}
where the last equation is simply a consequence of $x_\mu t^\mu = 0$.
Therefore $t^\mu$ describes a space-like $S^2$ with radius 
$r^{-2} \cosh^2(\eta)$.
Conversely, the above relations allow to express   $\theta^{\mu\nu}$ 
 in terms of the momenta $t^\mu$ as follows
\begin{align}
\begin{aligned}
 \theta^{ij} &= \frac{n r^3}{2 \cosh(\eta)} \varepsilon^{ijk} \, t^k  \;, \\
 \theta^{0i} &=  R r^2\tanh(\eta)\,  t^i \; .
 \end{aligned}
 \label{theta-in-terms-of-p}
\end{align}
By means of $R \sim \frac 12 n r$, one can summarize 
\eqref{theta-in-terms-of-p} neatly:
\begin{align}
 \theta^{\mu\nu} &= r^2 R\,\eta^{\mu\nu}_\a(x)\, t^\a \:, 
 \label{theta-P-relation} 
\end{align}
where $\eta^{\mu\nu}_\a(x)$ is a  $SO(3,1)$-invariant tensor field on 
$\cM^{3,1}$, which is analogs of the t'Hooft symbols.
Note that $\theta^{0i} \gg \theta^{ij}$ for late times $\eta \gg 1$;
this reflects the embedding of $H^4 \subset \R^{4,1}$ which approaches the light cone at late times.
Thus space is almost commutative, but space-time is not.
Nevertheless the effects of non-commutativity will be weakened due to the 
averaging on $S^2$.
Finally the constraint \eqref{theta-constraint} reads 
\begin{align}
  \g^{\a\b} \coloneqq \eta_{\mu\nu}\theta^{\mu\a}\theta^{\nu\b} 
 &= \frac{L_{NC}^4}{4}(\eta^{\a\b} + \frac 1{R^2} x^\a x^\b - R^2\, t^\a t^\b) 
 \label{gamma-on-M31}
\end{align}
which at the chosen reference point yields
\begin{align}
\begin{aligned}
  \g^{ij}  &= \frac{L_{NC}^4}{4}\,(\d^{ij} - R^2\, t^i t^j)  \; ,\\
  \g^{00}  &= \frac{L_{NC}^4}{4} \sinh^2(\eta)  \; ,\\
  \g^{0j}  &= 0 \; .
\end{aligned}
\end{align}
\paragraph{Averaging  and effective metric on 
\texorpdfstring{$\cM^{3,1}$}{M31}.}
An effective metric  for scalar fields $\phi(x)$ on $\cM^{3,1}$
can be defined by the quadratic action \eqref{eff-G-action}.
Looking at \eqref{gamma-on-M31}, we note that $\g^{\a\b}$ contains the term 
$t^\a t^\b$,
which is not constant on the fiber $S^2$.
By averaging over the fiber, one obtains the following result \cite{Steinacker:2017bhb}
\begin{align} 
\begin{aligned}
 [\g^{ij}]_0 &= \frac{L_{NC}^4}{4}\,\d^{ij} - [t^{i} t^j]_0 
    =  \frac{L_{NC}^4}{12}\,\big(3 - \cosh^2(\eta) \big) \d^{ij}  \; , \\
 [\g^{00}]_0  &= \frac{L_{NC}^4}{4} \sinh^2(\eta) , \qquad [\g^{0i}]_0  = 0 \; .
 \end{aligned}
 \label{gamma-M31-average}
\end{align}
Note the signature change at $\cosh^2(\eta) = 3$ which marks the Big-Bang in 
this model, and the large pre-factors which grow in time $\eta$. 
Taking into account  the conformal factor in the effective metric $G^{\mu\nu}$ 
\eqref{eff-metric-G}, 
one obtains the cosmic scale parameter $a(t) \propto t$ for late times, corresponding to a 
coasting universe \cite{Steinacker:2017bhb}.

However we have not yet shown that this metric $G^{\mu\nu}$  governs all of the 
low-energy physics, and that there are no tachyonic or ghost modes.
The large local symmetry of the model 
and the universal structure of the Yang--Mills action should help to elaborate 
the full dynamics.
Here we only take some steps in that direction: we  
establish a precise correspondence between the fluctuation modes as well as a 
close relation between the action of both spaces. 
\subsection{Wave-functions, higher-spin modes and constraints}
\label{sec:wavefunct-higherspin}
In this section we briefly comment on the  fluctuation modes on  $\cM^{3,1}$.
The space of functions $\End(\cH_n)$ on $\cM^{3,1}$ is the same as on $H^4_n$, 
meaning that the decomposition \eqref{End-decomposition-Cn} remains valid.
The modes  will still be considered as  
functions (or sections of higher-spin bundles) on $H^4$, such that a 
representation as in \eqref{phi-M-expand} is expected to hold.
Consequently, the modes can be interpreted as functions (or higher-spin modes) 
on\footnote{We will 
ignore the dependence on two sheets $\cM^\pm$ for simplicity.}
$\cM^{3,1}$ via $\Pi$ of \eqref{proj-class}.
The  $\phi_{ab}(x)$ etc.\ then define some higher-rank field on $\cM^{3,1}$.
In the following, we will only address a few basic points.
\subsection{Tangential fluctuation modes, relation with 
\texorpdfstring{$H^4$}{H4} and \texorpdfstring{$SO(4,1)$}{SO(4,1)}}
\label{sec:tang-fluct-M31}
Now consider  fluctuations $y^\mu = x^\mu + \cA^\mu$ around $\cM^{3,1}$. 
The first observation is that these four fluctuation modes $\cA^\mu, \ \mu = 
0,\ldots ,3$ are in one-to-one correspondence with the tangential fluctuations 
on $H^4$. To see this, recall 
that tangential fluctuations on $H^4$ satisfy by definition\footnote{A 
gauge-invariant constraint would be 
 $(X^a + \cA^a)(X_a + \cA_a) = -R^2$. For the present purpose, 
 its linearized form is what we want.} the constraint
\begin{align}
 \cA_a x^a  &= 0, \qquad   \cA_4 = - \frac{x_\mu}{x_4}\, \cA^\mu \ ,
 \label{constraints-A-M31}
 \end{align}
 with $\cA_a \in \End(\cH)$. To associate a general fluctuation mode 
on $\cM^{4,1}$ one simply drops $\cA^4$, and conversely $\cA^4$ can be 
recovered from $\cA^\mu$ via \eqref{constraints-A-M31}. 
 Hence there is a correspondence of tangential fluctuations
 \begin{align}
 \xymatrix{
    \ar[d]_{SO(4,1)}  \cA^a \ \text{on}\ H^{4}  \ar@{<->}[r] & \cA^\mu \ 
\text{on}\ \cM^{3,1}  \\
     \cA^a \ \text{on}\ H^{4}    \ar@{<->}[r]    & \cA^\mu \ \mbox{on}\ 
\cM^{3,1}  
} 
\end{align}
Since the maps are invertible, an $SO(4,1)$-action is defined on the 
fluctuations $\cA^\mu$ on $\cM^{3,1}$,
 which, however, is not an isometry and not unitary. Nevertheless it acts as a 
structural group, and organization developed for $H^4$ in the previous sections 
remains applicable.
As a consequence, configurations in the $\cM^{3,1}$ 
model can be mapped one-to-one to configurations in the $H^4$ model.
 Similarly, higher-rank tangential tensors on $H^4$ such as the gravitons
\begin{align}
  h_{ab} x^a = 0
\end{align}
can be mapped one-to-one to tensors $h_{\mu\nu}$ on $\cM^{3,1}$, and 
the missing components $h_{ab}$ are uniquely determined from the $h_{\mu\nu}$. 
In the same vein, all internal fluctuations on $S^2$ will be organized in a 
$SO(4,1)$-covariant way as on $H^4$.
This relation is somewhat analogous to a Wick rotation.
\paragraph{Action and dynamics.}
The matrix model provides again an action for the fluctuation modes $\cA^\mu$, which has the same structure
as in section \ref{sec:quad-action-H4}, 
\begin{align}
 S_\cM[\cA] &= \int \cA_\mu \left(\Box_\cM -2\cI+\frac{1}{2} \mu^2 \right) 
\cA^\mu  
 \label{action-M31-quad}
\end{align}
upon gauge fixing.
The  matrix Laplacian on $\cM^{3,1}$ is related to the one on $H^4$ through 
\begin{align}
 \Box_\cM  = \Box_H - [X^4,[X_4,\cdot]] 
 = [X^\mu,[X_\mu,\cdot]] \ \sim \ 
 -\frac{L_{NC}^4}{4}\ \g^{\mu\nu}\del_\mu\del_\nu + 
\ldots  \; .
\end{align}
We can utilize the same mode expansion in terms of $\cA^{(i)}_\mu$ as in 
section \ref{sec:H4-tang},
\begin{align}
\begin{aligned}
 \cA_\mu^{(1)} &= \eth_\mu \phi^{(s)} \ \in \cC^s  ,
 \qquad  \phi^{(s)} = \{x^a,\phi_a^{(s)}\} = 
\{x^\mu,\phi_\mu^{(s)}\}+\{x^4,\phi_4^{(s)}\}  \ \in \cC^s   \\
 \cA_\mu^{(2)} &= \theta^{\mu b}\eth_b \phi^{(s)}  \ = \{x^\mu,\phi^{(s)}\} 
                              \quad \in \cC^{s+1} \oplus \cC^{s-1}   \\
 \cA_\mu^{(3)} &= \phi_{\mu}^{(s)}  \ \in \cC^{s-1} ,  \\
 \cA_\mu^{(4)} &= \theta^{\mu b}\phi_b^{(s)}  \ \in \cC^s   \; ,
 \end{aligned}
 \label{A-M31-spins}
\end{align}
which is $SO(3,1)$-covariant. 
As explained in section \ref{sec:reducible}, the irreducibility constraints, 
i.e.\ transversality and tracelessness, can be implemented as appropriate for 
$\cM^{3,1}$ without changing the setup. 
The relation \eqref{x-phi-comm-spin1} still applies; for example
\begin{align} 
 \{x_\mu,\phi^{(1)}\}_0 = -\frac{2}3 \phi_\mu +  \frac{R^2}{3} \eth^c\eth_c \phi_\mu  \ .
  \label{X-phi-comm-M31}
\end{align}
Note that  $\eth^c\eth_c$ is the \emph{Euclidean} Laplace operator on $H^4$, 
even though we are working in the Minkowski case. Hence the right-hand side of 
\eqref{X-phi-comm-M31} amounts to some field redefinition. 
In the same vein, the higher-derivative terms in the action 
\eqref{kinetic-gaugefixed-general} for 
the rank $s$ tensor fields $\phi_{a_1\ldots a_2}$ amount to  field 
redefinitions.
Therefore one should expect that these  higher-derivative terms
do \emph{not} lead to new degrees of freedom or ghosts.

On the other hand it might be tempting to use a $SO(3,1)$-covariant formalism,
where e.g.\ $\{x^a,\phi_a^{(s)}\}$ in \eqref{A-M31-spins} is replaced by 
$\{x^\mu,\phi_\mu^{(s)}\}$. 
However then some identities are lost,
and it remains to be seen which formalism is more advantageous.
%
%
%
\section{Conclusion and outlook}
\label{sec:conclusion}
In this article we provide a careful and detailed analysis of the fluctuation 
modes on fuzzy $H^4_n$ as a background in Yang--Mills matrix models, focusing
mainly on the semi-classical case. 
While the results are largely analogous to the case of $S^4_N$ 
\cite{Sperling:2017gmy},
the present approach based on a suitable Poisson calculus is  more 
transparent and fairly close to a standard field-theory treatment. 
The intrinsic structure of these quantum spaces is responsible for obtaining
a higher-spin gauge theory, which is fully $SO(4,1)$-covariant.
The key feature is the equivariant bundle structure, which 
leads to a transmutation of would-be Kaluza--Klein modes into higher-spin 
modes.
\paragraph{Summary.}
Let us summarize the main points: 
A suitable set of representations for the construction of $H^4_n$ is
identified as the \emph{minireps} or \emph{doubleton} $\cH_n$, for which we 
recall the oscillator realization in section \ref{sec:fuzzy_geometry}. The 
first 
major step  is a classification of the fuzzy algebra of functions 
$\End(\cH_n)$, which relies on two pillars: (i) the construction of a spin Casimir 
invariant $\cS^2$ which measures the intrinsic angular momentum on the $S^2_n$ fiber, and
(ii) the statement that the quantization map \eqref{quantization-map} is 
surjective. 
This  provides the basis for the expansion \eqref{phi-M-expand} of a 
generic function in $\End(\cH_n)$, which is an expansion in
the generators associated to the fiber. More precisely, $\End(\cH_n)$  decomposes 
into a
set of higher-spin sectors $\cC^s$
\eqref{End-decomposition-Cn}, labeled by 
the spin Casimir. In the semi-classical limit, these become modules over the algebra of functions on $H^4$,
which are identified with tangential, divergence-tree, traceless rank $s$ 
tensor fields on $H^4$.

The second major step is the development of a suitable differential calculus 
built upon derivations $\eth_a$ defined via the Poisson bracket in \eqref{x-nabla-id}. This 
provides the tools to work explicitly with the generic spin $s$ modes on a 
non-compact space.

Having in mind  the IKKT matrix model, we observe that $H^4_n$ is a solution of 
Yang--Mills matrix models with mass term, and classify the fluctuation modes 
around an $H_n^4$ background in section 
\ref{sec:matrix_model}. Building on the understanding of $\End(\cH_n)$, we 
find four tangential \eqref{A-H4-spins} and one radial fluctuation modes for 
each spin $s\geq 1$. We find the explicit eigenmodes $\cB^{(i)}$ of the 
differential operator $\cD^2$ , see \eqref{vector-Laplacian}, which governs the 
fluctuations in the matrix model.
It turns out that the tangential modes are stable, due to positivity results on 
their kinetic terms.

Next, we identify the physical graviton \eqref{eq:def_phys_graviton} as linearized
fluctuation of the effective metric \eqref{eff-metric-G}  around the $H_n^4$ background, 
and we compute the associated graviton modes for spin $s=0,1,2$. 
The gravitons at spin $s=0,2$ naturally satisfy the de Donder gauge. However, it turns out that
the spin $2$ graviton behaves as an auxiliary field, at least at the classical level. 
A more interesting gravitational behavior should be obtained by including 
quantum corrections,
leading to induced gravity terms. We briefly discuss this scenario in 
section \ref{sec:induced_gravity}.

Considering $H_n^4$ as a starting point towards the fuzzy space-time
$\cM_n^{3,1}$, these issues are however less important.   
Since $\cM_n^{3,1}$ is obtained from  $H_n^4$ by a projection, 
the fuzzy algebra of functions for $\cM^{3,1}_n$ coincides with $\End(\cH_n)$, and
our results   provide a useful set of tools.  As first steps, we briefly discuss the geometry
and the organization of higher-spin modes of $\cM^{3,1}_n$, 
and establish a relation between tangential fluctuation on $H_n^4$ and $\cM^{3,1}_n$ in 
section \ref{sec:projection-Lorentzian}.
\paragraph{Discussion and outlook.}
From a physics point of view, the results may seem a bit disappointing 
in the sense that the spin 2 modes do not lead to a propagating graviton at the classical level.
However gravity could be restored in the quantum case, where induced 
gravity terms  arise.
The most encouraging result is that the tangential fluctuations are
stable and do not lead to  ghost-like modes. This is an improvement over 
GR where the off-shell conformal modes have the wrong sign, 
and arguably also over quadratic gravity where ghost-like modes 
arise at least superficially, cf.\ \cite{Salvio:2018crh,Alvarez-Gaume:2015rwa}. 
On the other hand, the radial modes are
unstable here, which however could be cured by a radial constraint.

There are several issues which deserve to be studied further. 
For example, the Poisson calculus developed here should be extended to the 
fully non-commutative case. Likewise, the relation of the present higher-spin 
gauge theory with Vasiliev theory should be clarified.
Some of the structural statemens in sections \ref{sec:wavefunct-spin-modes-H} 
and \ref{sec:wavefunct-semiclassical-H}
would deserve a more rigorous treatment.
Furthermore, the 1-loop computation in \cite{Steinacker:2016vgf} could 
 easily be adapted, since  $H^4$ is locally very similar to  $S^4$.
 This would allow to make more specific statements about the induced gravity terms, 
 although to obtain the  Einstein-Hilbert term may require a more refined approach.
Finally, the minimal case $n=0$ is very remarkable and special, because
it does not correspond to a quantized symplectic space.

The main physics motivation for the present work is the close relation to the
cosmological FLRW-type solutions $\cM^{3,1}$ of 
\cite{Steinacker:2017bhb}, which are obtained 
from a projection of $H^4_n$. 
The fluctuation analysis on  $\cM^{3,1}$ can largely proceed along the same 
lines, with some important differences.
 In particular, the radial modes will disappear while the signature becomes  Lorentzian.
  Furthermore,  field redefinitions such as \eqref{graviton-a4-explicit}, which
 are responsible for the non-propagating nature of the graviton on $H^4$, 
 should  no longer  cancel the propagator.
 Therefore $\cM^{3,1}_n$ is a very promising candidate for a quantum space-time
 with interesting gravitational physics in the framework of matrix models.
However, we  postpone a detailed analysis of $\cM^{3,1}$  to future work.
\paragraph{Acknowledgments.}
We would like to thank Stefan Fredenhagen, Bent Orsted, Peter Presnajder, 
Sanjaye Ramgoolam, Jan Rosseel, and Genkai Zhang for useful discussions and communications.
This work was supported by the Austrian Science Fund (FWF) grant
 P28590, and by the Action MP1405 QSPACE from the European Cooperation in Science and Technology (COST).
%
%
%
\appendix
\section{Some aspects of \texorpdfstring{$SO(4,2)$}{SO(4,2)}}
\label{app:so42}
The Lie algebra $\mso(4,2)$ is defined by
\begin{align}
\left[M_{ab}, M_{cd}\right] = \im \left(\eta_{ac}M_{bd} - \eta_{ad}M_{bc} 
+\eta_{bd}M_{ac}  -\eta_{bc}M_{ad}  \right) \,,
\label{SO42app}
\end{align}
where   $\eta_{ab}=\diag(-1,1,1,1,1,-1)$ with  
$\ a,b,\ldots  =0,1,2,3,4,5$.
Unitary representations of $SO(4,2)$ are given by Hermitian $\cM^{ab}$. 
The maximal compact subgroup of $SO(4,2)$ is $SU(2)_{L}\times SU(2)_{R}\times 
U(1)_{E}$, generated by the following generators:
\begin{align}
L_{m}&=\frac{1}{2} \left( \frac{1}{2} \varepsilon_{mnl} M_{nl}+M_{m4}\right)
\quad
\longrightarrow \ SU(2)_{L}\nn\cr
R_{m}&=\frac{1}{2} \left( \frac{1}{2} \varepsilon_{mnl} M_{nl}-M_{m4}\right)
\quad
\longrightarrow \ SU(2)_{R}
\end{align}
with $m,n,l=1,2,3$. They
satisfy
\begin{align}
\begin{aligned}
 [L_{m},L_{n}]&= \im \varepsilon_{mnl}L_{l} \; ,\\
[R_{m},R_{n}]&= \im \varepsilon_{mnl}R_{l} \; , \\
[L_{m},R_{n}]&= [E,L_{n}]=[E,R_{n}] =0 \; .
\end{aligned}
\end{align}
The $U(1)_{E}$ generator $E = M_{05}$ is the
conformal Hamiltonian, whose spectrum is positive in a positive energy representation.
Denoting the maximal compact Lie sub-algebra  of $SU(2)_{L}\times 
SU(2)_{R}\times U(1)_{E}$ as $\mathcal{L}^{0}$, the conformal algebra $\mg$ has 
a three-graded decomposition
\begin{align}
\mg = \mathcal{L}^{+} \oplus \mathcal{L}^{0} \oplus \mathcal{L}^{-},
\end{align}
with respect to $E$, such that 
\begin{align}
[\mathcal{L}^{0},\mathcal{L}^{\pm}] =  \mathcal{L}^{\pm}, \qquad
[E,\mathcal{L}^{\pm}]=\pm \mathcal{L}^{\pm} \; ,
\end{align}
and $\cL^{\pm}$ are the  non-compact generators. The six roots of $\mso(6)_\C$ 
decompose accordingly into two compact roots $X_{\b_i}^\pm$
and four non-compact roots $X_{\pm\hat\a_{ij}}$. The latter transform as 
$(2)_L\otimes (2)_R$ i.e.\ as complex vectors of $SO(4)$, and satisfy
  $(X_{\pm\hat\a_{ij}})^\dagger = - X_{\mp\hat\a_{ij}}$.

Spinorial representations of $SO(4,2)$ are obtained in terms of the $SO(3,1)$ 
gamma matrices $\gamma_{\mu}$ satisfying 
$\{ \gamma_{\mu},\gamma_{\nu}\}=-2\eta_{\mu\nu}$ for $\mu, \nu = 0, 1, 
2, 3$ and 
$\gamma_{4}\coloneqq\gamma_{0}\gamma_{1}\gamma_{2}\gamma_{3}$ as 
follows\footnote{To maintain a consistent notation for $SO(4,2)$, our 
$\g^4$ is what is usually called $\g^5$; this will not arise explicitly and should not cause confusion.}:
\begin{align}
\Sigma_{\mu\nu} \coloneqq\frac{1}{4\im}\left[\gamma_{\mu},\gamma_{\nu}\right]
\qquad
\Sigma_{\mu 4} \coloneqq -\frac{\im}{2}\gamma_{\mu}\gamma_{4}
\qquad
\Sigma_{\mu 5} \coloneqq -\frac{1}{2}\gamma_{\mu}
\qquad
\Sigma_{45} \coloneqq- \frac{1}{2}\gamma_{4} \ .
\label{Sigma-explicit}
\end{align}
We  adopt the gamma matrix conventions 
\begin{align}
\gamma_{0}=
\begin{pmatrix}
 \one_2 & 0 \\ 0 & -\one_2
\end{pmatrix}
\qquad
\gamma_{m}=
\begin{pmatrix}
 0 & -\sigma_m \\ \sigma_m & 0
\end{pmatrix}
\qquad \Rightarrow \qquad 
\gamma_{4} =
\im 
\begin{pmatrix}
 0 & \one_2 \\ \one_2 & 0
\end{pmatrix}
\end{align}
where $\sigma_m, \, \, m=1,2,3 $ are the usual Pauli matrices. 
They satisfy 
\begin{align}
\begin{aligned}
 \gamma_a^\dagger &= -\gamma_b \eta^{ba} = \g_0  \gamma_a\g_0^{-1},
 \qquad a = 0,1,2,3,4 \\
 \Sigma_{ab}^\dagger &=  \Sigma_{a'b'} \eta^{aa'}\eta^{bb'} = \g_0  \Sigma_{ab}  
\g_0^{-1},\qquad a,b = 0,1,2,3,4,5
 \end{aligned}
 \label{gamma-conj}
\end{align}
as it should be.
The universal covering group of $SO(4,2)$ is 
$SU(2,2)$, which is the group of $4\times 4$ complex matrices with 
\begin{align}
 U^{-1} = \g_0 U^\dagger \g_0^{-1}
\end{align}
which respects the indefinite sesquilinear form
\begin{align}
 \bar \psi_1 \psi_2 = \psi_1^\dagger \g^0 \psi_2 \;.
\end{align}
The 15-dimensional Lie algebra $\msu(2,2)= \mso(4,2)$ can thus be identified with the space of 
traceless complex $4\times 4$ matrices $Z^\a_\b$ with real structure
\begin{align}
 Z^\dagger = \g_0  Z \g_0^{-1} \ .
\end{align}
%
%
\section{Conventions and identities for Gamma matrices}
\label{app:gamma_matrices}
Using the sign conventions $\eta_{ab} = \diag(-1,1,1,1,1)$ and $\eta_{ab} 
=\diag (-1,1,1,1,1,-1)$,
the Gamma matrices of $\mso(4,1)$ are
\begin{align}
 \{\g_a,\g_b\} = - 2 \eta_{ab}, \qquad a,b=0,\ldots ,4
\end{align}
such that $\g_0^2 = \one$ and $\g_0^\dagger = \g_0$, and more generally
\begin{align}
 \g_a^\dagger = \g_0 \g_a\g_0^{-1} = - \eta_{ab} \g_b\ \eqqcolon - \g^a . 
 \label{gamma-dagger}
\end{align}
Then 
\begin{align}
 \g_a \g^a &= -5 \one  
 \label{gamma-tensor-id}
\end{align}
We can  evaluate the  $SO(4,1)$ intertwiner
\begin{align}
 \sum_{a,b\leq 4} \Sigma_{ab}\otimes \Sigma^{ab} 
  &= \cC^2[\mso(4,1)]_{(4)\otimes (4)} - 2 \cC^2[\mso(4,1)]_{(4)}  
\end{align}
acting on 
\begin{align}
 (4) \otimes (4) = ((10)_S \oplus (6)_{AS}\big)_{\mso(4,2)} 
  = ((10)_S \oplus (5)_{AS} \oplus (1)_{AS}\big)_{\mso(4,1)} 
\end{align}
Using the well-known eigenvalues of the quadratic Casimirs (which coincide with those of the 
compact group),
it follows that 
\begin{align}
 \left(\sum_{a,b\leq 4} \Sigma_{ab}\otimes \Sigma^{ab}\right)_S &= \one   
\label{SO41-sigma-sigma-id}  \; ,\\
  \left(\sum_{a,b\leq 5} \Sigma_{ab}\otimes \Sigma^{ab}\right)_S &= \frac 32 
\one  \; .
\label{SO42-sigma-sigma}
\end{align}
This implies
\begin{align}
 \sum_{a\leq 5}  \left(\g_a \otimes \g^a\right)_S =  -\one   
 \qquad \text{and} \qquad
  \sum_{a,b\leq 5} \Sigma_{ab}\Sigma^{ab} = 5  \; .
  \label{SO41-sigma-sigma-mult}
\end{align}
Similarly, there is an  $\mso(4,2)$ identity
\begin{align}
 \eta_{cc'}\big(\Sigma^{ac}\otimes \Sigma^{bc'} + \Sigma^{bc}\otimes \Sigma^{ac'}\big)_S &= \frac 12 \eta_{ab} \ .
\label{Sigma-Sigma-contract}
\end{align}
This holds because both sides are symmetric, therefore it acts on $(0,0,1)^{\otimes_S 2} = (0,0,2)$; 
the resulting symmetric tensor operator $\S^{ab}$
would have to be in
$(0,1,0)^{\otimes_S 2} = (0,2,0) + (0,0,0)$, but $(0,2,0)\notin \End(0,0,2)$, 
thus only $\eta_{ab}$ can occur.
We also note the following $\mso(4,2)$ identities:
 \begin{align}
  \big(\epsilon_{abcdef}\Sigma^{ab}\otimes \Sigma^{cd}\big)_{S}  &= 2 (\Sigma_{ef}\otimes\one + \one\otimes\Sigma_{ef})
  \label{so6-id-sigma}
 \end{align}
 and
\begin{align}
 \{\Sigma^{ab}, \Sigma^{cd}\}_+ &= (\eta^{ac}\eta^{bd} - \eta^{ad}\eta^{bc})\one 
 + \frac 12 \epsilon^{abcdef}  \Sigma^{ef} \;.
\label{sigma-sigma-id}
\end{align}
In particular,
\begin{align}
 \epsilon^{abcdef}\Sigma^{ab}\Sigma^{cd} &= 12  \Sigma^{ef} \; .
\label{eps-sig-sig-id-6d}
\end{align}
%
%
\section{Basic identities for fuzzy \texorpdfstring{$H_n^4$}{H4n}}
\label{sec:fuzzyH4-derivations}
We provide the proofs for the identities given in section \ref{sec:alg-id-H4n}.
First, \eqref{R2-fuzzy-explicit} is obtained from
\begin{align}
 4 X_a X^a
  &= r^2 \bar{\psi}\g^{a}\psi  \bar{\psi'}\g_{a}\psi'  \nn\\
  &=  r^2  \bar{\psi}\g^{a}\g_{a}\psi + r^2 \bar{\psi} \bar{\psi'}\g^{a}\otimes\g_{a}\psi \psi'\nn\\
  &= - 5 r^2  \bar{\psi}\psi - r^2  \bar{\psi} \bar{\psi'}\psi \psi'\nn\\
  &= -5 r^2  \bar{\psi}\psi -  r^2 \bar{\psi}(-\d+\psi\bar{\psi'}) \psi'\nn\\
  &= -\hat N  (\hat N+4)r^2  
 \label{XX-minirep-der}
\end{align}
using the  \eqref{gamma-tensor-id}.
Similarly, \eqref{C2-minirep} follows from
\begin{align}
 C^2[\mso(4,1)] &=
  \sum_{a<b\leq 4}\bar{\psi}\Sigma^{ab}\psi \bar{\psi'}\Sigma_{ab}\psi' \nn\\
  &= \sum_{a<b\leq 4} \Big( \bar{\psi}\Sigma^{ab}\Sigma_{ab}\psi
    + \bar{\psi}\bar{\psi'}\Sigma^{ab}\otimes \Sigma_{ab} \psi \psi'   \Big) \nn\\
 &= \frac 52 \hat N +\frac 12 \bar{\psi}\bar{\psi'} \psi \psi'  \nn\\
   &=
   \frac 12 \hat N(\hat N+4) 
   \label{C2-minirep-der}
\end{align}
and \eqref{C2-SO42-minirep} follows similarly
\begin{align}
 C^2[\mso(4,2)] &= \sum_{a<b\leq 5} \bar{\psi}\Sigma^{ab}\psi \bar{\psi'}\Sigma_{ab}\psi'\nn\\
   &= \sum_{a<b\leq 5} \Big( \bar{\psi}\Sigma^{ab}\Sigma_{ab}\psi
    + \bar{\psi}\bar{\psi'}\Sigma^{ab}\otimes \Sigma_{ab} \psi \psi'   \Big) \nn\\
   &= \frac {15}{4} \hat N +\frac 34 \bar{\psi}\bar{\psi'} \psi \psi'  \nn\\
   &=\frac{3}{4}\hat N(\hat N +4)   
\end{align}
using \eqref{SO42-sigma-sigma}. 
The identity \eqref{epsilonMM-id} is obtained as 
\begin{align}
\epsilon_{abcde}\cM^{ab} \cM^{cd} &= \epsilon_{abcde}\bar{\psi}\Sigma^{ab}\psi \bar{\psi'}\Sigma^{cd}\psi' \nn\\
  &=  \bar{\psi}\epsilon_{abcde}\Sigma^{ab} \Sigma^{cd}\psi  + \bar{\psi}\bar{\psi'}\epsilon_{abcde}\Sigma^{ab}\otimes \Sigma^{cd}\psi\psi'  \nn\\
  &=  12  \bar{\psi}\Sigma^{e6}\psi + 2 \bar{\psi}\bar{\psi'}(\Sigma^{e6}\otimes 1 + 1 \otimes \Sigma^{e6})\psi\psi'  \nn\nn\\
   &=  12  \bar{\psi}\Sigma^{e6}\psi + 4 (\hat N\bar{\psi}\Sigma^{e6}\psi - \bar{\psi}\Sigma^{e6}\psi ) \nn\\
  &=  4r^{-1}(\hat N + 2) X_e = 4 nr^{-1} X_e 
  \label{epsilonMM-id-app}
 \end{align}
 using \eqref{eps-sig-sig-id-6d} and  the Euclidean identities 
\eqref{so6-id-sigma}.
 Finally,  \eqref{selfdual-fuzzy} is obtained from
\begin{align}
 \epsilon_{abcde}\cM^{ab} X^{c} &= r\epsilon_{abcde}\bar{\psi}\Sigma^{ab}\psi \bar{\psi'}\Sigma^{c5}\psi' \nn \\
  &=  r\bar{\psi}\epsilon_{abc5de}\Sigma^{ab} \Sigma^{c5}\psi  + r\bar{\psi}\bar{\psi'}\epsilon_{abc5de}\Sigma^{ab}\otimes \Sigma^{c5}\psi\psi'  \nn\\
 &=  3r \bar{\psi}\Sigma_{de}\psi + \frac 12 r\bar{\psi}\bar{\psi'}(\Sigma_{de}\otimes 1 + 1 \otimes \Sigma_{de})\psi\psi' \nn\\ 
 &=  3 r\bar{\psi}\Sigma_{de}\psi + r(\hat N\bar{\psi}\Sigma_{de}\psi - \bar{\psi}\Sigma_{de}\psi ) \nn\\
 &= (\hat N+2) r\cM_{de} = n r \cM_{de}
\end{align}
 using $\epsilon_{abc5de}\Sigma^{ab}\Sigma^{c5} = 3  \Sigma^{de}$, 
 which follows from \eqref{eps-sig-sig-id-6d}. 
\subsection{Functions on \texorpdfstring{$\cH_n$}{Hn}, spin Casimir 
\texorpdfstring{$\cS^2$}{S2} and  quantization}
\label{sec:funct-spin-coherent}
%


Consider the spin operator \eqref{Spin-casimir}
\begin{align}
\cS^2 &\coloneqq C^2[\mso(4,1)] + \Box = \sum_{a<b\leq 4} 
[\cM_{ab},[\cM^{ab},\cdot]] + [X_a,[X^a,\cdot]] 
 \end{align}
acting on $\End(\cH)$.
We can write this in a Lie algebra basis adapted to $H^4$ 
at the reference point $\xi\in H^4$ 
as follows.
Let $\cM_{ij}, \ i,j=1,2,3,4$ 
be the $SO(4)$ generators, $E = \cM_{05}$ the energy and 
\begin{align}
 Z_j^\pm = \frac{1}{\sqrt{2}}(\cM_{j0}\pm \im \cM_{j5})
 \label{Zpm-generators}
\end{align}
be the non-compact root generators. Then 
\begin{align}
 \cS^2 &= (C^2[\mso(4)] - \d^{ij}\cM_{i0} \cM_{j0}) + (\d^{ij}\cM_{i5}\cM_{j5} - \cM_{05}\cM_{05}) \nn\\
  &= C^2[\mso(4)] - \d^{ij} Z_i^- Z_j^- + i\d^{ij}( \cM_{i0} \cM_{j5} + \cM_{i5} \cM_{j0}) - E^2\nn\\
  &= C^2[\mso(4)] - \d^{ij} Z_i^- Z_j^- - E^2
\end{align}
using \eqref{so6-MM-rel}
\begin{align}
   \d^{ij}(\cM_{i0} \cM_{j5} + \cM_{j5} \cM_{i0}) = \eta_{05} = 0 \ .
\end{align}
Now let $|0\rangle$ be the ground state of $\cH_n$, which satisfies 
\begin{align}
 Z_i^- \left|0\right\rangle = 0 \ .
\end{align}
Then 
\begin{align}
  \cS^2 \left|0 \right\rangle &=
  \left(2\frac{n}{2} \left( \frac{n}{2}+1\right) - 
E^2\right)\left|0\right\rangle 
   = \left(\frac{n}{2}+1\right)\left(\frac{n}{2}-1\right) \left|0\right\rangle  
\nn\\
   &\eqqcolon\cS^2_n \left|0\right\rangle 
\end{align}
using $C^2[\mso(4)]|0\rangle = 2\frac{n}{2}\left(\frac 
n2+1\right)\left|0\right\rangle$
and $E\left|0\right\rangle = \left(1+\frac{n}{2}\right)\left|0\right\rangle$, 
see \eqref{minireps-Fock}.
Now consider
\begin{align} 
 \cS^2\triangleright |0\rangle\langle 0| &= 2 \cS^2_n |0\rangle\langle 0| - \cM_{ij}|0\rangle\langle 0 |\cM^{ij} 
 + \d^{ij}(\cM_{i0}|0\rangle\langle 0|\cM_{j0} - \cM_{i5}|0\rangle\langle 0|\cM_{j5}) + 2E |0\rangle\langle 0 |E  \nn\\
 &=  2 \cS^2_n |0\rangle\langle 0| - \cM_{ij}|0\rangle\langle 0 |\cM^{ij} 
 + \frac 12 Z_{j}^+|0\rangle\langle 0 |Z_{j}^+ + \frac 12 Z_j^-|0\rangle\langle 
0 |Z_j^- + 2E |0\rangle\langle 0|E  \,, \nn
 \end{align} 
noting that the cross-terms cancel. Then $Z_j^-|0\rangle = 0 = \langle 0 |Z_{j}^+$, and 
for the minimal case $\cH_0$ we have moreover $\cM_{ij}|0\rangle = 0$ and 
$E|0\rangle =|0\rangle$. 
We conclude 
 \begin{align}
  \cS^2\triangleright |0\rangle\langle 0| &=  2 \cS^2_0 |0\rangle\langle 0| +2|0\rangle\langle 0| =0 
\end{align}
The same argument applies for any point $\xi\in H^4$, and 
 \eqref{quantization-map} implies that
 the image $\cQ()$ of the coherent state quantization map 
 contains only spin $\cS^2 =0$ states.

For $\cH_n$ with $n\geq 1$, we have to consider the entire $SO(4)$ orbit 
$g \triangleright |0\rangle\langle 0| = g \cdot |0\rangle\langle 0|\cdot g 
\eqqcolon|m\rangle\langle m|$ over $\xi$
where $|m\rangle = g \cdot |0\rangle$ for $g\in SO(4)$.
We can express $\cM_{ij}|m\rangle\langle m|\cM^{ij}$  in terms of the $SO(4)$ Casimir 
\begin{align} 
 - \cM_{ij}|m\rangle\langle m |\cM^{ij} &= 
 \left(C^2[\mso(4)] - 4 \frac n2 \left(\frac n2+1\right)\right)
 \left|m\right\rangle \left\langle m \right| \nn\\
 &= \left(\bigoplus\limits_{s=0}^{n} 2s(s+1) \one_{s} - n(n+2) 
\right)\left|m\right\rangle \left\langle m \right| \ .
\end{align}
 Here  $\bigoplus\limits_{s=0}^{n} 2s(s+1) \one_{s}$
is the decomposition of $C^2[\mso(4)]$
into spin $s$ irreps of $SU(2)$.
Moreover, 
\begin{align} 
 \d^{ij}(\cM_{i0}|m\rangle\langle m|\cM_{j0} - \cM_{i5}|m\rangle\langle m|\cM_{j5}) = 0
\end{align}   
by $SO(4)$ invariance. 
The same argument applies for any point $\xi\in H^4$.
Therefore 
the image of $\cQ(.) \subset \End(\cH_n)$
decomposes  into spin $s$ irreps as follows 
\begin{align}
 \cS^2_{\cQ(.)} &= \bigoplus\limits_{s=0}^{n} \big(2\cS_n^2 + 2s(s+1) 
\one_{s}
    - n(n+2) +2(\frac n2+1)^2 \big) \nn\\
  &=  \bigoplus\limits_{s=0}^{n}\, 2s(s+1)\one_s \ 
\end{align}
which implies \eqref{End-decomposition-Cn} and \eqref{spin-n-expand}.
This respects the structure of a  $\cC^0$ module,
hence $\cC$ can be viewed as a bundle over $H^4$ with fiber given by
the space of functions on the fuzzy sphere $S^2_n$, with
all multiplicities equal to one.

\subsection{Minimal fuzzy \texorpdfstring{$\cH_{n=0}$}{Hn=0}}
\label{sec:app-minimal}
%

For $n=0$, the coherent states fail to describe the underlying symplectic space, as the $SO(4,1)$
orbit acting on the lowest weight state $|\Omega\rangle=|1,0,0\rangle$ of $\cH_0$ is only 4-dimensional. Nevertheless, $\End(\cH_0)$
should be understood as quantized $\C P^{1,2}$.
A more detailed discussion of this interesting case will be provided elsewhere.

\section{Auxiliary identities for semi-classical \texorpdfstring{$H^4_n$}{H4n}}
\label{sec:semiclass-id-appendix}
For any tangential $\phi_a \in \cC$, the following identity holds:
\begin{align}
  x^a\eth_b \phi_a &= -\phi_a\eth_b x^a = -P^{ab} \phi_a = -\phi_b \;.
  \label{x-del-phi-id}
\end{align}
For any tangential, divergence-free $\phi_a\in\cC^0$, formula 
\eqref{M-ad-explicit} yields
\begin{align}
 \{\theta^{ab},\phi_b\} &= r^2(x^a\eth^b - x^b \eta^a)\phi_b = -r^2x^b 
\eta^a\phi_b = r^2 \phi^a \;.
 \label{theta-phi-id}
\end{align}
Moreover, one can verify for any $\phi$ that
\begin{align}
 \{x_c,\eth^c \phi \} &= -\frac{1}{r^2 R^2}\{x_c,\theta^{cd}\{x_d,\phi \}\} \nn\\
  &= -\frac{1}{r^2 R^2}\Big(\{x_c,\theta^{cd}\}\{x_d,\phi \} 
  + \theta^{cd}\{x_c,\{x_d,\phi\}\} \Big) \nn\\
   &= \frac{1}{r^2R^2}\big(4 r^2 x^d \{x_d,\phi\} - \frac{1}{2} \theta^{cd}\{\theta_{cd},\phi \}\big) \nn\\
    &= 0 \; .
  \label{x-del-id}
\end{align}
Furthermore, one computes
\begin{align}
\eth^d (\{x_d,\phi\}) =
\frac{1}{r^2 R^2}
 \theta_{ad}\{x^a,\{x^d,\phi\}\} = 
 \frac{1}{2r^2 R^2}\theta_{ad}\{\theta^{ad},\phi\} = 0
 \label{del-Q-id}
\end{align}
for any  $\phi\in \cC$, 
and 
\begin{align}
 \{x^b,\Box f\}
 &= -\{\{x^b,x^a\},\{x_a, f\}\} - \{x^a,\{\{x^b,x_a\}, f\}\}
  - \{x^a,\{x_a, \{x^b,f\}\}\}\nn\\
 &= -\{\theta^{ba},\{x_a, f\}\} - \{x^a,\{\theta^{ba}, f\}\}
  +  \Box (\{x^b,f\})\nn\\
 &= -\{\{\theta^{ba},x_a\}, f\}\} - 2\{x^a,\{\theta^{ba}, f\}\}
  +  \Box (\{x^b,f\})\nn\\
 &= -\{\{\theta^{ba},x_a\}, f\}\} - 2r^2\{x^a,(x^b \eth_a - x^a \eth_b)f \}
  +  \Box (\{x^b,f\})\nn\\
 &= -4r^2\{x^b, f\} - 2r^2\theta^{ab} \eth_a f +  \Box (\{x^b,f\})\nn\\
 &=  (\Box -2r^2) (\{x^b,f\})
 \label{x-box-f}
\end{align} 
for any scalar function $f\in \cC^0$, 
using \eqref{x-del-id}. 
Finally,
\begin{align}
 \Box(\theta^{ab}\cA_b) &= \theta^{ab}\Box\cA_b + (\Box \theta^{ab})\cA_b - 2 \{x^c,\theta^{ab}\}\{x^c,\cA_b\} \nn\\
  &= \theta^{ab}\Box\cA_b + (\Box \theta^{ab})\cA_b - 2 r^2(\eta^{ac}x^b - \eta^{bc} x^a)\{x^c,\cA_b\} \nn\\
  &= \theta^{ab}\Box\cA_b -2 r^2 \theta^{ab}\cA_b - 2 r^2(- \{x^a,x^b\}\cA_b - x^a \{x^c,\cA_c\} )\nn\\
  &= \theta^{ab}\Box\cA_b -2 r^2 \theta^{ab}\cA_b + 2 r^2(\theta^{ab}\cA_b + x^a \{x^c,\cA_c\} )\nn\\
  &= \theta^{ab}\Box\cA_b  + 2 r^2 x^a \{x^c,\cA_c\}  \,.
  \label{Box-thetaA}
\end{align}
For the reducible tensor contributions \eqref{spin2-puregauge-id}, we need 
\begin{align}
 \{x^b,\eth_a \phi_b\} &=  \theta^{bc}\eth_c\eth_a \phi_b
  = \theta^{bc}\eth_a\eth_c \phi_b + \theta^{bc}[\eth_c,\eth_a] \phi_b   \nn\\
   &=  \eth_a(\theta^{bc}\eth_c \phi_b) - (\eth_a\theta^{bc})\eth_c \phi_b
   -\frac{1}{r^2 R^2} \theta^{bc}\{\theta^{ca}, \phi_b\}    \nn\\
   &=  \eth_a \{x^b,\phi_b\} - \frac{1}{R^2}(\theta^{ac} x^b - \theta^{ab}x^c)\eth_c \phi_b +\{P^{ba}, \phi_b\}  
   + \frac{1}{r^2 R^2} \theta^{ca}\{\theta^{bc}, \phi_b\}  \nn\\
   &=  \eth_a \phi - \frac{1}{R^2} x^b \{x^a, \phi_b\} 
      + \frac{1}{R^2} (x^a  \{x^b,\phi_b\} - \theta^{ab} \phi_b ) 
      - \frac{1}{r^2 R^2} \theta^{ca}\cI(\phi_c) \nn\\
   &=  \eth_a \phi + \frac{1}{R^2} \theta^{ab} \phi_b 
   +  \frac{1}{R^2} (x^a  \phi - \theta^{ab} \phi_b) - \frac{1}{r^2 R^2} \theta^{ca}\cI(\phi_c) \nn\\
   &=  \eth_a \phi + \frac{1}{R^2} x^a \phi - \frac{1}{r^2 R^2} \theta^{ca}\cI(\phi_c) 
      \label{x-eth-phi-rel}
\end{align}
using \eqref{xx-phi-CR}, for any tangential $\phi_a$ with 
$\phi\coloneqq\{x^a,\phi_a\}$.
Finally, we provide a  proof for \eqref{del-del-CR}: 
We compute 
\begin{align}
 \eth^a\eth^c \phi &=  \frac{1}{r^4 R^4} x_b \{\theta^{ab} ,x_d \{\theta^{cd} ,\phi\}\} \nn\\
  &=  \frac{1}{r^4 R^4} \Big( x_b \{\theta^{ab} ,x_d\} \{\theta^{cd} ,\phi\} 
   +  x_b x_d \{\theta^{ab} ,\{\theta^{cd} ,\phi\}\} \Big) \nn\\
  &=  \frac{1}{r^4 R^4} \Big( r^2(R^2 \{\theta^{ca} ,\phi\} +  x^a x_d 
\{\theta^{cd} ,\phi\})  
   +  x_b x_d \{\theta^{ab} ,\{\theta^{cd} ,\phi\}\} \Big)
\end{align}
using \eqref{eq:xx=R}.
Hence
\begin{align}
  [\eth^a,\eth^c] \phi 
  &=  \frac{1}{r^4 R^4} \Big(2R^2 r^2 \{\theta^{ca} ,\phi\} 
  + r^2(x^a x_d \{\theta^{cd} ,\phi\} - (x^c x_d \{\theta^{ad} ,\phi\})  \nn\\
  &\quad +  x_b x_d( \{\theta^{ab} ,\{\theta^{cd} ,\phi\}\}-  \{\theta^{cd} ,\{\theta^{ab} ,\phi\}\})\Big)\nn\\
  &= \frac{1}{r^4 R^4} \Big(2R^2 r^2 \{\theta^{ca} ,\phi\} 
  + r^2(x^a x_d \{\theta^{cd} ,\phi\} - x^c x_d \{\theta^{ad} ,\phi\})  \nn\\
  &\quad +  x_b x_d \{\{\theta^{ab} ,\theta^{cd}\} ,\phi\}\Big)\nn\\
  &= \frac{1}{r^4 R^4} \Big(2R^2 r^2 \{\theta^{ca} ,\phi\} 
  + r^2\big(x^a x_d \{\theta^{cd} ,\phi\} - x^c x_d \{\theta^{ad} ,\phi\}  \nn\\
  &\quad + x_c x_d\{\theta^{ad},\phi\} + x_b x_a\{\theta^{bc},\phi\}
   - R^2\{\theta^{ac},\phi\} \big)\Big)\nn\\
  &= \frac{1}{r^4 R^4} \Big(2R^2 r^2 \{\theta^{ca} ,\phi\} 
  - R^2 r^2\{\theta^{ac},\phi\} \Big)\nn\\
  &= -\frac{1}{r^2 R^2} \{\theta^{ac} ,\phi\} \,.
\end{align}
\paragraph{Covariant derivative \texorpdfstring{$\nabla$}{nabla}.}
Recalling  $\nabla_a \nabla_c \phi = P_{cc'} \eth_a \eth_{c'} \phi$, we obtain
\begin{align}
 [\nabla_a,\nabla_c]\phi &=
 [\eth^a,\eth^c] \phi + \frac{1}{R^2} (x_a \eth_c - x_c \eth_a )\phi  
 +\frac{1}{ R^4} x^a x^c \left(  x^d  \eth_d   -  x^d  \eth_d \right)\phi  \nn
\\
 &\qquad +\frac{1}{r^4 R^6} x_b x_d 
 \left(
 x^c x^{c'} \{\theta^{ab},\{\theta^{c' d}, \phi \}\}
-x^a x^{a'} \{\theta^{cb},\{\theta^{a' d}, \phi \}\} 
 \right) \nn \\
&=
 [\eth^a,\eth^c] \phi + \frac{1}{R^2} (x_a \eth_c - x_c \eth_a )\phi  \nn
\\
 &\qquad +\frac{1}{r^4 R^6} x_b x_d 
 \left(
 -x^c x^{c'} \{\phi , \{\theta^{ab},\theta^{c' d} \}\}
+x^a x^{a'} \{ \phi ,\{\theta^{cb} ,\theta^{a' d} \}\} 
 \right) \nn  \\
&=
 [\eth^a,\eth^c] \phi + \frac{1}{R^2} (x_a \eth_c - x_c \eth_a )\phi  \nn\\ 
&=
-\frac{1}{r^2 R^2 } \left( \{\theta^{ac} , \phi  \}  - r^2 (x_a 
\eth_c - x_c \eth_a )\phi  \right)  \nn\\
&=
-\frac{1}{r^2 R^2 } P^{aa'} P^{cc'}\{\theta^{a'c'} , \phi  \} 
=P^{aa'} P^{cc'}[\eth^{a'},\eth^{c'}] \phi
\end{align}
Hence, the $\nabla_a$ commute on scalar functions. For generic tensor fields 
$\phi_{b_1 \ldots b_n} $ we have to be more careful and proceed as follows:
\begin{align}
 \nabla_a \nabla_c \phi_{b_1 \ldots b_n} 
 &= \nabla_a \left(  P^{b_1 b'_1}\cdots P^{b_n b'_n} \eth_c
\phi_{b'_1 \ldots b'_n} \right)  \nn \\
&=
P^{cc'} P^{b_1 b'_1}\cdots P^{b_n b'_n}  \eth_a
\left( P^{b'_1 b''_1}\cdots P^{b'_n b''_n}  \,
\eth_{c'}
\phi_{b''_1 \ldots b''_n} \right) \\
&= P^{cc'} P^{b_1 b'_1}\cdots P^{b_n b'_n} 
 \,  \eth_a \eth_{c'} \phi_{b'_1 \ldots b'_n} +
 P^{b_1 b'_1}\cdots P^{b_n b'_n}  \eth_a
\left( P^{b'_1 b''_1}\cdots P^{b'_n b''_n} \right)  \,
\eth_c
\phi_{b''_1 \ldots b''_n}  \nn
\end{align}
Inspecting the second term in more detail, we arrive at
\begin{align}
  \prod_{k=1}^n P^{b_k b'_k} 
  \eth_a
\left(
\prod_{j=1}^n P^{b'_j b''_j} 
\right)  \eth_c
\phi_{b''_1 \ldots b''_n}  
&= \prod_{k=1}^n P^{b_k b'_k} 
\sum_{j=1}^n \left(  (\eth_a P^{b'_j b''_j})   \prod_{i\neq j} P^{b'_i 
b''_i} \right) \eth_c
\phi_{b''_1 \ldots b''_n} \nn \\
&= \frac{1}{R^2} 
\sum_{j=1}^n  \left(  P^{a b_j} x^{b''_j}    \prod_{i\neq j} 
P^{b_i b''_i} \right) \eth_c
\phi_{b''_1 \ldots b''_n} \nn \\
&=  \frac{1}{R^2} 
\sum_{j=1}^n  \bigg(  P^{a b_j}     \prod_{i\neq j} 
P^{b_i b''_i}   
\underbrace{x^{b''_j} \eth_c \phi_{b''_1 \ldots b''_n}}_{
-\phi_{b''_1 \ldots b''_{j-1}  c  b''_{j+1} b''_n}  }  \bigg) \nn \\
&= -\frac{1}{R^2} 
\sum_{j=1}^n    P^{a b_j}   
\phi_{b_1 \ldots b_{j-1}  c  b_{j+1} b_n}   \; .
\end{align}
Consequently, the commutator looks as follows:
\begin{align}
 [\nabla_a, \nabla_c] \phi_{b_1 \ldots b_n} 
&= \prod_{j=1}^n P^{b_j b'_j}
  P^{aa'} P^{cc'} \left(  [\eth_{a'}, \eth_{c'}] \phi_{b'_1 \ldots b'_n} 
\right)  \\
 &\qquad
 -\frac{1}{R^2} 
\sum_{j=1}^n   \left(  P^{a b_j}   \phi_{b_1 \ldots b_{j-1}  c  b_{j+1} b_n}
-P^{c b_j}   \phi_{b_1 \ldots b_{j-1}  a  b_{j+1}\ldots b_n} \right) \nn \,.
\end{align}
With a little relabeling, we obtain
\begin{align}
 [\nabla_a, \nabla_c] \phi_{b_1 \ldots b_n} &=
-\frac{1}{r^2R^2} \prod_{j=1}^n P^{b_j b'_j}
  P^{aa'} P^{cc'} \{\theta^{a'c'}, \phi_{b'_1 \ldots b'_n} \} \nn \\
 &\qquad -\frac{1}{R^2} 
\sum_{j=1}^n   \left(  P^{a b_j}   \phi_{b_1 \ldots b_{j-1}  c  b_{j+1}\ldots 
b_n}
-P^{c b_j}   \phi_{b_1 \ldots b_{j-1}  a  b_{j+1}\ldots b_n} \right) \nn \\
&= \frac{1}{R^2} \prod_{j=1}^n P^{b_j b'_j}
  P^{aa'} P^{cc'} \{m^{a'c'}, \phi_{b'_1 \ldots b'_n} \} \nn \\
 &\qquad 
 -\frac{1}{R^2} 
\sum_{j=1}^n   \left(  P^{a b_j} P^{cd}  
-P^{c b_j} P^{ad}   \right)
\phi_{b_1 \ldots b_{j-1}  d  b_{j+1} \ldots
b_n} \\
&\equiv \cR_{ac} \phi_{b_1 \ldots b_n} \,. \nn
\label{curvature-general}
\end{align}
For an ordinary tensor field $\phi_{b_1 \ldots b_n} \in\cC^0$, the first term 
coincides with $\nabla_{[\eth_a,\eth_c]}$ 
due to \eqref{del-del-CR}, which means that curvature coincides with that of the 
Levi--Civita connection on $H^4$,
\begin{align}
 \cR_{ab} \phi_{b_1 \ldots b_n} 
 &=\sum_{j=1}^n  \cR_{ac;b_jd}\phi_{b_1 \ldots b_{j-1}  d  b_{j+1} \ldots b_n} \;,  \nn\\
 \cR_{ac;bd} &= -\frac{1}{R^2} \left(  P_{a b} P_{cd}  -P_{c b} P_{ad} \right) 
\,.
\end{align}
As a further check, consider $\phi_{b_1 b_2}=\theta^{b_1 b_2} \in \cC^1$, where
both contributions in \eqref{curvature-general} are non-vanishing  but cancel:
\begin{align}
[\nabla_a, \nabla_c] \theta^{b_1 b_2} &= -\frac{1}{r^2R^2}  P^{b_1 b'_1}P^{b_2 
b'_2}
  P^{aa'} P^{cc'} \{\theta^{a'c'}, \theta^{b'_1 b'_2} \} \nn \\
  &\qquad -\frac{1}{R^2} 
   \left(  P^{a b_1} P^{cd}  
-P^{c b_1} P^{ad}   \right)
\theta^{ d  b_{2}}
 -\frac{1}{R^2} 
   \left(  P^{a b_2} P^{cd}  
-P^{c b_2} P^{ad}   \right)
\theta^{b_1  d }
  \nn \\
 &= \frac{1}{R^2}P^{b_1 b'_1}P^{b_2 b'_2}  P^{aa'} P^{cc'} 
  \left( \eta_{a'b'_1} \theta^{c'b'_2} - \eta_{a'b'_2} \theta^{c'b'_1}
  -\eta_{c'b'_1} \theta^{a'b'_2} +\eta_{c'b'_2} \theta^{a'b'_1} \right) \nn \\
  &\qquad 
  -\frac{1}{R^2} 
   \left(  P^{a b_1} \theta^{ c  b_{2}}  -P^{c b_1}\theta^{ a b_{2}}  
    +  P^{a b_2}   \theta^{b_1  c } -P^{c b_2} \theta^{b_1  a }\right)
  \nn \\
  &=0 \; 
\end{align}
as it must, since $\nabla \theta^{b_1 b_2}=0$.
Similarly, we can check
\begin{align}
 [\nabla_a, \nabla_c] x^b &= -\frac{1}{r^2 R^2} P^{aa'}P^{bb'}P^{cc'} 
\{ \theta^{a'c'}, x^{b'}\}  
-\frac{1}{R^2} (P^{ab} P^{cd} - P^{cb} P^{ad})x^d \nn \\
&= \frac{1}{ R^2} P^{aa'}P^{bb'}P^{cc'} 
 \left( \theta^{a'b'} x^{c'} - \theta^{c'b'} x^{a'} \right) =0 \; .
\end{align}
\paragraph{Identities for spin 1 fields.}
In order to derive \eqref{x-phi-comm-spin1},  consider
\begin{align}
 2\{x_a,\phi^{(1)}\} &= \{x_a,\cF_{bc} \theta^{bc}\} 
  = \cF_{bc} \{x_a,\theta^{bc}\}  +  \{x_a,\cF_{bc}\} \theta^{bc}  \nn\\
   &=  r^2\cF_{bc}(\eta_{ab}x^c - \eta_{ca} x^b)  + \{x_a,\cF_{bc}\} \theta^{bc} 
  \nn\\
   &=  r^2 (\cF_{ac}x^c -\cF_{ba} x^b)  + \{x_a,\cF_{bc}\} \theta^{bc}   \nn\\
   &= 2r^2 \phi_a + \{x_a,\cF_{bc}\} \theta^{bc} \nn\\
   &= 2r^2 \phi_a + \eth_d\cF_{bc} \theta^{ad}\theta^{bc} \ ,
   \label{eq:aux_calc_spin1_1}
\end{align}
where one recalls $\theta^{bc} = -r^2 \cM^{bc}$.
The averaged second term can be evaluated as follows
\begin{align}
 \eth_d\cF_{bc} [\theta^{ad}\theta^{bc}]_0
  &= \frac{R^2r^2}{3}\eth_d\cF_{bc}(P_{ab}P_{cd} -  P_{ac}P_{bd}
 + \frac{x^e}{R}\varepsilon_{adbce}) \nn\\
  &=  \frac{2 R^2r^2}{3} P_{ab}\eth^c\cF_{bc}  
  + \frac{Rr^2}{3}x^e\varepsilon_{adbce}\eth_d\cF_{bc}    \nn\\
  &=  \frac{2R^2r^2}{3} P_{ab}\eth^c(\eth_c \phi_b - \eth_b\phi_c)
  + \frac{2Rr^2}{3}x^e\varepsilon_{adbce}\eth_d\eth_c \phi_b \nn\\
  &=  \frac{2R^2r^2}{3} P_{ab}\eth^c\eth_c \phi_b 
    - \frac {1}{3} \Big(2P_{ab}\{\theta_{bc},\phi_c\} 
  - \frac{x^e}{R} \varepsilon_{adcbe} \{\theta_{dc}, \phi_b\}\Big) \nn\\
 &=  \frac{2R^2r^2}{3} P_{ab}\eth^c\eth_c \phi_b 
    - \frac {2}{3}r^2 \phi_a
    \label{eq:aux_calc_spin1_2}
\end{align}
using \eqref{theta-phi-id},  \eqref{del-del-CR} and $\eth^c\phi_c=0$, 
self-duality \eqref{theta-del-SD} and the identity \eqref{SD-id-del-phi}.
Noting that 
\begin{align}
 x^b \eth^c\eth_c \phi_b =  \eth^c\eth_c x^b \phi_b - 2 \eta^{bc} \eth_c \phi_b 
= 0
 \label{xdeldelphi}
\end{align}
we obtain
\begin{align}
 P_{ab}\eth^c\eth_c \phi_b = \eth^c\eth_c \phi_a \; ,
\end{align}
i.e. $\Box$ respects divergence-free tangential vector fields.
Collecting all the pieces, one obtains \eqref{x-phi-comm-spin1}.
\paragraph{Identities for spin $s$ fields.}
The following identity holds for any tangential traceless divergence-free spin $s$ field $\phi_a \in \cC$:
\begin{align}
 \int P^{ac}\eth_b \phi_a \eth^b \phi_c &= \int \eth_b \phi_a \eth^b \phi^a + \frac 1{R^2} (x^a\eth^b\phi_a) (x^c\eth_b\phi_c) \nn\\
 &= \int \eth_b \phi_a \eth^b \phi^a + \frac 1{R^2} \phi_b \phi^b  \nn\\
 \int \eth^a f \eth_a g &= - \int  f \eth^a\eth_a g  \nn\\
 \int  \eth^d \phi_a  \eth^a\phi_d &= - \int \phi_a   \eth^d \eth^a\phi_d - \frac{4}{R^2}\int x^d \phi_a  \eth^a\phi_d \nn\\
  &= - \int \phi_a  [\eth^d, \eth^a]\phi_d + \frac{4}{R^2}\int \eth^a x^d \phi_a  \phi_d  \nn\\
  &= -\frac{1}{r^2 R^2} \int \phi_a  \{\theta^{ad},\phi_d\} + \frac{4}{R^2}\int   \phi^a  \phi_a  \nn\\
 \int \frac{x^e}{R} \varepsilon^{abcde} \eth_b \phi_a \eth_d \phi_c   
  &= - \int \frac{x^e}{R} \varepsilon^{abcde} \phi_a  \eth_b\eth_d \phi_c   
   = \int \frac{x^e}{2r^2R^3} \varepsilon^{abcde} \phi_a \{\theta_{bd}, \phi_c\}   
   \label{int-tensor-id}
\end{align}
and $x^a\eth_a\phi = 0$. Here $\eth\cdot\eth$
is the Euclidean Laplacian on $H^4$.
Further, using the self-duality
\begin{align}
  \frac 1{2R} \varepsilon_{adcbe} \{x^e\theta_{dc}, \phi_b\}
   =  \{\theta_{ab}, \phi_b\}
  \label{theta-del-SD}
\end{align}
we have the identity
\begin{align}
P_{ab}\{\theta_{bc},\phi_c\} - \frac{x^e}{2R} \varepsilon_{adcbe} \{\theta_{dc}, \phi_b\}
 &= \frac 1{2R}P_{aa'} \varepsilon_{a'dcbe} \theta_{dc}\{x^e, \phi_b\}  \nn\\
  &= \frac 1{2R} P_{aa'}\varepsilon_{a'dcbe} \theta_{dc} \theta^{ef}\eth_f \phi_b \nn\\
  &= - r^2 P_{aa'}(g^{a'f} x^b -  g^{bf} x^{a'} )\eth_f \phi_b \nn\\
  &= - r^2 P_{af} x^b \eth_f \phi_b =  r^2 \phi_a
  \label{SD-id-del-phi}
\end{align}
using irreducibility, \eqref{eq:xx=R} and
\begin{align}
 \varepsilon_{abcde} \theta^{cd} \theta^{ef} &= 
 \varepsilon_{abcde} \theta^{cd} \{x^e , x^f\}  \notag \\
 &= \varepsilon_{abcde} \left( 
  \{ \theta^{cd} x^e , x^f\} - x^e  \{ \theta^{cd} , x^f\}  
 \right)  \notag \\
 &= 2R \{\theta^{ab} , x^f \} +r^2 \varepsilon_{abcde} x^e (\eta_{cf} x^d - 
\eta_{fd} x^c) \notag \\
&=-2r^2 R (\eta_{af} x^b - \eta_{bf} x^a) \notag \\
  \varepsilon_{eadcb} \theta^{dc} \theta^{ea} &= -8 r^2 R x^b 
  \label{epsilonMM-id-3}
\end{align}
which is \eqref{epsilonMM-id-2}. 
Note that \eqref{SD-id-del-phi} holds for any divergence-free, tangential 
$\phi_b \in \cC$.
\paragraph{Graviton identity}
The following identity will be useful
\begin{align}
 \eth_a H^{ab}[\cA] &= \eth_a(\theta^{c a}\{\cA_c,x^b\} +\theta^{c 
b}\{\cA_c,x^a\})  \nn\\
   &= (\eth_a\theta^{c a})\{\cA_c,x^b\} + \theta^{c a}\eth_a\{\cA_c,x^b\} 
   +(\eth_a\theta^{c b})\{\cA_c,x^a\}) +\theta^{c b}\eth_a \{\cA_c,x^a\}  \nn\\
   &=  \eth_a\theta^{c a}\{\cA_c,x^b\} + \{x^c,\{\cA_c,x^b\}\} 
   +(\eth_a\theta^{c b})\{\cA_c,x^a\}  \nn\\
   &=  - \{\cA_c,\{x^b,x^c\}\} - \{x^b,\{x^c,\cA_c\}\} + 
\frac{1}{R^2}(\theta^{ac} x^b - \theta^{ab} x^c) \{\cA_c,x^a\}  \nn\\
   &= \cI(\cA_b)- \{x^b,\{x^c,\cA_c\}\} - r^2 ( x^b \eth^c\cA_c - x^c 
\eth_b\cA_c)  \nn\\
   &= \tilde\cI(\cA_b)- r^2 \cA_b  - \{x^b,\{x^c,\cA_c\}\} 
   \label{del-Hab-id}
\end{align}
for tangential $\cA_a$, using \eqref{del-Q-id} and \eqref{I-radial-id};
note that $\eth_a$ respects the projection $[\cdot]_0$.

\subsection{Casimirs, positivity, and eigenvalues of 
\texorpdfstring{$\cD^2$}{D2}}
\label{sec:positivity}
In order to show that the kinetic term is positive, we need some 
positivity results.
A first result for spin 1 is the following. Assume that $\phi^{(1)}$ is Hermitian and determined by the  tangential  
divergence-free vector field $\phi_a$ as in \eqref{phi-1-def}. Then
\begin{align}
 0 \leq \int  \phi^{(1)}  \phi^{(1)} &= \int \{X_a,\phi^a\} \phi^{(1)} = -\int \phi^a \{X_a,\phi^{(1)}\} \nn\\
 &= \frac{r^2}3\int  \phi^a (-2 - R^2\eth\cdot\eth) \phi_a 
 \label{positivity-spin1}
\end{align}
This implies that  $(\Box -2 r^2)\phi_a$
is positive for divergence-free square-integrable tangential tensor fields, cf. \eqref{phi-spin1-normalization}.
In particular, this gives 
\begin{align}
 \hat\a_1 \phi_a  = \frac{r^2}3 (-R^2 \eth\cdot\eth -2) \phi_a , \qquad \phi_a\in\cC^1, \qquad \hat\a_1 \geq 0 \ .
\end{align}
We also observe 
\begin{align}
 \int \phi^{(s)} \Box\phi^{(s)} \propto \int \eth_a\phi^{(s)} \eth^a\phi^{(s)} \ \geq 0, \qquad \phi^{(s)} \in \cC^s
\end{align}
using \eq{partint}, since $x\cdot \eth = 0$, i.e.\ $\eth\phi$ has no 
radial components, hence the metric is Euclidean.
Therefore $\Box$ is a positive operator on any square-integrable $\phi\in\cC^s$.

For higher spin, we  need the following intertwining property of the vector fluctuations 
\begin{align}
 r^2 C^2[\mso(4,1)]^{\rm (full)} \cA^a[\phi^{(s)}] &= -(\Box  + 2 \cI - r^2(\cS^2 +4)) \cA_a[\phi^{(s)}] \nn\\
  &=  \cA^a[r^2 C^2[\mso(4,1)]\phi^{(s)}]  \nn\\
  &=  \cA^a[r^2 C^2_{\rm full}[\mso(4,1)]\phi^{(s)}_a] = \ldots  \nn\\
  &= \cA^a[r^2 C^2_{\rm full}[\mso(4,1)]\phi_{a_1\ldots  a_s}]
  \label{C2-A-phi-general}
\end{align}
using \eqref{C2-total-id}.
The various forms on the right-hand side can be evaluated using 
the quadratic Casimir acting on the spin $s$ field $\phi^{(s)}$ in its various realizations:
\begin{align}
 -r^2C^2 \phi^{(s)} &= (\Box - r^2\cS^2) \phi^{(s)} \nn\\
 &= (\Box - 2r^2s(s+1)) \phi^{(s)}
\end{align}
and
 \begin{align}
 -r^2C^2_{\rm full} \phi^{(s)}_a &= (\Box  + 2 \cI - r^2(\cS^2 +4)) \phi^{(s)}_a \nn\\
  &= (\Box  + 2r^2(2-s) - r^2(2s(s-1) +4))\phi^{(s)}_a \nn\\
 &= (\Box  - 2r^2s^2)\phi^{(s)}_a 
 \end{align}
and similarly
\begin{align}
- r^2 C^2_{\rm full} \phi_{a_1\ldots a_s} &= (\Box  + 2 \cI - 
r^2C^2[\mso(4,1)]_{(s,0)}) \phi_{a_1\ldots a_s}]  \nn\\
 &= (\Box  + 2 r^2 s - r^2 s(s+3))\phi_{a_1\ldots a_s} \nn\\ 
  &=  (\Box -  r^2 s(s+1))\phi_{a_1\ldots a_s}
 \label{BoxA4-2}
\end{align}
because $\cS^2 = 0$ on $\phi_{a_1\ldots a_s}\in\cC^0$.
Here we define a generalized intertwiner $\cI$ for spin $s$ tensor fields 
\begin{align}
 \cI(\phi_{a_1\ldots a_s}) &\coloneqq\{\theta^{aa_1},\phi_{a_1\ldots a_s}\} + 
\ldots  + \{\theta^{aa_s},\phi_{a_1\ldots a_s}\}
  = s \{\theta^{aa_1},\phi_{a_1\ldots a_s}\} \nn\\
  &= r^2 s\, \phi_{a_1\ldots a_s}
\end{align}
for symmetric $\phi_{a_1\ldots a_s}$, and the Casimir of $SO(4,1)$ on its 
indices in $(5)^{\otimes_S s}$ is
\begin{align}
 C^2[\mso(4,1)]_{(s,0)} = s(s+3) \ .
\end{align}
This is consistent with \eqref{BoxA4-1} for $s=1$.
In particular, the action of $\Box$  on various realizations 
of the same spin $s$ field is related as follows 
\begin{align}
    \cA[(\Box -2 r^2 s(s+1)) \phi^{(s)}] =  \cA[(\Box -2r^2s^2) \phi^{(s)}_a] 
    =  \cA[(\Box - r^2 s(s+1))\phi_{a_1\ldots a_s}] \ .
    \label{Box-Phi-intertwiner}
\end{align}
Now we can evaluate \eqref{C2-A-phi-general} for the individual spin $s$ modes. 
For $\cB^{(2)}$, we obtain
\begin{align}
 (\Box  + 2 \cI - r^2(\cS^2 +4)) \cB^{(2)}_a[\phi^{(s)}_a] 
  &= \cB^{(2)}_a[(\Box  + 2 \cI - r^2(\cS^2 +4)) \phi^{(s)}_a] \;, \nn\\
(\Box  - 2r^2(s+1)^2 \cB^{(2)}_a[\phi^{(s)}_a] &= 
\cB^{(2)}_a[(\Box-2r^2s^2)\phi^{(s)}_a] \; ,\nn\\
\cD^2 \cB^{(2)}_a[\phi^{(s)}_a]  =(\Box -2\cI+4r^2) \cB^{(2)}_a[\phi^{(s)}_a]
  &= \cB^{(2)}_a[(\Box + 2r^2s)\phi^{(s)}_a]  \; .
\end{align}
For $\cB^{(4)}$, we obtain
 \begin{align}
 (\Box  + 2 \cI - r^2(\cS^2 +4)) \cB^{(4)}_a[\phi^{(s)}_a] &= \cB^{(4)}_a[(\Box 
 + 2 \cI - r^2(\cS^2 +4)) \phi^{(s)}_a]  \; ,\nn\\
 (\Box  - 2r^2 s^2) \cB^{(4)}_a[\phi^{(s)}_a] &= \cB^{(4)}_a[ (\Box  - 
2r^2s^2)\phi^{(s)}_a] \; , \nn\\
 \cD^2 \cB^{(4)}_a[\phi^{(s)}_a] =
 (\Box - 2 \cI +4r^2) \cB^{(4)}_a[\phi^{(s)}_a] &= \cB^{(4)}_a[(\Box - 2r^2  s) 
\phi^{(s)}_a] \; .
\label{BoxA4-1}
 \end{align}
For $\cB^{(1)}$ we obtain similarly
\begin{align}
 (\Box  + 2 \cI - r^2(\cS^2 +4)) \cB^{(1)}_a[\phi^{(s)}_a] &= 
\cB^{(1)}_a[(\Box + 2 \cI - r^2(\cS^2 +4)) \phi^{(s)}_a] \; , \nn\\
 \cD^2  \cB^{(1)}_a[\phi^{(s)}] = (\Box + 2r^2(1+s)) \cB^{(1)}_a[\phi^{(s)}] 
           &=\cB^{(1)}_a[(\Box+ 2r^2(3s+2)) \phi^{(s)}_a] \; ,
 \end{align}
and finally for  $\cB^{(3)}$
\begin{align}
 \cD^2 \cB^{(3)}_a[\phi^{(s)}_a] = (\Box -2\cI+4r^2) \cB^{(3)}_a[\phi^{(s)}_a] 
     &= \cB^{(3)}_a[(\Box +2r^2 s) \phi^{(s)}_a] \; .
\end{align}
%
%
\bibliographystyle{JHEP}
\bibliography{papers}
\end{document}